\iffalse
\documentclass[aps,prl,onecolumn,12pt,
			   groupedaddress,superscriptaddress,
			   amsfonts,amssymb,amsmath,
			   citeautoscript,
			   a4paper]{revtex4-1}
\else
\documentclass[aps,pra,reprint,
			   groupedaddress,superscriptaddress,notitlepage,
			   amsfonts,amssymb,amsmath,
			   citeautoscript,
			   a4paper]{revtex4-1}
\fi

\usepackage[pdftex]{hyperref}
\hypersetup{colorlinks,
			linkcolor={blue!75!black!80!yellow},
			citecolor={blue!75!black!80!yellow},
			urlcolor={blue!75!black!80!yellow},
			pdfstartview=FitH}
\usepackage{enumitem}
\usepackage{graphicx}
\usepackage{mathrsfs}
\usepackage{amsthm}
\usepackage{physics}
\usepackage{xspace}
\usepackage{braket}
\usepackage{xr}
\usepackage{gensymb} 
\usepackage{xcolor,soul}
\usepackage{stmaryrd} 
\usepackage[UKenglish]{babel}
\usepackage{placeins} 
\usepackage{siunitx}
\DeclareSIUnit[number-unit-product=]\percent{\char`\%} 
\usepackage{makecell}
\usepackage{diagbox}
\usepackage{colortbl}

\usepackage{txfonts}  
\usepackage{txfontsb} 

\makeatletter
\def\mathcolor#1#{\@mathcolor{#1}}
\def\@mathcolor#1#2#3{%
  \protect\leavevmode
  \begingroup
    \color#1{#2}#3%
  \endgroup
}
\makeatother

\newcommand*\diff{\mathop{}\mathrm{d}}

\newcommand{\iu}{\mathrm{i}}
\newcommand{\e}{\mathrm{e}}

\newcommand{\appropto}{\mathrel{\vcenter{
			\offinterlineskip\halign{\hfil$##$\cr
				\propto\cr\noalign{\kern2pt}\sim\cr\noalign{\kern-2pt}}}}}

\newcommand{\matr}[1]{\boldsymbol{#1}}     

\newcommand{\sigmaz}{\sigma_z}

\newcommand{\ie}{i.e.\@\xspace}  

\newcommand{\eg}{e.g.\@\xspace}

\makeatletter
\newcommand*{\addFileDependency}[1]{
  \typeout{(#1)}
  \@addtofilelist{#1}
  \IfFileExists{#1}{}{\typeout{No file #1.}}
}
\makeatother

\usepackage{scalerel}

\usepackage{textcomp} 
\usepackage{xifthen}
\usepackage{etoolbox}
\newboolean{togglecomments}
\newboolean{togglechanges} 

\setboolean{togglecomments}{false}
\setboolean{togglechanges}{true}

\newcommand{\comment}[2]{%
    \ifbool{togglecomments}%
    {\textcolor{blue!70!black}{\small\textsf{%
    \textsuperscript{\textsc{\textsf{\MakeLowercase{#1}}}}%
    [#2]}}} 
    {}}     
\newcommand{\swap}[2]{\ifbool{togglechanges}
    {#2}  
    {\textcolor{red!70!black}{[#1]}\textrightarrow{}\textcolor{green!50!black}{[#2]}}}
\newcommand{\remove}[1]{\ifbool{togglechanges}
    {}    
    {\textcolor{red!70!black}{#1}}}
\newcommand{\inset}[1]{\ifbool{togglechanges}
    {#1}  
    {\textcolor{green!50!black}{#1}}}
\newcommand{\optional}[1]{\ifbool{togglechanges}
    {#1}  
    {\textcolor{gray!30!orange!80!gray}{#1}}}
\newcommand{\citeremind}[1]{%
    [\textcolor{blue!75!black!80!yellow}{
        $\blacksquare$%
           \ifthenelse{\isempty{#1}}
               {}
               {\textsuperscript{\tiny\textsf{#1}}}%
        }]\xspace}
\newcommand{\todo}[1]{
    \textcolor{orange!80!yellow!95!black}{\textbf{[}%
        \ifthenelse{\isempty{#1}}%
        {\text{$\blacksquare$}}%
        {{\small\textsf{#1}}}%
        \textbf{]}}}
        
\begin{document}

\title{Non-Abelian physics in light and sound}

\author{Yi~Yang}
\email{yiyg@hku.hk}
\affiliation{Department of Physics, The University of Hong Kong, Pokfulam, Hong Kong, China}

\author{Biao Yang}
\email{yangbiaocam@nudt.edu.cn}
\affiliation{College of Advanced Interdisciplinary Studies, National University of Defense Technology, Changsha 410073, China}

\author{Guancong Ma}
\email{phgcma@hkbu.edu.hk}
\affiliation{Department of Physics, Hong Kong Baptist University, Kowloon Tong, Hong Kong, China}

\author{Jensen Li}
\affiliation{Department of Physics, Hong Kong University of Science and Technology, Clear Water Bay, Hong Kong, China}

\author{Shuang Zhang}
\affiliation{Department of Physics, The University of Hong Kong, Pokfulam, Hong Kong, China}
\affiliation{Department of Electrical \& Electronic Engineering, The University of Hong Kong, Pokfulam, Hong Kong,
China}

\author{C.~T. Chan}
\affiliation{Department of Physics, Hong Kong University of Science and Technology, Clear Water Bay, Hong Kong, China}

\date{\today} 

\begin{abstract}
    There has been a recent surge of interest in using light and sound as platforms for studying non-Abelian physics. Through a kaleidoscope of physical effects, light and sound provide diverse ways to manipulate their degrees of freedom to constitute the Hilbert space for demonstrating non-Abelian phenomena. The review aims to provide a timely and comprehensive account of this emerging topic. Starting from the foundation of matrix-valued geometric phases, we cover non-Abelian topological charges, non-Abelian gauge fields, non-Abelian braiding, non-Hermitian non-Abelian phenomena, and their realizations with photonics and acoustics. This topic is fast evolving at the intersection of atomic, molecular, optical physics, condensed matter physics, and mathematical physics, with fascinating prospects ahead.
\end{abstract}

\maketitle

Non-Abelian phenomena are ubiquitous among different branches of physics, ranging from the non-commutative rotations of a classical rigid body in three dimensions to the non-Abelian anyonic excitations in quantum systems. 
Light and sound are becoming an ideal playground for exploring non-Abelian phenomena because they contain many degrees of freedom that can be effectively engineered across different frequency regimes.
Non-Abelian geometric phases, the matrix generalization to the better-known, scalar Berry phase, lie at the heart of this emerging topic.


Consider an $n$-dimensional eigenstate $\ket{\matr{\psi}}=\ket{\psi_1,\psi_2,\cdots,\psi_n}$ of a classical or quantum dynamic system. We can define an $n\times n$ matrix-valued connection $\matr{A}$ along a path $\mathbf{r}$ in parameter space as $\matr{A}\equiv\iu\braket{\matr{\psi}(\matr{r})|\partial_{\matr{r}}|\matr{\psi}(\matr{r})}$, which is known as the Wilczek--Zee or Mead--Berry connection~\cite{wilczek1984appearance,mead1992geometric,bohm2013geometric}.
For this connection, a two-form curvature can be defined as~\cite{bohm2013geometric}
$
F_{\mu\nu}= \partial_\mu A_\nu - \partial_\nu A_\mu - \iu\left[A_\mu,A_\nu\right],
$
where $\mu$ and $\nu$ can be position coordinates, momenta, or general parameters. The first two terms of the curvature represent the conventional part that resembles the magnetic field in Maxwell's equations, while the last term is the manifestation of non-Abelian physics due to non-commutative actions between two different components of the connection.
Notably, unlike the scalar Berry curvature, this matrix-valued curvature becomes gauge covariant. 
To obtain the non-Abelian geometric phase, one needs to perform parallel transport along a closed path $\matr{r}$ via integrating the connection $\matr{A}$: $\matr{W} \equiv \mathcal{P}\exp \iu \oint \matr{A}\diff\mathbf{r}$, where $\mathcal{P}$ indicates path-ordered integral because the connection $\matr{A}$ is matrix-valued.
Its trace, $W \equiv \mathrm{Tr}\,\matr{W}$, is the gauge-invariant Wilson loop~\cite{wilson1974confinement}.


One of the most widely known applications of the formulation above is the multi-band description of topological band theory, where bands can become fully degenerate or touch at certain points in the Brillouin zone. In momentum space, the eigenvalues of the Wilson-loop operator $\matr{W}$ are exponentials of the multiband Berry phases, which have been widely used for topology analysis for both fermionic and bosonic systems~\cite{vanderbilt2018berry,alexandradinata2020crystallographic,christensen2022location,gupta2022wannier}.
The formulation also applies to high-dimensional momentum space, where the non-Abelian Berry curvature plays an essential role in describing the non-Abelian Yang monopoles and the associated second Chern number~\cite{sugawa2018second,lu2018topological,ma2021linked}. 
However, the topological classification of matter has remained Abelian integers~\cite{chiu2016classification} until the recent discovery of non-Abelian topological charges described by matrices~\cite{RN6}.
Such novel topology has been realized in photonic and acoustic transmission-line networks and metamaterials, where site connectivity and constituent relations can be intricately tuned~\cite{RN21,RN11,RN9}. 
In those experiments, the entanglement among multiple bandgaps and the rich geometric configurations of the degenerate points were demonstrated as the consequence of the underlying non-Abelian topological invariants.


The same mathematical formulation can be applied in real space or, more generally, in parameter space, when other internal degrees of freedom are leveraged.
To see this, heuristically consider a spinful particle of mass $m$ and momentum $\matr{p}$ immersed in real-space non-Abelian vector potentials $\matr{A}(\matr{r})$: $H = \left[\matr{p}-\matr{A}(\matr{r})\right]^2/2m$. 
Evidently, non-Abelian magnetic fields (\ie curvatures) and the loop operators can both be analogously defined.
Such gauge potentials can be synthesized by various means in optics and acoustics, for example, using metamaterials with anisotropic and gyrotropic responses~\cite{chen2019non}, the splitting and degeneracy between transverse-electric (TE) and transverse-magnetic (TM) modes of polaritonic planar cavities and crystals~\cite{terccas2014non}, and gyrotropic and time-varying components in fiber optics~\cite{yang2019synthesis}. While the Abelian part of the magnetic fields generates a cyclotron motion, the non-Abelian part generates oscillatory motion, known as zitterbewegung, due to its action on both the trajectory and pseudospin~\cite{chen2019non,sedov2018zitterbewegung,polimeno2021experimental,lovet2022observation,whittaker2021optical}.
The effect of the non-commutative operations along different paths can be quite drastic, possibly ending up in totally different final states, leading to non-Abelian Aharonov--Bohm effect~\cite{osterloh2005cold,yang2019synthesis,wu2022non}, non-Abelian mode braiding~\cite{noh2020braiding,chen2022classical,zhang2022non}, and non-Abelian Thouless pumping~\cite{brosco2021non,sun2022non,you2022observation}.

This review comprises four major parts to summarize the recent advances in non-Abelian topological charges, non-Abelian gauge fields, non-Abelian mode pumping, and non-Abelian non-Hermitian phenomena on photonic and acoustic platforms.
Judiciously achieving an expanded Hilbert space, particularly via internal degrees of freedom, is a crucial prerequisite for studying non-Abelian physics in any system, as evident from the formulation of the matrix-valued geometric phases above. To this end, we start the review by summarizing various photonic and acoustic approaches in Box 1.

\begin{figure*}[htbp]
 \includegraphics[width=0.95\linewidth]{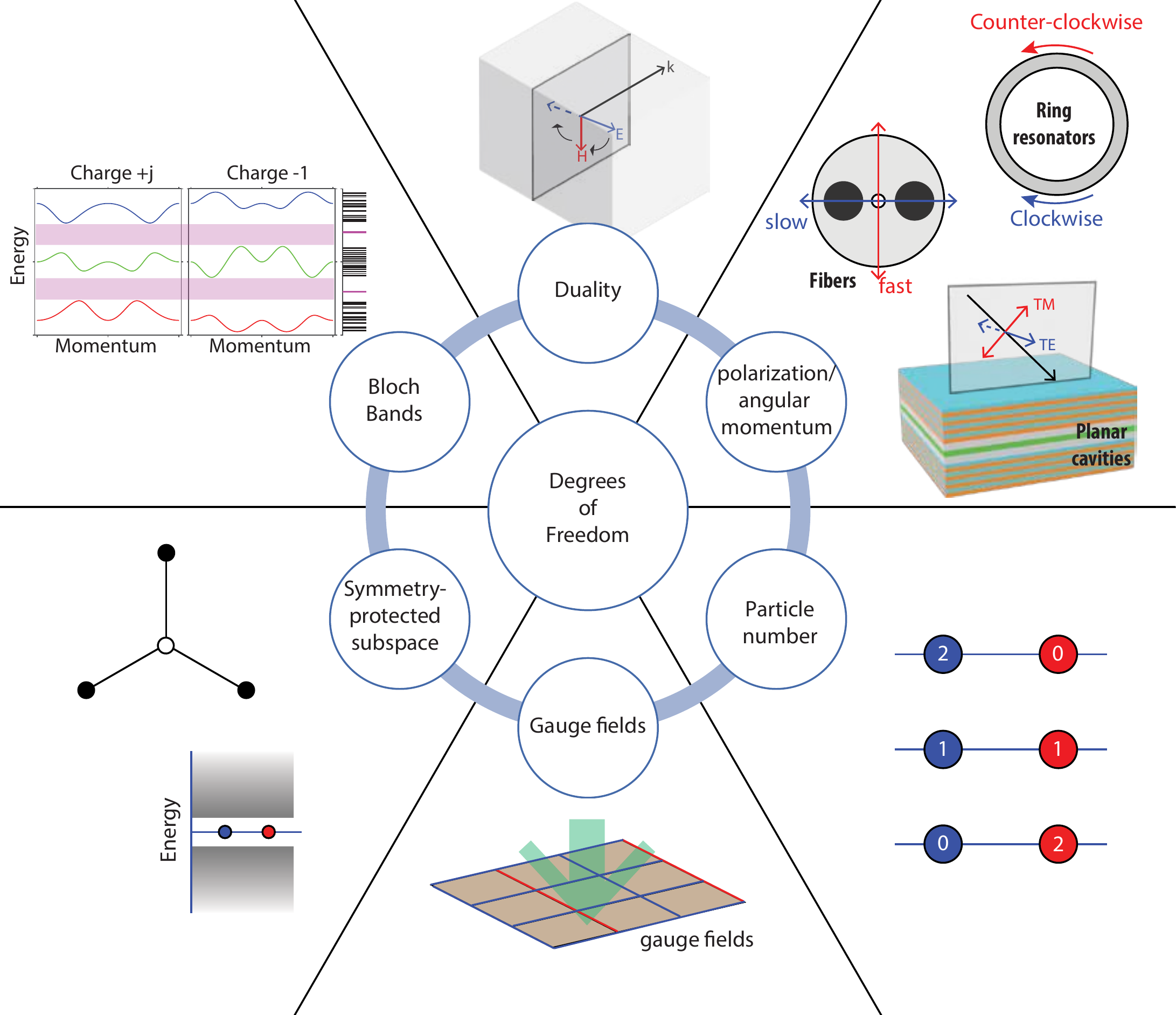}
        \caption{
        \textbf{Box: Degrees of freedoms of light and sound for non-Abelian phenomena.} 
        Many degrees of freedom (DoFs) are available in light and sound for expanding the Hilbert space.
        \textbf{Duality} - In vacuum, a strict electromagnetic duality holds between the electric and magnetic fields, \ie $\matr{E\rightarrow H}$ and $\matr{H\rightarrow -E}$. In metamaterials, this duality can be kept intact if the effective permittivity $\epsilon$ and permeability $\mu$ tensors satisfy $\epsilon=\mu$ (or proportional), which has been widely employed for creating photonic topological insulators~\cite{khanikaev2013photonic,he2016photonic,silveirinha2017p}. Similarly, this DoF can be used for building homogeneous non-Abelian metamaterials, whose requirements in 2D reduce to an in-plane duality, \ie a $2\times2$ degenerate subspace of $\epsilon$ and $\mu$~\cite{liu2015gauge,chen2019non}.
        \textbf{Polarizations and angular momentum} - Many optical structures support quasi-degenerate modes of orthogonal polarizations (also known as spin angular momentum). These include polarization-maintaining fibers and planar polaritonic microcavities. The former features quasi-degenerate modes along the slow and fast axes because of the azimuthal symmetric breaking, while the latter also has quasi-degenerate but splits transverse-electric (TE) and transverse-magnetic (TM) modes due to the 2D nature of the geometry. The coupling between the quasi-degenerate modes can be controlled, thereby enabling synthetic non-Abelian gauge fields.
        In waveguide resonators, clockwise and counter-clockwise modes, described by the azimuthal angular momentum mode number $\pm m$, can also label a pseudospin~\cite{hafezi2011robust}; the associated gauge-field scheme [SU(2) but Abelian] has been widely used in integrated photonics~\cite{hafezi2013imaging,mittal2018topological,mittal2019photonic,dai2022topologically}.
        This scheme may also be applied in the new ring-resonator platforms like the photonic-crystal and M\"{o}bius--strip micro-rings~\cite{lu2022high,wang2023experimental}.
        \textbf{Particle number} - The bosonic nature of light and sound enables convenient manipulation of their quantum particle number, a typical example of which is boson sampling. The particle number can encode spatially coupled optical modes to generate non-Abelian holonomies~\cite{neef2023three}.
        \textbf{Bloch bands} - Photonic and acoustic crystal's Bloch bands can be used for creating non-Abelian topological charges that are related to the Zak phases of multiple bandgaps (shaded purple) sandwiched by multiple bands.
        Their domain-wall states follow the unique non-Abelian quotient relation between bulks of distinct non-Abelian charges~\cite{RN21}.
        \textbf{Symmetry-protected subspace} - Symmetries can stabilize degenerate subspace that corresponds to higher-dimensional irreducible representations. For example, degenerate zero modes can be created if the number of sites in different sublattices differs (see the figure panel: one in sublattice A vs. three in sublattice B). For non-Abelian mode operations, chiral and rotational symmetries have been employed to realize this type of degeneracy in meta-atoms, electric circuits, coaxial cable networks, and coupled waveguides; see Refs.~\cite{kremer2019optimal,wu2022non,cheng2023artificial,sun2022non}.
        \textbf{Gauge-field-enabled degeneracy} - The existence of gauge fields in real space can enable the appearance of nonsymmorphic symmetries with momentum dependence, and can thus projectively alter the symmetry of a system~\cite{zhao2020z,zhao2021switching,yang2022non,xue2022projectively,li2022acoustic,meng2023spinful}. This projective symmetry could therefore provide a viable way to synthesize fermionic-like behaviors based on spatial gauge-field engineering for bosons.        
        }
\label{fig:box}
\end{figure*}

\section{Non-Abelian topological charges}

\begin{figure*}[htbp]
 \includegraphics[width=\linewidth]{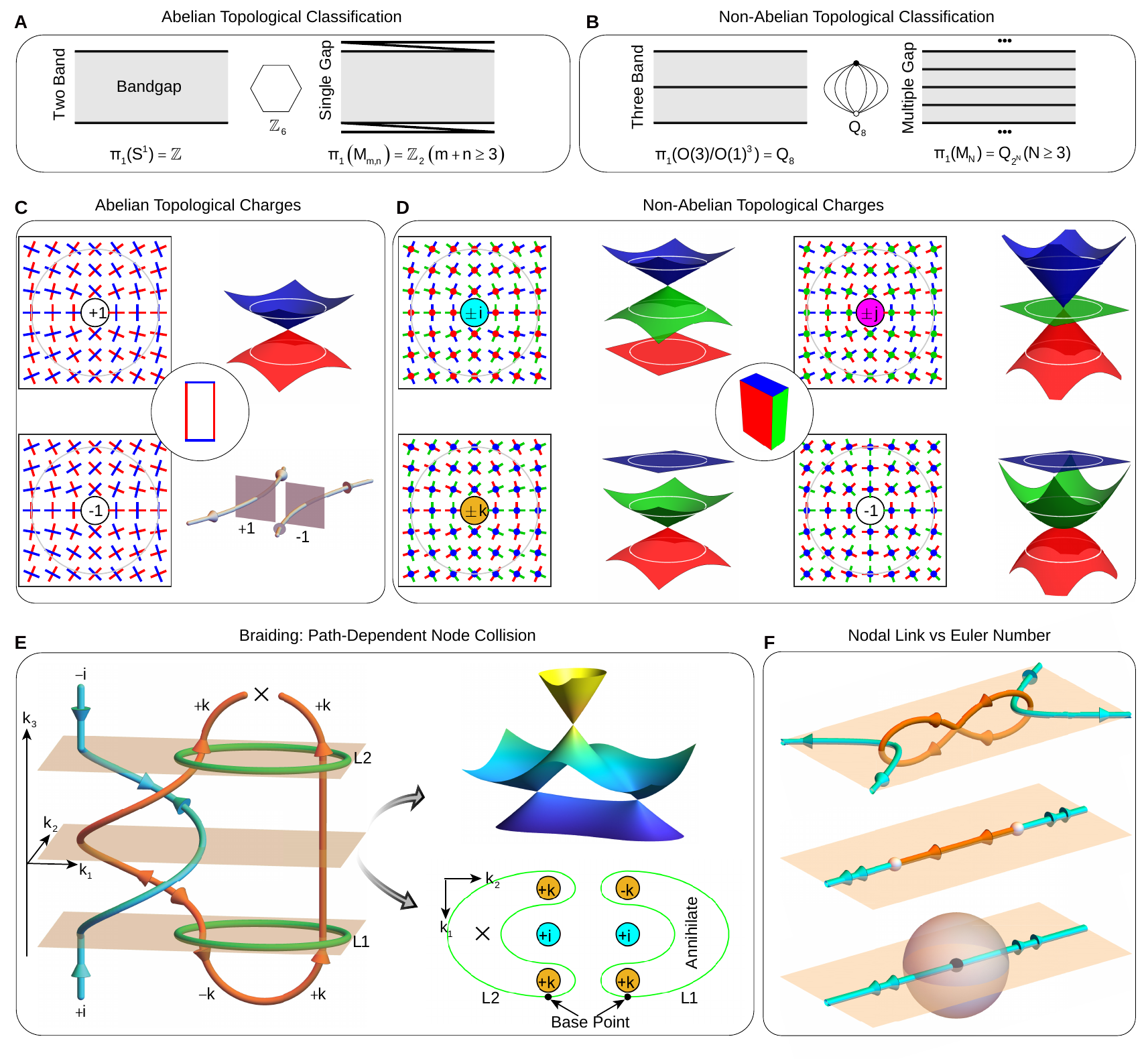}
        \caption{
        \textbf{Theoretical aspects of non-Abelian topological charges.}
        (\textbf{A}) Abelian topological classification of 1D topological bands with one single bandgap~\cite{RN6}.
        (\textbf{B}) Non-Abelian topological classification of 1D topological bands with two or more bandgaps~\cite{RN6}.
        (\textbf{C-D}) 2D extended band structure and the corresponding eigenstate frame rotations for Abelian/non-Abelian topological charges. The position and type of band degeneracies in the extended 2D systems can predict the Abelian/non-Abelian topological edge states of the 1D subsystems that are unit circles~\cite{RN21}. The insets correspond to a rectangle (2D) and a cuboid (3D), respectively, indicating the number of orthogonal eigenstates.
        (\textbf{E}) Topological constraints on multi-gap nodal line configurations and the braiding of band nodes~\cite{RN8}. 
        (\textbf{F}) Monopole charge (Euler class) and the linking structure.
        Partially adapted from Refs.~\cite{RN21,RN8}. 
        }
\label{fig:band}
\end{figure*}

\begin{figure*}[htbp]
 \includegraphics[width=\linewidth]{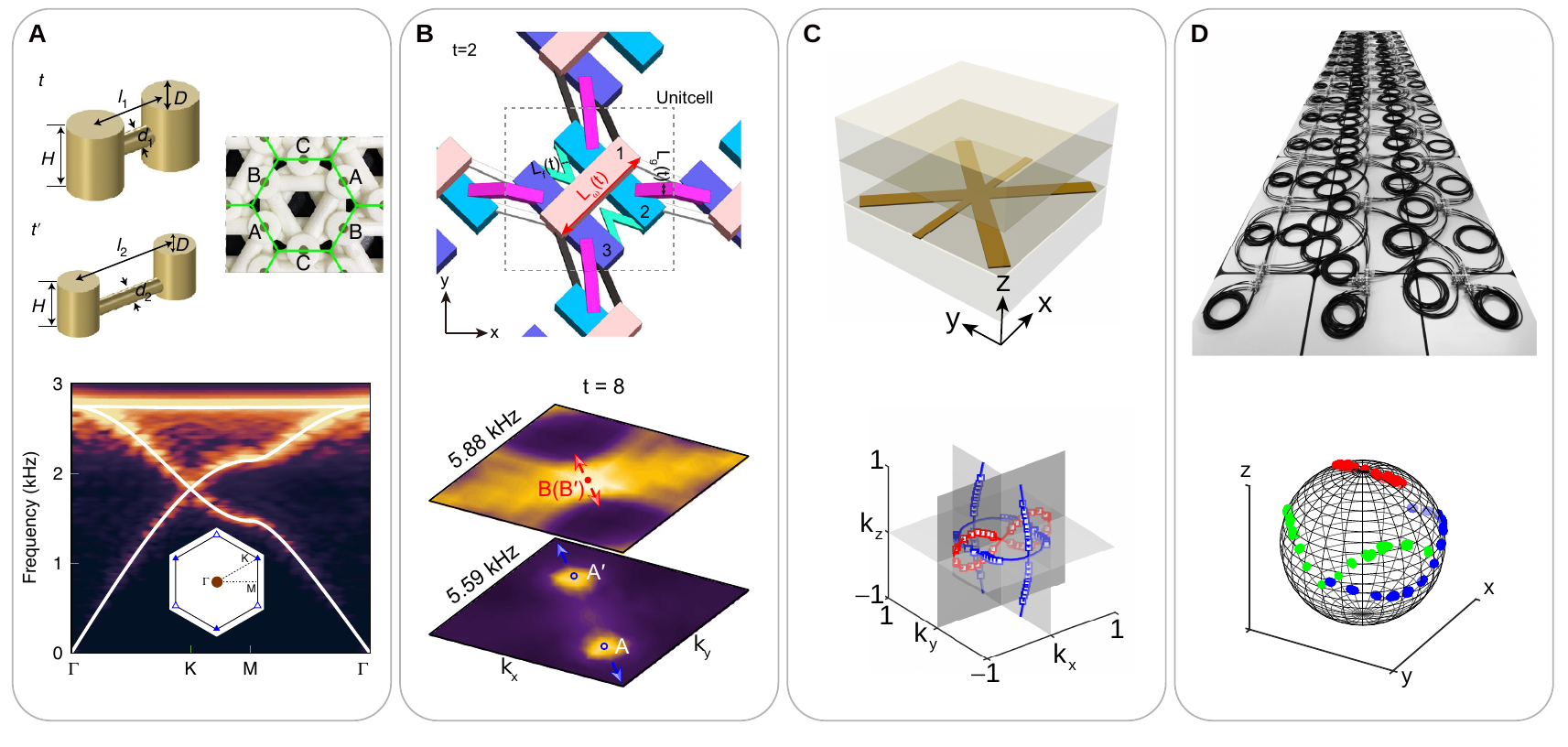}
        \caption{
        \textbf{Acoustic and photonic realizations of non-Abelian topological charges.}
        (\textbf{A}) Cylindrical acoustic resonators forming three-band kagome lattices and mapped acoustic band structures~\cite{RN9}.
        (\textbf{B}) Unit-cell structure of acoustic metamaterial and experimental characterization of the collision stability of nodes~\cite{RN10}.
        (\textbf{C}) The metallic metamaterials and experimentally probed non-Abelian nodal links~\cite{RN11}.
        (\textbf{D}) Transmission line network and experimentally mapped eigenstate frame spheres~\cite{RN21}.
        Partially adapted from Refs.~\cite{RN9,RN10,RN11,RN21}.
        }
\label{fig:exp_band}
\end{figure*}

The discovery of the integer quantum Hall effect~\cite{RN1} and the subsequent topological interpretation~\cite{RN2} ushered in a novel epoch in the field of condensed matter physics. At present, the concept of topological phases has expanded significantly beyond its initial scope within condensed matter physics, exerting considerable influence on the exploring and understanding of various topological matters~\cite{RN3,RN4}. Wherein the bulk properties of topological matters can be comprehensively categorized by employing a topological invariant, including but not limited to Chern numbers or winding numbers, which map to Abelian integer groups~\cite{RN5}. Very recently, it has been found that symmetry-protected topological phases can go beyond the Abelian classifications~\cite{RN6}, which takes topological phase classification to another level. Within this framework, bulk materials are classified by non-Abelian entities that behave like matrices (such as “quaternions”). With multiple bandgaps considered together, the non-Abelian topological invariants reveal the underlying braiding structures of topological bands. This leads to interesting observables, including trajectory-dependent Dirac collisions in two-dimensional planes~\cite{RN6,RN7,RN8,RN9,RN10,PhysRevB.105.085115,PengNC}, admissible nodal line configurations in three-dimension~\cite{RN6,RN8,RN11,RN12,RN13,RN14,RN15,PhysRevB.105.214108}, braiding of Weyl nodes or conversions between Weyl nodes and nodal loops~\cite{RN8,RN16,PhysRevB.105.L081117}, relations between monopole charge (Euler class) and the linking structure~\cite{RN7,RN12,RN17,RN18,RN19,RN20,bouhon2022multigap}, breakdown of Nielsen-Ninomiya theorem in twisted bilayer graphene~\cite{RN7}, interesting non-Abelian topological edge states~\cite{RN6,RN21,RN22}, knots and braiding structure of non-Hermitian topological bands~\cite{RN23}, and so on.

The non-Abelian topological charges were first found useful in the classification of $PT$ (the combination of parity and inversion) symmetric systems. As both $P$ and $T$ can flip the momentum, the $PT$ operator is antiunitary that preserves the momentum~\cite{RN24}. In the spinless system, the $PT$ operator can be represented by complex conjugation $K$ when a suitable basis is chosen~\cite{RN25}. Hence, under $PT$ symmetry, the Hamiltonian can be gauged to be real at all momenta $k$, i.e., $H(k)=H^*(k)$. For comparison, we first introduce the simplest Abelian topological charges protected by $PT$ symmetry, i.e. a real two-band Hamiltonian as shown in the left panel of Fig.~\ref{fig:band}A. Without loss of generality, the Hamiltonian takes the form of $H=h_x (k) \sigma_x+h_z (k) \sigma_z$. After the band flattening (which preserves the band topology), the order parameter space can be described by a normalized two-component real vector $(h_x,h_z)$ with $h_x^2+h_z^2=1$ being a circle ($S^1$). The fundamental group $\pi_1 (S^1 )=\mathbb{Z}$ characterizes the Hamiltonian by the integer group. 
The corresponding eigenstate distributions are shown in Fig.~\ref{fig:band}C left, where the eigenstates rotate clockwise or counterclockwise along the circle. One can also see such topological defects in real space in human fingerprints. The positive or negative charges define the directions of charge flow in three dimensions (see Fig.~\ref{fig:band}C right panel).
When there are multiple bands but we consider only one single bandgap as shown in Fig.~\ref{fig:band}A right, the topological classification turns out to be $\pi_1 (M_{m,n})=\mathbb{Z}_2$ with $M_{m,n}=O(m+n)/O(m) \times O(n)$, where $O(m)$ is the orthogonal group and $m$ ($n$) indicates the number of conduction  (valence) bands. The $\mathbb{Z}_2$ group still belongs to Abelian topological charges~\cite{RN25}: A corresponding real space example is the defect line in uniaxial liquid crystal, which is classified by $\pi_1 (M_{1,2})= \mathbb{Z}_2$.

When multiple bandgaps ($n\geq2$) in a multiband system are considered together, the system can be characterized by non-Abelian topological charges. For a $PT$-symmetric three-band system as shown in the left panel of Fig.~\ref{fig:band}B, the corresponding order parameter space of Hamiltonians is then $M_3=O(3)⁄(O(1)^3$. The $O(3)$ identifies the rotation of the frame formed by the three eigenstates and the quotient $O(1)^3$ originates from the fact that the sign of an eigenstate is arbitrary. The fundamental homotopy group of the Hamiltonian space is $\pi_1 (M_3)=Q_8$, where $Q_8=(+1,\pm i,\pm j,\pm k,-1)$ forms the non-Abelian quaternion group~\cite{RN6,RN26,RN27} (see Fig.~\ref{fig:band}D). This group consists of three anticommuting imaginary units satisfying $ij=k,jk=i,ki=j$ and $i^2=j^2=k^2=-1$ where the non-commuting relation reveals the braiding features of topological bands. For multiple bandgaps with $n>3$~(see Fig. \ref{fig:band}B right panel), the non-Abelian topological charges turn to be generalized quaternion groups~\cite{RN6}.

Non-Abelian topological charges have been used to study the geometry, topology, and physics of defects from a homotopy perspective. In the realm of material science, real space non-Abelian topological charges have been elegantly applied to describe the disclination line defects in biaxial nematic liquid crystals~\cite{RN6,RN26,RN28}. They are assemblies of brick-like (see the inset of Fig.~\ref{fig:band}D) molecules that self-organize to form mesophases~\cite{RN26}. At each point, the molecules collectively orient themselves along a specific direction, which locally defines an orientational order~\cite{RN29}. The topological defects consist of regions where the order locally breaks down, as shown in Fig.~\ref{fig:band}D. The ellipsoidal biaxial nematic molecule has three different principal axes related to the height, width, and length; And the three axes (red, green, and blue) define a frame that indicates the orientations of the molecule~\cite{RN6,RN21}. Let us consider a loop enclosing a defect. The molecule frame rotates along the loop, and the frame must return to itself after going around the loop. The frame rotation can be $\pi$ or $2\pi$. The charge of $+i$ corresponds to the frame rotation of $\pi$ around the red axis~\cite{RN6,RN21}. Similarly, the other two non-Abelian topological charges $+j$ and $+k$ indicate the frame rotation of $\pi$ around the green and blue axes, respectively. Therefore, the non-Abelian topological charges are also called non-commutative ``frame-rotation charge'', whose underlying topology can be explained using Dirac’s belt trick~\cite{RN30}. However, when the width and thickness of the molecule are equal, i.e. the biaxial nematic molecule becomes uniaxial, the topological defects are described by $\pi_1 (M_{1,2} )=\mathbb{Z}_2$ as mentioned above~\cite{RN6,RN26} (like rods orienting in three dimensions).

As mentioned earlier, topological defects appear not only in nematics but also in topological bands. Here the molecules (or frames) in liquid crystals can be directly mapped onto the eigenstate frames of the topological bands in three-band $PT$-symmetric systems. The topological structure of the one-dimensional bands (Fig.~\ref{fig:band}A and B) are clearer after extending the 1D Hamiltonian $H(k)$ onto a 2D plane~\cite{RN21}. Figures~\ref{fig:band}C and D show the extended 2D band structure. The original Hamiltonian $H(k)$ exactly locates on the unit circle $k_1^2+k_2^2=1$ (white/grey solid circle) of the extended Hamiltonian. Each nontrivial topological charge represents a band degeneracy in the 2D system which exhibits Dirac cone dispersion, in the range $k_1^2+k_2^2<1$ as one can see in Figs.~\ref{fig:band}C and D. They represent an obstruction that cannot be removed unless topological phase transition happens with the degeneracy point moving out of the unit circle corresponding to bandgap closing and re-opening. For Abelian topological descriptions, we consider only one bandgap. The 2D Hamiltonian carries one Dirac point. While for a multiple bandgap descriptions, different non-Abelian topological charges (Fig.~\ref{fig:band}D) correspond to the band degeneracies (Dirac points) appearing in different bandgaps accordingly. As such, frame rotations and topological charges of 1D Hamiltonians are closely related to band degeneracies in a high dimensional extended Hamiltonian.

Abelian topological invariants exhibit additive operations. The induced edge state~\cite{RN31,RN32,RN33,RN34} also inherits the Abelian nature through the celebrated ``bulk-edge correspondence'', which itself is an elegant theory that connects the properties of an infinite periodic system and those of an exposed edge of a truncated bulk. For 1D systems, the formation of edge states can be visualized from the 2D extended plane, where the edge states of the 1D system are inherently related to the topological degeneracy points encircled by the 1D Hamiltonian, i.e., the existence of enclosed Dirac points directly predicts the appearance of edge states. From this viewpoint, the bulk-edge correspondence of non-Abelian topological charges has been explained heuristically~\cite{RN21}. In contrast to the Abelian bulk-edge correspondence that applies to an individual bandgap, the non-Abelian one predicts both the position and number of the edge states for multiple bandgaps.

In three-dimensional crystals, a nodal line is a 1D curve in momentum space that arise due to a gap closing in the eigenvalue spectrum. When considering multiple bands, non-Abelian topological charges can be used to characterize the topological link structure of the multiple nodal-line systems, with an example shown in the left panel of Fig.~\ref{fig:band}E. The linked nodal lines threading through each other are formed by the crossings between adjacent bands. As shown by Wu et~al.~\cite {RN6}, non-Abelian charges can be used to explain various topological constraints on the nodal-link configurations. For example, in Fig.~\ref{fig:band}E: A pair of nodal lines of different colors cannot move across each other; A closed nodal ring can only encircle an even number of nodal lines of the other color; Nodes formed by consecutive pairs of bands anti-commute, while all nodes formed by more distant pairs of bands commute. Another interesting feature is that the sign of the charge assigns an orientation to the nodal lines (see Fig.~\ref{fig:band}E left), and the sign of a nodal line is reversed each time it goes under a nodal line of the adjacent bands. This is due to the non-trivial braiding rules arising from the non-commutativity of the quaternion charge~\cite{RN6}.

The local band structure near the nodal-line degeneracy is topologically equivalent to a two-dimensional Dirac cone. The right panel of Fig.~\ref{fig:band}E shows the two-dimensional band structure on the cutting plane at a fixed $k_3$, where three Dirac points can be clearly seen (two are in the first gap while one is in the second gap). The non-Abelian topological charges of nodal links are related to how they are braided together, and determine the outcome of the path-dependent node collision process~\cite{RN6,RN7,RN8,RN9,RN10}. The two Dirac points formed between the first and second bands cannot be removed along the path $L_2$ as shown in the left panel of Fig.~\ref{fig:band}E because the non-Abelian topological charge is $C(L_2)=(+k)\times(+k)=-1$. On the other hand, the charge of $L_1$ loop in Fig.~\ref{fig:band}E left panel is $C(L_1)=(+k)\times(-k)=1$, indicating trivial topology, the loop can hence be shrunk to a point, and thus the two lower-band Dirac points can annihilate by bringing them together along a path enclosed by $L_1$. For a system consisting of three bands, the annihilation of two Dirac points formed between the two lower bands depends on the positions of Dirac points between two higher bands. The above statement also indicates that two nodal ring linking is not allowed~\cite{RN6} (see the second topological constraint of nodal-link configuration as mentioned above). More generally, on a plane protected by $C_2T$ symmetry, the arguments of 2D Dirac points can be applied on to Weyl points as well. Therefore, one also can braid Weyl points on the $C_2T$ invariant plane~\cite{RN8}. It is worth mentioning that the non-Abelian topological charges are also related to triple degeneracies in 2D systems~\cite{RN15}, which were found to support zero-refractive-index propagation in photonics~\cite{RN35}. Further evolution of the nodal-link structure leads to $PT$-symmetric triple degeneracy in three dimensions~\cite{RN19,RN20}, as shown in Fig.~\ref{fig:band}F. The underlying topology of the triple degeneracy can be well described by the Euler number~\cite{RN19,RN20,PhysRevB.105.064301}, e.g. defined via integrating Euler curvature on the sphere surrounding the triple point, which leads to a universal higher-order bulk-boundary correspondence~\cite{RN36}. Researchers have proposed the possibility of exploring the new topological charges beyond the paradigmatic Chern insulators~\cite{RN37,RN38}. Interplaying with other crystalline symmetry, photonic, acoustic, and cold-atomic setups will further fuel the excitement in this research direction. Very recently, the second Euler number in four-dimensional synthetic matters has been discussed~\cite{bouhon2023second}.

Non-Abelian topological charges demonstrate the novel non-commuting properties that may enable new ways to manipulate wave packets and may inspire new applications in information transmission and processing. Based on the advantages of high degrees of freedom, these theoretical model has been widely realized in both acoustics and photonics, as shown in Figs.~\ref{fig:exp_band}A-D. The path-dependent node collision of Dirac points has been observed in acoustic systems using Kagome lattice formed by cylindrical resonators~(Fig.~\ref{fig:exp_band}A) and ideal metamaterials with three variable geometry parameters~(Fig.~\ref{fig:exp_band}B). The non-Abelian nodal links have been experimentally demonstrated in acoustic crystals \cite{RN15} and photonic biaxial hyperbolic metamaterials (Fig.~\ref{fig:exp_band}C), which illustrates the constraints imposed by non-Abelian charges. The photonic biaxial hyperbolic metamaterials offer a natural platform for implementing the three-band continuum models directly deriving from Maxwell's equations~\cite{RN6,RN11}. The three-band system serves as the minimal non-Abelian topological model, on which Guo et~al.~\cite{RN21} designed artificial sub-lattices and mapped the non-Abelian topological charges via rotating the eigenstate frame (Fig.~\ref{fig:exp_band}D).

\begin{figure*}[htbp]
 \includegraphics[width=\linewidth]{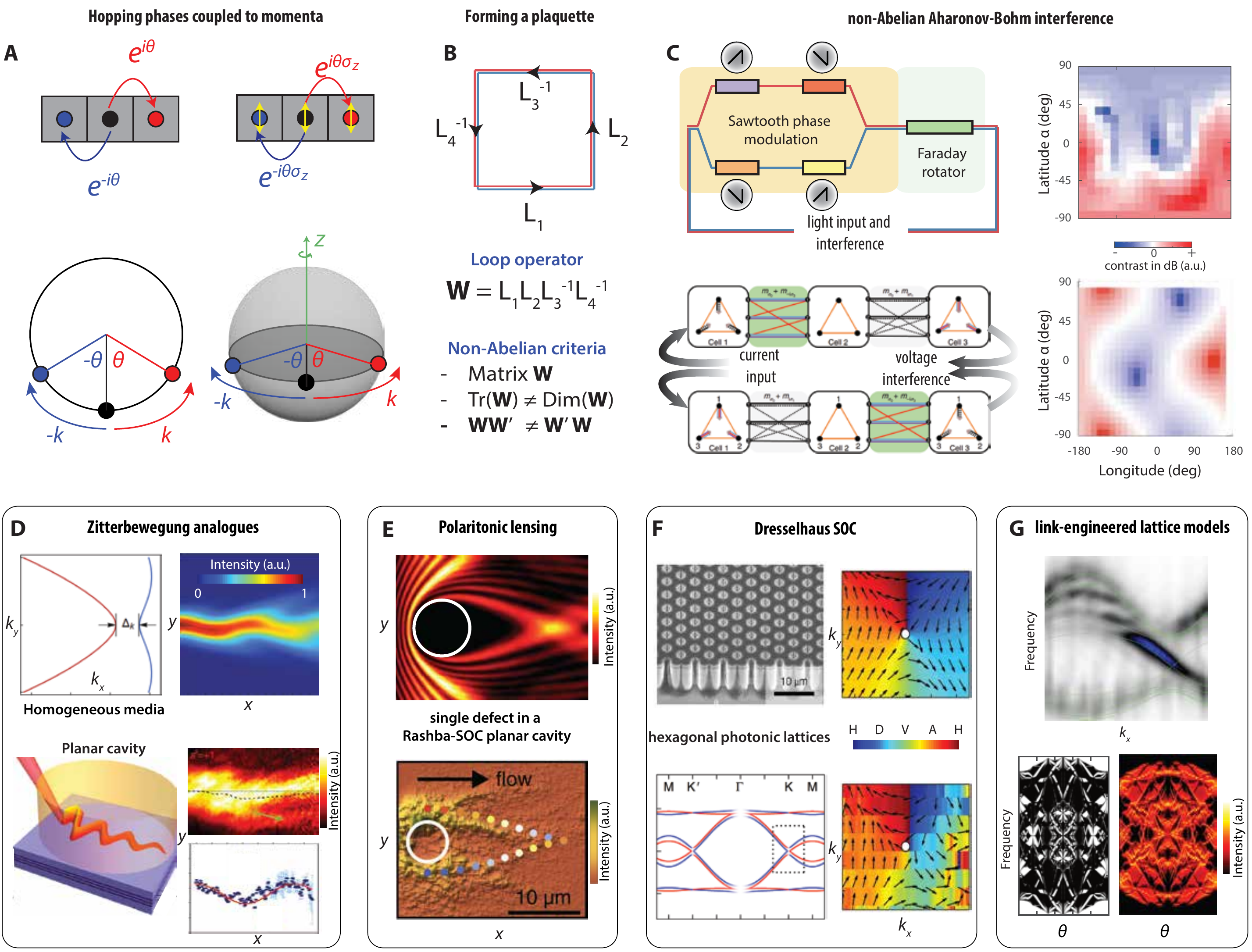}
        \caption{
        \textbf{Non-Abelian gauge fields.}
        (\textbf{A}) Lattice illustration of U(1) Abelian gauge fields (top left) and SU(2) non-Abelian gauge fields (top right; yellow arrows representing a spinful internal DoF). They lead to momentum-dependent rotation in the Hilbert space, \ie a unit circle (bottom left) and a unit sphere (bottom right), respectively. The former is always commutative, while the latter can be non-commutative around different axes. 
        (\textbf{B}) Non-Abelian gauge flux and non-Abelian criteria are defined from a matrix-valued loop operator. Tr, trace; Dim, dimension.
        (\textbf{C}) Non-Abelian Aharonov--Bohm interference to examine gauge-field commutativity in fiber optics and electric circuits~\cite{yang2019synthesis,wu2022non}.
        (\textbf{D}) Zitterbewegung analogs can be realized by coherent superposition of eigenstates of a homogeneous anisotropic non-Abelian medium (top) or planar quantum-well cavities (bottom)~\cite{chen2019non,nalitov2015spin,polimeno2021experimental,lovet2022observation}.
        (\textbf{E}) Wave focusing around an impenetrable defect (white circles) embedded in a non-Abelian medium of Rashba spin-orbit coupling~\cite{terccas2014non,polimeno2021experimental}.  
        (\textbf{F}) Polaritonic graphene (top left) features Drssselhaus-type non-Abelian gauge fields near the $K$ and $K'$ points, causing pseudospin switching on opposite sides of the degeneracy point (see color changes of the bands at the bottom left) and the dipolar magnetic field texture (top right: theory; bottom right: experiment)~\cite{whittaker2021optical}.
        (\textbf{G}) Probing chiral edge states of the QWZ/half-BHZ model realized with non-Abelian hopping phases in electric circuits (top)~\cite{wu2022non}. Proposal for probing the non-Abelian Hofstadter spectra using averaged photon transmission spectroscopy (bottom)~\cite{yang20202d,cheng2023artificial}.
        Partially adapted from Refs.~\cite{yang2019synthesis,wu2022non,chen2019non,nalitov2015spin,polimeno2021experimental,lovet2022observation,terccas2014non,polimeno2021experimental,whittaker2021optical,yang20202d,cheng2023artificial}.
        }
\label{fig:gauge}
\end{figure*}

\section{Non-Abelian gauge fields}

Next, we address non-Abelian synthetic gauge fields in light and sound, primarily focusing on synthetic non-Abelian magnetic fields that associate with the vector potentials; we will briefly discuss synthetic non-Abelian electric fields towards the end of this section. The recent development of this topic has drawn inspiration from other platforms, in particular cold atomic systems, for which we would like to draw readers' attention to the relevant reviews, \eg Ref.~\cite{dalibard2011colloquium,goldman2014light}, to facilitate a deeper understanding of the discussions here.

We illustrate the basic concept of synthetic gauge fields using the picture of a particle hopping among lattices, where the particle picks up a phase $\theta \propto \int \mathbf{A}\cdot \diff\mathbf{r}$ during the hopping, where $\mathbf{A}$ is the real-space gauge fields.
U(1) gauge fields couple to spinless particles whose Hilbert space forms a unit circle (Fig.~\ref{fig:gauge}A left). Consequently, all geometric phases accumulated during the hopping are Abelian because all the rotation around the unit circle is commutative.
In contrast, non-Abelian gauge fields couple to particles of internal degrees of freedom living in an enlarged Hilbert space, \eg a unit sphere for SU(2) gauge fields (Fig.~\ref{fig:gauge}A right). Because rotations around different axes of the sphere are non-commutative, the gauge fields and their geometric phases become non-Abelian.
Crucially, synthetic gauge fields couple to the momentum that is associated with a sign flip of geometric phases when particles propagate towards opposite directions (Fig.~\ref{fig:gauge}A bottom)---it is a requirement for maintaining Hermiticity, which is also the key to the time-reversal symmetry breaking for the U(1) gauge fields. This sign flip also distinguishes synthetic gauge fields from other related types of Hilbert space operations, such as single qubit gates that typically suffice along a single direction.

Similar to their Abelian counterparts, closed loops are needed to define real-space curvature, \ie the magnetic field. To see this, we define matrix-valued link variables $L \equiv \mathcal{P}\exp \iu \oint \matr{A}\cdot\diff\mathbf{r}$, where $\mathcal{P}$ denotes path-ordered integral. A real-space loop operator can thus be defined as $\matr{W} \equiv \mathcal{P}\prod_\Box L$. For a square-lattice plaquette, a counterclockwise loop operator beginning from its bottom left corner can be explicitly expressed as $\matr{W}=L_1L_2L_3^{-1}L_4^{-1}$ (Fig.~\ref{fig:gauge}B). The real-space curvature, \ie magnetic field, can thus be defined by $\matr{B} \equiv -\iu\log \matr{W}$ for this plaquette. In the continuum limit, $\matr{B}$ reduces to $\matr{B}=\nabla\times \matr{A} -\iu \matr{A}\times \matr{A}$, an exact real-space counterpart to the multi-band Berry curvature in momentum space, which is introduced at the beginning of the review.

This loop operator enables several criteria for identifying non-Abelian gauge fields~\cite{goldman2014light}. Generally, a matrix-valued $\mathbf{A}$ or its non-commutative components could be loosely used to indicate non-Abelian gauge fields. 
Another criterion is based on the concept of the real-space Wilson loop $W\equiv\mathrm{Tr}(\matr{W})$ that is a gauge-invariant quantity; it is required that the Wilson loop should differ from the dimensionality of the Hilbert space in order for the gauge fields to be non-Abelian. This criterion works well for many circumstances but has its caveat---a system with decoupled spins can still couple to gauge fields that are Abelian, like a single $\e^{\iu\theta\sigmaz}$ term; however, its Wilson loop $W\neq2$ satisfied the non-Ablian condition.
So far, the most rigorous definition of non-Abelian gauge fields relies on examining the commutativity among different loop operators. It requires the existence of two different loop operators $\matr{W}$ and $\matr{W}'$ that satisfy the non-commutativity condition $\matr{W}\matr{W}'\neq\matr{W}'\matr{W}$.
%
Such a non-commutativity criterion has been applied to analytically examine the genuine non-Abelian conditions in non-Abelian Hofstadter models~\cite{yang20202d}.
Although introduced in a lattice context, the criterion above works equally well for continuum systems where the loop operators can be replaced with curvatures. 

The non-Abelian Aharonov--Bohm interference is particularly useful for experimentally detecting non-Abelian gauge fields and the resulting geometric phases. 
This effect describes a spinful particle moving along two paths, where the path integrals are reversely ordered to each other. The two paths form a closed loop, enabling interference measurements of the spin population that reflect the non-commutativity of the underlying gauge fields. 
This effect has stimulated longstanding theoretical interests on various physical platforms~\cite{wu1975concept,horvathy1986non,alford1990discrete,chaichian2002aharonov,osterloh2005cold,dalibard2011colloquium,jacob2007cold,zhang2008detecting,chen2019non,fruchart2019dualities}, and realized using nonreciprocal fiber optics and electric circuits~\cite{yang2019synthesis,wu2022non} (Fig.~\ref{fig:gauge}C).
In the experiments, projection measurements were performed along a certain basis to extract their associated population contrast at the interference. Since the input spin can be accurately controlled by optical and electrical means, the contrast can be measured around the entire surface of the Bloch sphere, where the creation, evolution, and annihilation of zeros and poles indicate the appearance of non-Abelian gauge fields. 


%
Similar to the Abelian counterpart, non-Abelian gauge fields $\mathbf{A}$ get incorporated in a Hamiltonian system via the minimal coupling described by the Peierls substitution $H(\mathbf{p})\rightarrow H(\mathbf{p}-\mathbf{A})$, where $\mathbf{A}$ can be spatially and temporally dependent.
Below, we present a summary of phenomena under this umbrella, categorized based on the degree of symmetry present in the systems, which spans from homogeneous media to lattice models with intentionally engineered link variables.


Zitterbewegung (ZB), initially proposed as a quantum mechanical interference between the solutions to the Dirac equation by Schrodinger in 1930, has been generalized into a universal wave phenomenon over the past century: It describes a trembling motion resulting from wave interference of quasi-degenerate modes~\cite{zawadzki2011zitterbewegung,vaishnav2008observing,hasan2022wave,zhang2008extremal,wang2009zitterbewegung,longhi2010photonic,zhang2008observing,dreisow2010classical}.
Among other established approaches, non-Abelian gauge fields have recently emerged as a new way to synthesize the ZB effect.
Homogeneous anisotropic non-Abelian optical media with electromagnetic duality are sufficient to this end~\cite{chen2019non} (Fig.~\ref{fig:gauge}D top).
In this approach, an anisotropic medium, with a single gauge-field contribution (\ie Abelian) in the permittivity and permeability tensors, exhibits two branches of modes forming a one-dimensional Dirac point in the isofrequency contour. The addition of another gauge-field contribution introduces the coupling between the two modes, lifts the Dirac point, and enables the definition of the ZB beat wave number proportional to the Dirac mass of the particle under the trembling motion.
ZB was also predicted for exciton polaritons in planar semiconductor microcavities~\cite{shelykh2009polariton} based on the TE--TM splitting (see Box 1), which provides an effective magnetic field that causes polaritonic precession~\cite{sedov2018zitterbewegung}. Meanwhile, polaritonic graphene was theoretically shown to exhibit the ZB effect in the $p$ bands near the $K$ point~\cite{whittaker2021optical}. The ZB effect was recently probed in hybrid organic-inorganic perovskites and GaAs/AlGaAs quantum wells~\cite{polimeno2021experimental,lovet2022observation}. In the angle- and polarization-resolved photoluminescence measurement, ZB was observed under a circular resonant pump that excites both polarization branches while diminished under a single polarization excitation (Fig.~\ref{fig:gauge}D bottom).


In the planar polaritonic cavities, a TE-TM crossing can appear at a nonzero critical momentum when the static in-plane field and the TE-TM field compensate each other~\cite{terccas2014non}. At this crossing momentum, the polaritonic Hamiltonian can be reformulated into a minimal-coupling gauge-field Hamiltonian with the Rashba-type spin-orbit interaction.
Notably, the quantum metric, the real part of the quantum geometric tensor, diverges at the TE-TM crossing~\cite{bleu2018measuring}. The crossing can be gapped under external magnetic fields, permitting measurement of nontrivial Berry curvature and quantum metric simultaneously~\cite{gianfrate2020direct}. 
Near the TE-TM crossing, a lensing effect appears that can be interpreted as ZB in the presence of a defect. In particular, when one of the polaritonic modes hits a defect, the spin-orbit interaction induces opposite group velocities for the scattered polaritons towards opposite directions, leading to the focusing effect in the total field intensity.
This lensing effect was observed in the perovskite polaritonic platform~\cite{polimeno2021experimental}, where a linear-polarized laser excites a polariton flow that hits a potential, splits into circular-polarized flows, and refocuses guided by
the non-Abelian magnetic fields.
A similar phenomenon should also occur for pure photonic systems, \eg a homogeneous non-Abelian media where the required mode dispersion (blue curve in Fig.~\ref{fig:gauge}D top left) can also be created.
%
%
For polaritonic cavities, creating the Dresselhaus-type non-Abelian gauge fields requires extra symmetry breaking (Fig.~\ref{fig:gauge}F). It has been theoretically predicted that hexagonal photonic graphene~\cite{jacqmin2014direct,sala2015spin} hosts the Dresselhaus-type fields at its $K$ and $K'$ points~\cite{nalitov2015spin}, which was confirmed experimentally~\cite{whittaker2021optical} by the dipolar pseudospin winding at $K$ and $K'$ (instead of the quadrupolar winding at $\Gamma$ similar to that of unpatterned cavities~\cite{leyder2007observation}) and the associated optical spin Hall effect.
Aside from quantum-well structures, other anisotropic materials like perovskites~\cite{spencer2021spin,polimeno2021experimental,lempicka2022electrically,li2022manipulating,polimeno2021tuning}, perylene~\cite{ren2021nontrivial,ren2022realization}, and liquid crystals~\cite{rechcinska2019engineering,lempicka2022electrically,li2022manipulating} are also emerging for the versatile engineering of spin-orbit coupling via artificial non-Abelian gauge fields. 

The controllability of optical and acoustic systems enables the realization of various lattice models with non-Abelian gauge fields. The Qi-Wu-Zhang (QWZ) or half-Bernevig-Hughes-Zhang (half-BHZ) model~\cite{qi2006topological}, a celebrated model for Chern insulators, can be written in real space where sites are connected by non-Abelian hopping links. This model has been realized with electric circuits, allowing the visualization of chiral edge states~\cite{wu2022non} (Fig.~\ref{fig:gauge}G top).
So far, all the phenomena addressed above deal with spatially homogeneous non-Abelian gauge fields, and interests in inhomogeneous ones are also spawning. A typical example is a class of non-Abelian Hofstadter models featuring linearly varying gauge fields~\cite{yang20202d}. Such link arrangements give rise to nonsymmorphic chiral symmetries of nontrivial symmetry algebra~\cite{yang2022non}. Proposals have been made towards realizing the models based on photonic synthetic dimensions~\cite{cheng2023artificial} (Fig.~\ref{fig:gauge}G bottom).   

Synthetic non-Abelian electric fields can be equally created and manipulated.
In the non-Abelian setting, the synthetic electric fields are given by~\cite{goldman2014light,chen2019non}
$\mathbf{E} = -\nabla\varphi - {\partial{\mathbf A}}/{\partial t}-\iu\left[\varphi,\mathbf{A}\right]$, where $\varphi$ is the scalar potential.
The first two terms of the expression are inherited from the Abelian counterpart (as in electromagnetism), meaning that synthetic electric fields can be created from the spatial and temporal gradients of the scalar and vector potentials, respectively.
Meanwhile, a unique third term, \ie the commutator between $\varphi$ and $\mathbf{A}$, also appears, which indicates that non-Abelian electric fields are even possible under temporally static and spatially uniform non-Abelian gauge potentials, a characteristic shared by the synthetic magnetic fields (via the $\mathbf{A}\times\mathbf{A}$ term). So far, non-Abelian electric fields have not been synthesized experimentally, according to our knowledge.


Synthetic dimensions enable the realization of non-Abelian gauge fields (\ie connection, and the associated curvature) in higher-dimensional space.
A quintessential example is the four-dimensional quantum Hall effect~\cite{ozawa2016synthetic,lohse2018exploring,zilberberg2018photonic,chen2021acoustic} featured by the second Chern number, which may be obtained by summing over the products over the first Chern numbers of different sub-dimensions~\cite{zilberberg2018photonic}.
In a fiber setting, a rotation angle complements the three-dimensional momenta, which permits the construction of a non-trivial second Chern number from non-Abelian Berry curvature, indicating one-way transport~\cite{lu2018topological}.
Moreover, synthetic non-Abelian Yang monopoles were created in 5D synthetic space using hyperfine ground states of rubidium~\cite{sugawa2018second} and photonic bianisotropic semimetal metamaterials~\cite{ma2021linked}.
Recently, non-trivial second Chern numbers in hyperbolic lattices were also realized in artificial circuit networks, featuring non-Abelian translational operations~\cite{zhang2023hyperbolic}.

\section{non-Abelian pumping}

\begin{figure*}[htbp]
 \includegraphics[width=\linewidth]{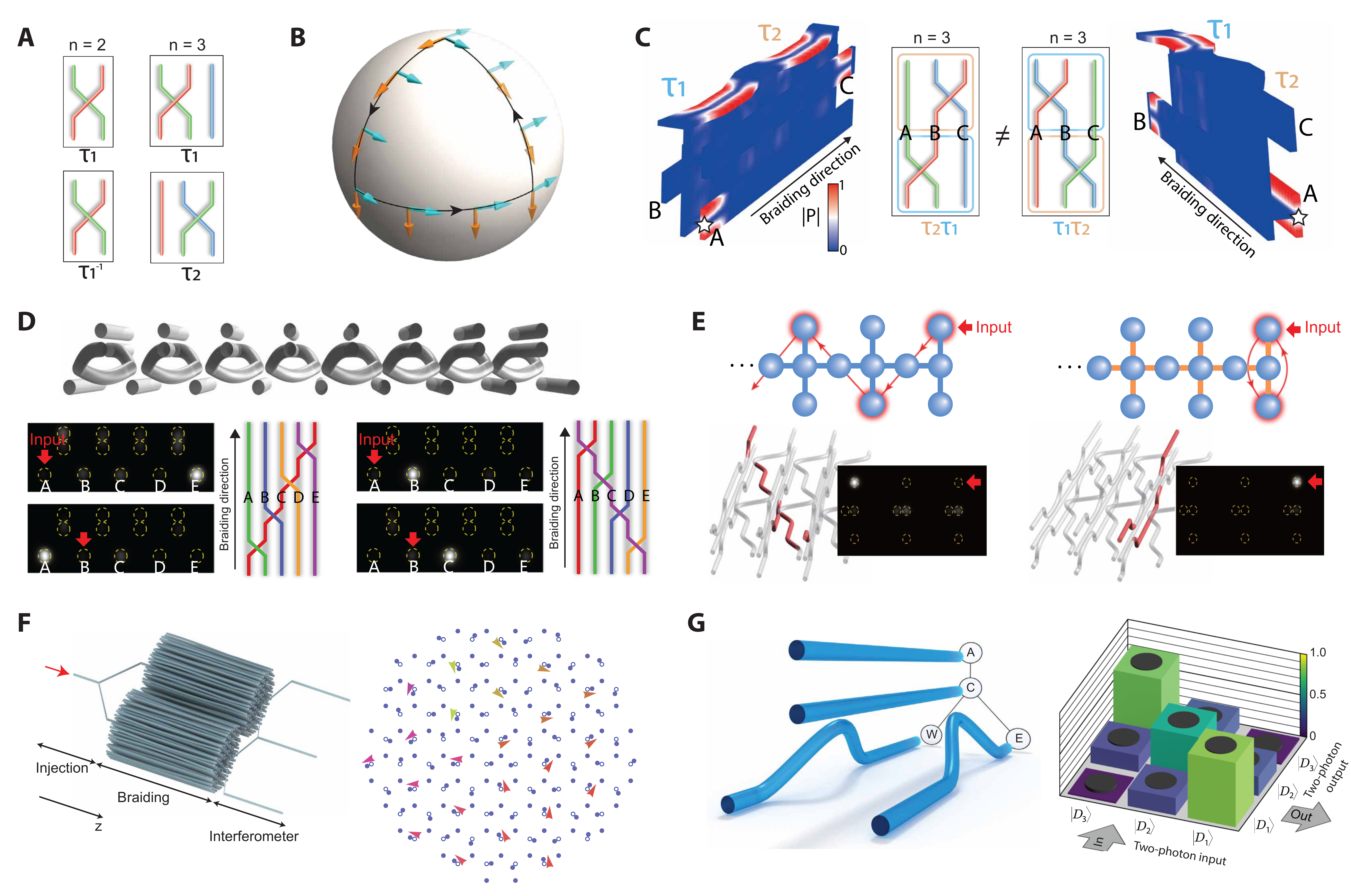}
        \caption{
        \textbf{Non-Abelian pumping.}
        (\textbf{A}) Braiding operations of two and three strands. 
        (\textbf{B}) Two-strand braiding manifests as the SO(2) local rotation of two orthogonal vectors induced by the SO(3) global rotation.
        (\textbf{C}) Non-Abelian braiding realized in coupled acoustic waveguides. Herein, two different sections of coupled waveguides are connected in different orders. The acoustic input in both cases is at waveguide A, but the output is detected at waveguide C (left) and B (right), respectively, which confirms the non-Abelian characteristics~\cite{chen2022classical}.
        (\textbf{D}) Five-mode non-Abelian braiding realized in coupled photonic waveguides. The upper panel is a schematic of the waveguide array. The lower panels are the measured results at the outputs, with bright spots indicating strong optical intensity. The red arrows indicate the injection position at the inputs~\cite{zhang2022non}.
        (\textbf{E}) Non-Abelian Thouless pumping in generalized Lieb lattices~\cite{sun2022non}. Left: under a particular gauge sequence, an optical zero mode is pumped to hop across the lattice. Right: by switching the gauge sequence, the optical mode stays at the same unit cell.
        (\textbf{F}) Braiding of two topological zero modes in bounded by Kekulé vortices in photonic waveguides~\cite{noh2020braiding}.
        (\textbf{G}) Realization of U(3) gauge structure in photonic waveguides and non-Abelian two-photon holonomy~\cite{neef2023three}.
        Partially adapted from Refs.~\cite{chen2022classical,zhang2022non,sun2022non,noh2020braiding,neef2023three}.
        }
\label{fig:braid}
\end{figure*}

The effect of non-Abelian gauge fields can manifest in the dynamic adiabatic evolution, or pumping, of an expanded Hilbert space. Because such pumping simultaneously involves multiple states, they are connected by Berry-Wilczek-Zee (BWZ) phase matrix~\cite{wilczek1984appearance}, which gives rise to dynamic transition processes among multiple eigenstates described by non-Abelian holonomies. 

One interesting case is the realization of non-Abelian mode braiding via pumping. Braiding is the operation that sequentially permutes two neighboring strands. The braiding of $n$ strands mathematically describes infinite discrete groups called braid groups, denoted $B_n$. $B_n$ has a set of $2(n-1)$ generators, denoted $\tau_i$ with $i\in\left\{1,2,\cdots,n-1\right\}$ and their inverses $\tau_i^{-1}$.  $\tau_i$ executes the exchange of the $i$-th and $i+1$-th strands with $i+1$ over crossing $i$, and the corresponding inverse element denotes under crossing. The generators follow $\tau_i \tau_j \tau_i=\tau_j \tau_i \tau_j$ if $\abs{i-j}=1$, and $\tau_i \tau_j=\tau_j \tau_i$ if $\abs{i-j}\geq2$. It is straightforward to see that $B_n$ is non-Abelian for all $n>2$, since $\tau_i \tau_j\neq\tau_j \tau_i$. These properties are schematically shown in Fig.~\ref{fig:braid}A using $B_2$ and $B_3$.

The generators of $B_n$ can map to $n\times n$ matrices. Take $B_2$ as an example, there is $\tau_1\rightarrow G_1 (2)=-i\sigma_y$ and $\tau_1^{-1} \rightarrow G_1^{-1} (2)=i\sigma_y$, with $\sigma_y$ being the second Pauli matrix. It follows that the initial and end states of an arbitrary braid can be connected using such matrices. Because braid groups have natural inclusion characteristics, the matrix representation for the generators of $B_n$ is transparent, e.g., for $B_3$, 
$\tau_1\rightarrow G_1 (3)= \left[G_1 (2), 0;0, 1\right]$
and
$
\tau_2\rightarrow G_2 (3)= \left[ 1, 0;0, G_1 (2) \right]$.
The non-Abelian characteristics of $B_3$ naturally emerge, since $G_1 (3) G_2 (3) \neq G_2 (3) G_1 (3)$. 

Note that all $G_i (n)$ are orthogonal matrices with unity determinants, which suggests that they are elements in SO($n$). Therefore, it is possible to emulate braiding using saliently designed rotation in an $n$-dimensional space. One route is the consider the adiabatic evolution of degenerate states. For example, consider a Hamiltonian 
$
H=
\left(\mathbf{0}_{M\times M}, \mathbf{t}^T;\mathbf{t}, \mathbf{0}_{N\times N}\right)
$
with $\mathbf{t}\in\mathbb{R}^{M\times N}$, a total of $n=\abs{M-N}$ eigenstates are pinned at zero energy because of the sublattice symmetry $C^{-1} HC=-H$, with 
$C=
\left(
\mathbf{-1}_{M\times M}, 0; 0, \mathbf{1}_{N\times N}\right)
$
These degenerate states form an $n$-dimensional subspace. When the entries in $\mathbf{t}$ are driven by external parameters, the degenerate states undergo adiabatic pumping. Such a multi-state evolution is captured by an $n$-dimensional BWZ phase matrix. Because all eigenvectors are real, the BWZ matrix belongs to $SO(n)$, which can vary the composition of the states within the subspace. For a two-dimensional subspace, the matrix is $O(\Omega)=\left(\cos\Omega,-\sin\Omega; \sin\Omega,\cos\Omega\right)$, where $\Omega$ is the solid angle enclosed by the loop of in the parameter space spanned by $\mathbf{t}$. It becomes clear that $G_1 (2)=O(\pi/2)$, which realizes the generating operation of $B_2$ (Fig.~\ref{fig:braid}B). The generalization to $B_n$ can be realized by using the natural inclusion property.

Waveguide systems are a good platform for realizing the adiabatic evolution of the abovementioned $H$. For example, Chen et al. constructed a set of coupled acoustic waveguides consisting of identical rectangular waveguides coupled by air bridges~\cite{chen2022classical}. The coupling magnitudes among the waveguides are tunable by adjusting the positions of the air bridges. The positions are slowly varied along the guiding direction, such that the guiding modes are adiabatically pumped as they propagate down the waveguides. The two generators of $B_3$ were successfully realized in two waveguide arrays with different pumping profiles. The braiding effects manifest as the swapping of dwelling waveguides between the input and output ports. Furthermore, by connecting the two waveguide arrays in different orders, it was observed that the same input mode was converted to different output modes, which confirms the non-Abelian characteristics of $B_3$ (Fig.~\ref{fig:braid}C). Based on a similar principle, Zhang et al. performed photonic experiments by fabricating an array of meandering optical waveguides etched in glass substrates using femtosecond laser writing~\cite{zhang2022non}, wherein the waveguides are evanescently coupled so the coupling strengths are tuned by changing their separations (Fig.~\ref{fig:braid}D). A similar approach can also be used to braid topological edge modes in a Y-junction formed by a one-dimensional topological lattice~\cite{boross2019poor,wu2020double}.

The braiding effect can be incorporated with topological pumping to realize non-Abelian pumping. Lattice systems with a dispersionless bulk band are ideal for this demonstration. In an open-boundary lattice, such a flat band consists of a set of highly degenerate modes that has zero group velocity. For example, optical waveguides can be arranged into a one-dimensional generalized Lieb lattice~\cite{brosco2021non}, in which a flat band exists at zero energy. A position-dependent gauge field drives the spatial evolution of the guiding modes and makes them sequentially hop from one unit cell to the next despite they have zero group velocity in the transverse directions. The transverse motion is thus a generalized form of Thouless pumping. Furthermore, it is experimentally observed that changing the spatial order of the gauge field produces different hopping sequences, meaning that the pumping is non-Abelian in character (Fig.~\ref{fig:braid}E). Such non-Abelian Thouless pumping is successfully realized in optics~\cite{sun2022non} using on-chip photonic waveguides. A demonstration in acoustics is also reported~\cite{you2022observation}.

Currently, most studies on non-Abelian holonomies have focused on adiabatic evolution in systems with perfectly degenerate bands everywhere in the parameter space. However, the requirement for perfectly degenerate bands is usually only approximately satisfied in realistic systems. Recently, it was shown that, in contrast to conventional wisdom, non-Abelian holonomy could also exist in systems with isolated degeneracies between multiple energy bands. Two groups independently showed that the transition between different states could be arbitrarily controlled by introducing abrupt turns when the evolution path traverses isolated degeneracies~\cite{you2022observation,brown2022direct}. These works suggest that U(N) holonomy may be used to describe the evolution of states in physical systems with $N$ bands connected by a finite number of isolated degeneracies across the entire parameter space. This new approach greatly broadens the choice of parameter spaces to achieve non-Abelian holonomy.

Braiding operations can also emerge in specially designed special-unitary and unitary operations. A primary candidate of this approach is the Majorana zero modes~\cite{nayak2008non}. One example is the topological modes bounded by a gauge vortex. For example, a honeycomb lattice under Kekul\'{e} modulation, which takes the form of a real-space vortex gauge,  sustains Majorana-like zero modes bounded to the vortex core~\cite{hou2007electron}. Such modulated graphene and the zero modes have been realized in two-dimensional photonic~\cite{gao2020dirac} or phononic crystals~\cite{chen2019mechanical}. Multiple spatially well-separated vortices can simultaneously bound multiple zero modes at different locations. A way to realize braiding is simply to slowly interchange the positions of the vortices in real space, which swap the relevant zero modes. To realize such effects in optics, waveguide arrays based on the modulated lattices were constructed and the evolution of the vortices are encoded in the propagation directions~\cite{iadecola2016non}. A photonic experiment successfully demonstrated the viability of this approach~\cite{noh2020braiding} (Fig.~\ref{fig:braid}F).%
More recently, non-Abelian U(3) quantum holonomy was realized with indistinguishable photons in coupled waveguide systems~\cite{kremer2019optimal,neef2023three}, indicating the possibility for realizing more complex braiding structures (Fig.~\ref{fig:braid}G).
Proposals also suggest the realization of Majorana-like zero modes in the classical-mechanical analog of Kitaev model~\cite{prodan2017dynamical}, which are also candidates for realizing braiding operations~\cite{barlas2020topological}. 

\section{Non-Abelian characteristics of non-Hermitian systems}

\begin{figure*}[htbp]
 \includegraphics[width=\linewidth]{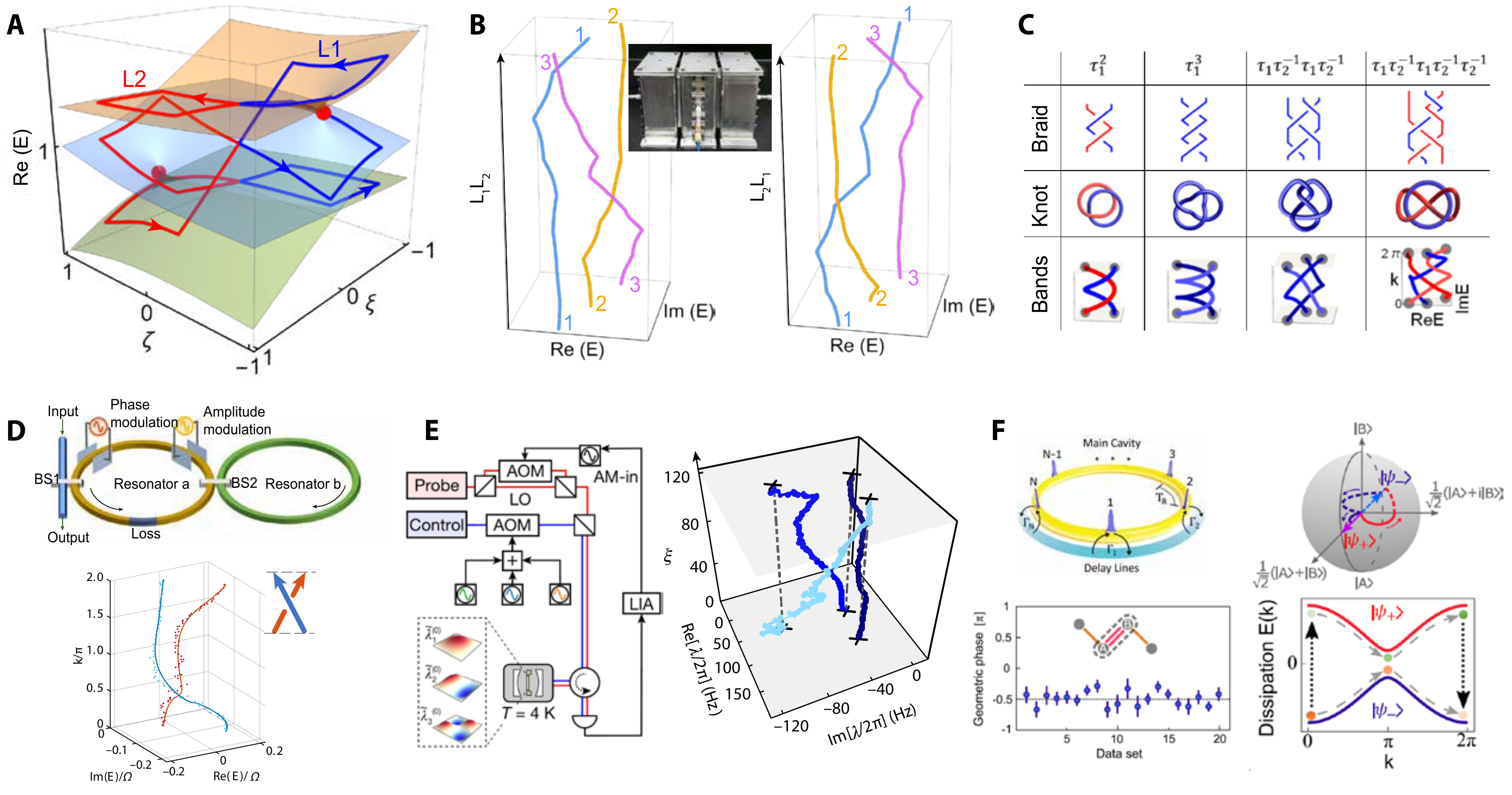}
        \caption{
        \textbf{Non-Abelian characteristics emerging in non-Hermitian systems.}
        (\textbf{A}) The real parts of the eigenvalue Riemann surfaces of a three-band non-Hermitian system. The red dots mark two order-2 EPs. The blue and red curves are the eigenvalue trajectories when the two EPs are encircled individually by the parametric loops labeled $L_1$ and $L_2$, respectively.  
        (\textbf{B})  Distinct eigenvalue braids produced by concatenating the two loops in different orders, $L_1 L_2$ for (left) and $L_2 L_1$ (right), and the coupled acoustic cavities (middle) that realized the non-Aeblian eigenvalue braids and non-Abelian state exchanges~\cite{tang2020exceptional,tang2022experimental}. 
        (\textbf{C}) The eigenvalue braids and the corresponding knot structures appear in non-Hermitian Bloch bands~\cite{hu2021knots}.
        (\textbf{D}) Synthetic frequency dimension in optical ring resonators and the measured complex bands accompanied by the corresponding braid diagram~\cite{RN45}.
        (\textbf{E}) A Si$_3$N$_4$ membrane controlled by cavity optomechanics and the measured braiding eigenvalue spectrum~\cite{RN42}.
        (\textbf{F}) Non-Abelian Wilson line of isolated bands via dissipative coupling in synthetic time dimension~\cite{parto2023non}. 
        Partially adapted from Refs.~\cite{tang2020exceptional,hu2021knots,RN45,RN42,parto2023non}.
        }
\label{fig:nh}
\end{figure*}

So far, the review has been focusing on Hermitian systems, while non-Hermitian systems are another realm in which non-Abelian effects can emerge. The non-Hermitian formalism is used to describe open systems in which energy exchange with external environments is permitted. Unlike Hermitian systems, their spectra, i.e., the eigenvalues, are generically complex functions of system parameters. Such characteristics have profound consequences. In the Hermitian case, bands are naturally ordered according to their energy. This is not the case for complex energy bands. As such, the complex spectrum presents an additional layer of topology. Remarkably, non-Hermitian spectral topology can be non-trivial even for a one-dimensional single-band system without symmetry constraints because the energy is a map from a one-dimensional parameter space, e.g., Bloch wavenumber, to a complex plane, on which non-trivial winding can readily emerge~\cite{kawabata2019symmetry}. Here, we mainly focus on the non-Abelian phenomena in non-Hermitian systems. For comprehensive accounts of non-Hermitian topology, the readers are referred to existing reviews such as Refs.~\cite{ashida2020non,bergholtz2021exceptional,RN23}.

Using the fundamental homotopy group, the space of $N$-dimensional non-Hermitian Hamiltonians are topologically classified as braid groups $B_N$~\cite{wojcik2020homotopy,li2021homotopical}, which are non-Abelian groups for $N>2$ [as discussed in section III]. Different from the non-Abelian topological classification of multiple bands in section I that considers eigenvector rotations, this unique topology mainly rests upon the geometry of non-Hermitian spectral manifolds, which are self-intersecting complex Riemannian sheets instead of isolated surfaces. Fig.~\ref{fig:nh}A shows an example of a three-band non-Hermitian system, where the first and second bands (green and blue), and the second and third bands (blue and orange) coalesce at different parameters, forming two order-2 EPs. It can be seen that when a closed parametric loop encloses spectral branch-point singularities, called exceptional points (EPs), the eigenvalues of different states smoothly connect to one another by crossing the intersection curves called branch cuts. Hence encircling an EP essentially braids the eigenvalues. For example, the loops $L_1$ and $L_2$ encircle the two different EPs in Fig.~\ref{fig:nh}A, swapping the eigenvalues in the process. It is then clear that encircling the two EPs in different order produces different eigenvalue braids, as shown in Fig.~\ref{fig:nh}B. Such non-Abelian braids have been observed in acoustic systems (Fig.~\ref{fig:nh}B)~\cite{tang2020exceptional,tang2022experimental}.

The nontrivial braids can also map to different knot structures when the eigenvalue braids are projected to a three-dimensional Euclidean space. Hu and Zhao studied the transition between different eigenvalue knots appearing in non-Hermitian Bloch bands (Fig.~\ref{fig:nh}C)~\cite{hu2021knots}. Wang et al. promoted synthetic platforms with ring resonator coupling and modulation designs, as shown in Fig.~\ref{fig:nh}D, and studied the single- or two-band knots~\cite{RN45,RN46}. The multi-band non-Abelian braids have also been realized in cavity optomechanics (Fig.~\ref{fig:nh}E)~\cite{RN42}. Eigenvalue knots are also experimentally realized in acoustic systems~\cite{zhang2023observation,li2023eigenvalue}.

Another key difference between non-Hermitian and Hermitian systems is that the eigenvectors of the former are not orthogonal. This feature, coupled with the fact that non-Hermitian eigenvectors are fiber bundles sticking to the spectral manifolds, means that eigenvectors can also smoothly evolve into one another in parallel transport. This effect was readily captured by the fractional geometric phase produced when an EP is encircled~\cite{dembowski2001experimental,dembowski2004encircling,gao2015observation,ding2016emergence,schindler2017winding,tang2021direct}, i.e., the state evolution on the eigenvalue manifold is not holonomic even when the parametric loop is closed, and multiple loops in the parameter space are required for the states to recover with a quantized geometric phase. It is also the origin of the non-Abelian exchange of non-Hermitian states~\cite{tang2022experimental,zhong2018winding}. An example in acoustics is reported in Ref.~\cite{tang2022experimental}, wherein two order-2 EPs formed by two coalescing states, are found on the spectral manifold of a three-state non-Hermitian system. Because these order-2 EPs are formed by different states, encircling them in different sequences produces different state-permutation outcomes. The non-Abelian characteristics are confirmed by stroboscopic measurement of the acoustic wavefunctions under a constant gauge.

Mode dynamics presents another intriguing scenario. Because of the intrinsic instability problem for states with higher imaginary eigenvalues and the non-orthogonality of the non-Hermitian eigenmodes, non-Hermitian mode evolutions inevitably leave the adiabatic path and move to the state with higher loss~\cite{berry2011slow}. Such a non-adiabatic effect has been observed and leveraged for asymmetric switching of waveguide modes~\cite{doppler2016dynamically,xu2016topological}. 
This non-adiabaticity also causes interesting non-Abelian Wilson line evolution of untouched dissipative bands in time-multiplexed photonic networks~\cite{parto2023non}.
If non-adiabatic transition can be precisely controlled, it may also function as a particular operation for tailoring non-Abelian mode dynamics in non-Hermitian systems.

\section{Conclusion and outlook}

Non-Abelian topological charges for a set of bandgaps are characterized by matrix-like entities, which complement and enrich the established integer classifications for a single bandgap. The non-commuting properties imply novel phenomena such as non-unique topological phase transition paths~\cite{RN21}. 
We expect further realizations of non-Abelian topological charges with photonics and acoustics in the near future, thereby enabling mutual cross-fertilization between different research fields.
We will see that non-Abelian defects may also play pivotal roles in morphogenesis as well as cosmology, where singularities called cosmic strings seem to be in correspondence with the defect lines~\cite{RN29}. %
Currently, the non-Abelian topological phase based on eigen-frame rotation mainly focuses on $PT$-symmetric Bloch systems, the expansions towards other directions deserve further exploration and study, e.g., the Floquet multi-gap topology~\cite{slager2022floquet}. 


Non-Abelian gauge fields have so far mostly been studied in Hermitian systems, and their interplay with non-Hermiticity deserves further exploration. In fact, the presence of non-Hermiticity in lattice systems has been treated as imaginary Abelian gauge fields in some studies of non-Hermitian skin effect (\eg Refs.~\cite{jin2019bulk,lee2019anatomy}), whose further interplay with non-Abelian gauge fields is thus anticipated. A recent endeavor in this direction shows that non-Abelian gauge fields can drive non-Hermitian topological phase transition despite the lack of gauge flux in one dimension~\cite{pang2023synthetic}. 


Perhaps the most attractive application of braiding is in quantum computation. Non-Abelian braiding is one of the essential components of universal quantum logic. The realization of non-Abelian braiding in light and sound, therefore, not only expands our capability for wave manipulations but may also bring new toolsets for implementing topological quantum-logic operations based on bosonic platforms~\cite{burrello2010non,keilmann2011statistically,kapit2012non,yuan2017creating}. 
Towards this goal, non-Abelian gauge fields could play an important role in synergy with nonlinearity for generating many-body photonic effects~\cite{carusotto2013quantum}.


More optical and acoustic degrees of freedom can be leveraged for non-Abelian phenomena. 
Orbital angular momenta have been proposed and widely realized as a synthetic spatial dimension, \ie to label lattice sites, in frequency-degenerate optical cavities and quantum walks~\cite{luo2015quantum,zhou2017dynamically,cardano2015quantum,cardano2017detection,wang2018experimental,yang2022topological,yang2023realization}. Nevertheless, to our knowledge, OAMs have not been used as a pseudospin degree of freedom for non-Abelian operations, which should also be possible. For example, one can couple angular momentum modes $m=\pm1$ with a $q=1$ q-plate to form the pseudospin in a cavity, where suitable dispersion should be created to minimize the leakage to other undesired high-order angular momenta.

Many of the realizations reviewed here are in the low-frequency domain. Therefore, the miniaturization of non-Abelian effects toward the Terahertz and optical regimes will be of interest, in particular, based on the integrated platforms where new topological building blocks are emerging~\cite{lu2022high,wang2023experimental,zhang2022spin}. This process could stimulate a variety of application opportunities, such as channel-multiplexed devices, path-dependent topological mode converters, and nonreciprocal optoelectronics~\cite{sounas2017non,guo2019nonreciprocal,yang2023self}.

\bibliographystyle{apsrev4-2}
\bibliography{sources}

\providecommand{\noopsort}[1]{}\providecommand{\singleletter}[1]{#1}%
\begin{thebibliography}{184}%
\makeatletter
\providecommand \@ifxundefined [1]{%
 \@ifx{#1\undefined}
}%
\providecommand \@ifnum [1]{%
 \ifnum #1\expandafter \@firstoftwo
 \else \expandafter \@secondoftwo
 \fi
}%
\providecommand \@ifx [1]{%
 \ifx #1\expandafter \@firstoftwo
 \else \expandafter \@secondoftwo
 \fi
}%
\providecommand \natexlab [1]{#1}%
\providecommand \enquote  [1]{``#1''}%
\providecommand \bibnamefont  [1]{#1}%
\providecommand \bibfnamefont [1]{#1}%
\providecommand \citenamefont [1]{#1}%
\providecommand \href@noop [0]{\@secondoftwo}%
\providecommand \href [0]{\begingroup \@sanitize@url \@href}%
\providecommand \@href[1]{\@@startlink{#1}\@@href}%
\providecommand \@@href[1]{\endgroup#1\@@endlink}%
\providecommand \@sanitize@url [0]{\catcode `\\12\catcode `\$12\catcode
  `\&12\catcode `\#12\catcode `\^12\catcode `\_12\catcode `\%12\relax}%
\providecommand \@@startlink[1]{}%
\providecommand \@@endlink[0]{}%
\providecommand \url  [0]{\begingroup\@sanitize@url \@url }%
\providecommand \@url [1]{\endgroup\@href {#1}{\urlprefix }}%
\providecommand \urlprefix  [0]{URL }%
\providecommand \Eprint [0]{\href }%
\providecommand \doibase [0]{https://doi.org/}%
\providecommand \selectlanguage [0]{\@gobble}%
\providecommand \bibinfo  [0]{\@secondoftwo}%
\providecommand \bibfield  [0]{\@secondoftwo}%
\providecommand \translation [1]{[#1]}%
\providecommand \BibitemOpen [0]{}%
\providecommand \bibitemStop [0]{}%
\providecommand \bibitemNoStop [0]{.\EOS\space}%
\providecommand \EOS [0]{\spacefactor3000\relax}%
\providecommand \BibitemShut  [1]{\csname bibitem#1\endcsname}%
\let\auto@bib@innerbib\@empty
\bibitem [{\citenamefont {Wilczek}\ and\ \citenamefont
  {Zee}(1984)}]{wilczek1984appearance}%
  \BibitemOpen
  \bibfield  {author} {\bibinfo {author} {\bibfnamefont {F.}~\bibnamefont
  {Wilczek}}\ and\ \bibinfo {author} {\bibfnamefont {A.}~\bibnamefont {Zee}},\
  }\href {https://doi.org/10.1103/PhysRevLett.52.2111} {\bibfield  {journal}
  {\bibinfo  {journal} {Phys. Rev. Lett.}\ }\textbf {\bibinfo {volume} {52}},\
  \bibinfo {pages} {2111} (\bibinfo {year} {1984})}\BibitemShut {NoStop}%
\bibitem [{\citenamefont {Mead}(1992)}]{mead1992geometric}%
  \BibitemOpen
  \bibfield  {author} {\bibinfo {author} {\bibfnamefont {C.~A.}\ \bibnamefont
  {Mead}},\ }\href {https://doi.org/10.1103/RevModPhys.64.51} {\bibfield
  {journal} {\bibinfo  {journal} {Rev. Mod. Phys.}\ }\textbf {\bibinfo {volume}
  {64}},\ \bibinfo {pages} {51} (\bibinfo {year} {1992})}\BibitemShut {NoStop}%
\bibitem [{\citenamefont {Bohm}\ \emph {et~al.}(2013)\citenamefont {Bohm},
  \citenamefont {Mostafazadeh}, \citenamefont {Koizumi}, \citenamefont {Niu},\
  and\ \citenamefont {Zwanziger}}]{bohm2013geometric}%
  \BibitemOpen
  \bibfield  {author} {\bibinfo {author} {\bibfnamefont {A.}~\bibnamefont
  {Bohm}}, \bibinfo {author} {\bibfnamefont {A.}~\bibnamefont {Mostafazadeh}},
  \bibinfo {author} {\bibfnamefont {H.}~\bibnamefont {Koizumi}}, \bibinfo
  {author} {\bibfnamefont {Q.}~\bibnamefont {Niu}},\ and\ \bibinfo {author}
  {\bibfnamefont {J.}~\bibnamefont {Zwanziger}},\ }\href@noop {} {\emph
  {\bibinfo {title} {The Geometric Phase in Quantum Systems: Foundations,
  Mathematical Concepts, and Applications in Molecular and Condensed Matter
  Physics}}}\ (\bibinfo  {publisher} {Springer Science \& Business Media},\
  \bibinfo {year} {2013})\BibitemShut {NoStop}%
\bibitem [{\citenamefont {Wilson}(1974)}]{wilson1974confinement}%
  \BibitemOpen
  \bibfield  {author} {\bibinfo {author} {\bibfnamefont {K.~G.}\ \bibnamefont
  {Wilson}},\ }\href {https://doi.org/10.1103/PhysRevD.10.2445} {\bibfield
  {journal} {\bibinfo  {journal} {Phys. Rev. D}\ }\textbf {\bibinfo {volume}
  {10}},\ \bibinfo {pages} {2445} (\bibinfo {year} {1974})}\BibitemShut
  {NoStop}%
\bibitem [{\citenamefont {Vanderbilt}(2018)}]{vanderbilt2018berry}%
  \BibitemOpen
  \bibfield  {author} {\bibinfo {author} {\bibfnamefont {D.}~\bibnamefont
  {Vanderbilt}},\ }\href@noop {} {\emph {\bibinfo {title} {Berry phases in
  electronic structure theory: electric polarization, orbital magnetization and
  topological insulators}}}\ (\bibinfo  {publisher} {Cambridge University
  Press},\ \bibinfo {year} {2018})\BibitemShut {NoStop}%
\bibitem [{\citenamefont {Alexandradinata}\ \emph {et~al.}(2020)\citenamefont
  {Alexandradinata}, \citenamefont {H{\"o}ller}, \citenamefont {Wang},
  \citenamefont {Cheng},\ and\ \citenamefont
  {Lu}}]{alexandradinata2020crystallographic}%
  \BibitemOpen
  \bibfield  {author} {\bibinfo {author} {\bibfnamefont {A.}~\bibnamefont
  {Alexandradinata}}, \bibinfo {author} {\bibfnamefont {J.}~\bibnamefont
  {H{\"o}ller}}, \bibinfo {author} {\bibfnamefont {C.}~\bibnamefont {Wang}},
  \bibinfo {author} {\bibfnamefont {H.}~\bibnamefont {Cheng}},\ and\ \bibinfo
  {author} {\bibfnamefont {L.}~\bibnamefont {Lu}},\ }\href
  {https://doi.org/10.1103/PhysRevB.102.115117} {\bibfield  {journal} {\bibinfo
   {journal} {Physical Review B}\ }\textbf {\bibinfo {volume} {102}},\ \bibinfo
  {pages} {115117} (\bibinfo {year} {2020})}\BibitemShut {NoStop}%
\bibitem [{\citenamefont {Christensen}\ \emph {et~al.}(2022)\citenamefont
  {Christensen}, \citenamefont {Po}, \citenamefont {Joannopoulos},\ and\
  \citenamefont {Solja{\v{c}}i{\'c}}}]{christensen2022location}%
  \BibitemOpen
  \bibfield  {author} {\bibinfo {author} {\bibfnamefont {T.}~\bibnamefont
  {Christensen}}, \bibinfo {author} {\bibfnamefont {H.~C.}\ \bibnamefont {Po}},
  \bibinfo {author} {\bibfnamefont {J.~D.}\ \bibnamefont {Joannopoulos}},\ and\
  \bibinfo {author} {\bibfnamefont {M.}~\bibnamefont {Solja{\v{c}}i{\'c}}},\
  }\href {https://doi.org/10.1103/PhysRevX.12.021066} {\bibfield  {journal}
  {\bibinfo  {journal} {Physical Review X}\ }\textbf {\bibinfo {volume} {12}},\
  \bibinfo {pages} {021066} (\bibinfo {year} {2022})}\BibitemShut {NoStop}%
\bibitem [{\citenamefont {Gupta}\ and\ \citenamefont
  {Bradlyn}(2022)}]{gupta2022wannier}%
  \BibitemOpen
  \bibfield  {author} {\bibinfo {author} {\bibfnamefont {V.}~\bibnamefont
  {Gupta}}\ and\ \bibinfo {author} {\bibfnamefont {B.}~\bibnamefont
  {Bradlyn}},\ }\href {https://doi.org/10.1103/PhysRevA.105.053521} {\bibfield
  {journal} {\bibinfo  {journal} {Physical Review A}\ }\textbf {\bibinfo
  {volume} {105}},\ \bibinfo {pages} {053521} (\bibinfo {year}
  {2022})}\BibitemShut {NoStop}%
\bibitem [{\citenamefont {Sugawa}\ \emph {et~al.}(2018)\citenamefont {Sugawa},
  \citenamefont {Salces-Carcoba}, \citenamefont {Perry}, \citenamefont {Yue},\
  and\ \citenamefont {Spielman}}]{sugawa2018second}%
  \BibitemOpen
  \bibfield  {author} {\bibinfo {author} {\bibfnamefont {S.}~\bibnamefont
  {Sugawa}}, \bibinfo {author} {\bibfnamefont {F.}~\bibnamefont
  {Salces-Carcoba}}, \bibinfo {author} {\bibfnamefont {A.~R.}\ \bibnamefont
  {Perry}}, \bibinfo {author} {\bibfnamefont {Y.}~\bibnamefont {Yue}},\ and\
  \bibinfo {author} {\bibfnamefont {I.}~\bibnamefont {Spielman}},\ }\href
  {https://doi.org/10.1126/science.aam9031} {\bibfield  {journal} {\bibinfo
  {journal} {Science}\ }\textbf {\bibinfo {volume} {360}},\ \bibinfo {pages}
  {1429} (\bibinfo {year} {2018})}\BibitemShut {NoStop}%
\bibitem [{\citenamefont {Lu}\ \emph {et~al.}(2018)\citenamefont {Lu},
  \citenamefont {Gao},\ and\ \citenamefont {Wang}}]{lu2018topological}%
  \BibitemOpen
  \bibfield  {author} {\bibinfo {author} {\bibfnamefont {L.}~\bibnamefont
  {Lu}}, \bibinfo {author} {\bibfnamefont {H.}~\bibnamefont {Gao}},\ and\
  \bibinfo {author} {\bibfnamefont {Z.}~\bibnamefont {Wang}},\ }\href
  {https://doi.org/10.1038/s41467-018-07817-3} {\bibfield  {journal} {\bibinfo
  {journal} {Nature communications}\ }\textbf {\bibinfo {volume} {9}},\
  \bibinfo {pages} {5384} (\bibinfo {year} {2018})}\BibitemShut {NoStop}%
\bibitem [{\citenamefont {Ma}\ \emph {et~al.}(2021)\citenamefont {Ma},
  \citenamefont {Bi}, \citenamefont {Guo}, \citenamefont {Yang}, \citenamefont
  {You}, \citenamefont {Feng}, \citenamefont {Sun},\ and\ \citenamefont
  {Zhang}}]{ma2021linked}%
  \BibitemOpen
  \bibfield  {author} {\bibinfo {author} {\bibfnamefont {S.}~\bibnamefont
  {Ma}}, \bibinfo {author} {\bibfnamefont {Y.}~\bibnamefont {Bi}}, \bibinfo
  {author} {\bibfnamefont {Q.}~\bibnamefont {Guo}}, \bibinfo {author}
  {\bibfnamefont {B.}~\bibnamefont {Yang}}, \bibinfo {author} {\bibfnamefont
  {O.}~\bibnamefont {You}}, \bibinfo {author} {\bibfnamefont {J.}~\bibnamefont
  {Feng}}, \bibinfo {author} {\bibfnamefont {H.-B.}\ \bibnamefont {Sun}},\ and\
  \bibinfo {author} {\bibfnamefont {S.}~\bibnamefont {Zhang}},\ }\href
  {https://doi.org/10.1126/science.abi7803} {\bibfield  {journal} {\bibinfo
  {journal} {Science}\ }\textbf {\bibinfo {volume} {373}},\ \bibinfo {pages}
  {572} (\bibinfo {year} {2021})}\BibitemShut {NoStop}%
\bibitem [{\citenamefont {Chiu}\ \emph {et~al.}(2016)\citenamefont {Chiu},
  \citenamefont {Teo}, \citenamefont {Schnyder},\ and\ \citenamefont
  {Ryu}}]{chiu2016classification}%
  \BibitemOpen
  \bibfield  {author} {\bibinfo {author} {\bibfnamefont {C.-K.}\ \bibnamefont
  {Chiu}}, \bibinfo {author} {\bibfnamefont {J.~C.}\ \bibnamefont {Teo}},
  \bibinfo {author} {\bibfnamefont {A.~P.}\ \bibnamefont {Schnyder}},\ and\
  \bibinfo {author} {\bibfnamefont {S.}~\bibnamefont {Ryu}},\ }\href
  {https://doi.org/10.1103/RevModPhys.88.035005} {\bibfield  {journal}
  {\bibinfo  {journal} {Reviews of Modern Physics}\ }\textbf {\bibinfo {volume}
  {88}},\ \bibinfo {pages} {035005} (\bibinfo {year} {2016})}\BibitemShut
  {NoStop}%
\bibitem [{\citenamefont {Wu}\ \emph {et~al.}(2019)\citenamefont {Wu},
  \citenamefont {Soluyanov},\ and\ \citenamefont {Bzdušek}}]{RN6}%
  \BibitemOpen
  \bibfield  {author} {\bibinfo {author} {\bibfnamefont {Q.}~\bibnamefont
  {Wu}}, \bibinfo {author} {\bibfnamefont {A.~A.}\ \bibnamefont {Soluyanov}},\
  and\ \bibinfo {author} {\bibfnamefont {T.}~\bibnamefont {Bzdušek}},\ }\href
  {https://doi.org/10.1126/science.aau8740} {\bibfield  {journal} {\bibinfo
  {journal} {Science}\ }\textbf {\bibinfo {volume} {365}},\ \bibinfo {pages}
  {1273} (\bibinfo {year} {2019})}\BibitemShut {NoStop}%
\bibitem [{\citenamefont {Guo}\ \emph {et~al.}(2021)\citenamefont {Guo},
  \citenamefont {Jiang}, \citenamefont {Zhang}, \citenamefont {Zhang},
  \citenamefont {Zhang}, \citenamefont {Yang}, \citenamefont {Zhang},\ and\
  \citenamefont {Chan}}]{RN21}%
  \BibitemOpen
  \bibfield  {author} {\bibinfo {author} {\bibfnamefont {Q.}~\bibnamefont
  {Guo}}, \bibinfo {author} {\bibfnamefont {T.}~\bibnamefont {Jiang}}, \bibinfo
  {author} {\bibfnamefont {R.-Y.}\ \bibnamefont {Zhang}}, \bibinfo {author}
  {\bibfnamefont {L.}~\bibnamefont {Zhang}}, \bibinfo {author} {\bibfnamefont
  {Z.-Q.}\ \bibnamefont {Zhang}}, \bibinfo {author} {\bibfnamefont
  {B.}~\bibnamefont {Yang}}, \bibinfo {author} {\bibfnamefont {S.}~\bibnamefont
  {Zhang}},\ and\ \bibinfo {author} {\bibfnamefont {C.~T.}\ \bibnamefont
  {Chan}},\ }\href {https://doi.org/10.1038/s41586-021-03521-3} {\bibfield
  {journal} {\bibinfo  {journal} {Nature}\ }\textbf {\bibinfo {volume} {594}},\
  \bibinfo {pages} {195} (\bibinfo {year} {2021})}\BibitemShut {NoStop}%
\bibitem [{\citenamefont {Yang}\ \emph
  {et~al.}(2020{\natexlab{a}})\citenamefont {Yang}, \citenamefont {Yang},
  \citenamefont {You}, \citenamefont {Chan}, \citenamefont {Mao}, \citenamefont
  {Guo}, \citenamefont {Ma}, \citenamefont {Xia}, \citenamefont {Fan},
  \citenamefont {Xiang},\ and\ \citenamefont {Zhang}}]{RN11}%
  \BibitemOpen
  \bibfield  {author} {\bibinfo {author} {\bibfnamefont {E.}~\bibnamefont
  {Yang}}, \bibinfo {author} {\bibfnamefont {B.}~\bibnamefont {Yang}}, \bibinfo
  {author} {\bibfnamefont {O.}~\bibnamefont {You}}, \bibinfo {author}
  {\bibfnamefont {H.-C.}\ \bibnamefont {Chan}}, \bibinfo {author}
  {\bibfnamefont {P.}~\bibnamefont {Mao}}, \bibinfo {author} {\bibfnamefont
  {Q.}~\bibnamefont {Guo}}, \bibinfo {author} {\bibfnamefont {S.}~\bibnamefont
  {Ma}}, \bibinfo {author} {\bibfnamefont {L.}~\bibnamefont {Xia}}, \bibinfo
  {author} {\bibfnamefont {D.}~\bibnamefont {Fan}}, \bibinfo {author}
  {\bibfnamefont {Y.}~\bibnamefont {Xiang}},\ and\ \bibinfo {author}
  {\bibfnamefont {S.}~\bibnamefont {Zhang}},\ }\href
  {https://doi.org/10.1103/PhysRevLett.125.033901} {\bibfield  {journal}
  {\bibinfo  {journal} {Physical Review Letters}\ }\textbf {\bibinfo {volume}
  {125}},\ \bibinfo {pages} {033901} (\bibinfo {year}
  {2020}{\natexlab{a}})}\BibitemShut {NoStop}%
\bibitem [{\citenamefont {Jiang}\ \emph
  {et~al.}(2021{\natexlab{a}})\citenamefont {Jiang}, \citenamefont {Bouhon},
  \citenamefont {Lin}, \citenamefont {Zhou}, \citenamefont {Hou}, \citenamefont
  {Li}, \citenamefont {Slager},\ and\ \citenamefont {Jiang}}]{RN9}%
  \BibitemOpen
  \bibfield  {author} {\bibinfo {author} {\bibfnamefont {B.}~\bibnamefont
  {Jiang}}, \bibinfo {author} {\bibfnamefont {A.}~\bibnamefont {Bouhon}},
  \bibinfo {author} {\bibfnamefont {Z.-K.}\ \bibnamefont {Lin}}, \bibinfo
  {author} {\bibfnamefont {X.}~\bibnamefont {Zhou}}, \bibinfo {author}
  {\bibfnamefont {B.}~\bibnamefont {Hou}}, \bibinfo {author} {\bibfnamefont
  {F.}~\bibnamefont {Li}}, \bibinfo {author} {\bibfnamefont {R.-J.}\
  \bibnamefont {Slager}},\ and\ \bibinfo {author} {\bibfnamefont {J.-H.}\
  \bibnamefont {Jiang}},\ }\href {https://doi.org/10.1038/s41567-021-01340-x}
  {\bibfield  {journal} {\bibinfo  {journal} {Nature Physics}\ }\textbf
  {\bibinfo {volume} {17}},\ \bibinfo {pages} {1239} (\bibinfo {year}
  {2021}{\natexlab{a}})}\BibitemShut {NoStop}%
\bibitem [{\citenamefont {Chen}\ \emph
  {et~al.}(2019{\natexlab{a}})\citenamefont {Chen}, \citenamefont {Zhang},
  \citenamefont {Xiong}, \citenamefont {Hang}, \citenamefont {Li},
  \citenamefont {Shen},\ and\ \citenamefont {Chan}}]{chen2019non}%
  \BibitemOpen
  \bibfield  {author} {\bibinfo {author} {\bibfnamefont {Y.}~\bibnamefont
  {Chen}}, \bibinfo {author} {\bibfnamefont {R.-Y.}\ \bibnamefont {Zhang}},
  \bibinfo {author} {\bibfnamefont {Z.}~\bibnamefont {Xiong}}, \bibinfo
  {author} {\bibfnamefont {Z.~H.}\ \bibnamefont {Hang}}, \bibinfo {author}
  {\bibfnamefont {J.}~\bibnamefont {Li}}, \bibinfo {author} {\bibfnamefont
  {J.~Q.}\ \bibnamefont {Shen}},\ and\ \bibinfo {author} {\bibfnamefont
  {C.~T.}\ \bibnamefont {Chan}},\ }\href
  {https://doi.org/10.1038/s41467-019-10974-8} {\bibfield  {journal} {\bibinfo
  {journal} {Nature communications}\ }\textbf {\bibinfo {volume} {10}},\
  \bibinfo {pages} {3125} (\bibinfo {year} {2019}{\natexlab{a}})}\BibitemShut
  {NoStop}%
\bibitem [{\citenamefont {Ter{\c{c}}as}\ \emph {et~al.}(2014)\citenamefont
  {Ter{\c{c}}as}, \citenamefont {Flayac}, \citenamefont {Solnyshkov},\ and\
  \citenamefont {Malpuech}}]{terccas2014non}%
  \BibitemOpen
  \bibfield  {author} {\bibinfo {author} {\bibfnamefont {H.}~\bibnamefont
  {Ter{\c{c}}as}}, \bibinfo {author} {\bibfnamefont {H.}~\bibnamefont
  {Flayac}}, \bibinfo {author} {\bibfnamefont {D.}~\bibnamefont {Solnyshkov}},\
  and\ \bibinfo {author} {\bibfnamefont {G.}~\bibnamefont {Malpuech}},\ }\href
  {https://doi.org/10.1103/PhysRevLett.112.066402} {\bibfield  {journal}
  {\bibinfo  {journal} {Physical review letters}\ }\textbf {\bibinfo {volume}
  {112}},\ \bibinfo {pages} {066402} (\bibinfo {year} {2014})}\BibitemShut
  {NoStop}%
\bibitem [{\citenamefont {Yang}\ \emph {et~al.}(2019)\citenamefont {Yang},
  \citenamefont {Peng}, \citenamefont {Zhu}, \citenamefont {Buljan},
  \citenamefont {Joannopoulos}, \citenamefont {Zhen},\ and\ \citenamefont
  {Solja{\v{c}}i{\'c}}}]{yang2019synthesis}%
  \BibitemOpen
  \bibfield  {author} {\bibinfo {author} {\bibfnamefont {Y.}~\bibnamefont
  {Yang}}, \bibinfo {author} {\bibfnamefont {C.}~\bibnamefont {Peng}}, \bibinfo
  {author} {\bibfnamefont {D.}~\bibnamefont {Zhu}}, \bibinfo {author}
  {\bibfnamefont {H.}~\bibnamefont {Buljan}}, \bibinfo {author} {\bibfnamefont
  {J.~D.}\ \bibnamefont {Joannopoulos}}, \bibinfo {author} {\bibfnamefont
  {B.}~\bibnamefont {Zhen}},\ and\ \bibinfo {author} {\bibfnamefont
  {M.}~\bibnamefont {Solja{\v{c}}i{\'c}}},\ }\href
  {https://doi.org/10.1126/science.aay3183} {\bibfield  {journal} {\bibinfo
  {journal} {Science}\ }\textbf {\bibinfo {volume} {365}},\ \bibinfo {pages}
  {1021} (\bibinfo {year} {2019})}\BibitemShut {NoStop}%
\bibitem [{\citenamefont {Sedov}\ \emph {et~al.}(2018)\citenamefont {Sedov},
  \citenamefont {Rubo},\ and\ \citenamefont
  {Kavokin}}]{sedov2018zitterbewegung}%
  \BibitemOpen
  \bibfield  {author} {\bibinfo {author} {\bibfnamefont {E.}~\bibnamefont
  {Sedov}}, \bibinfo {author} {\bibfnamefont {Y.}~\bibnamefont {Rubo}},\ and\
  \bibinfo {author} {\bibfnamefont {A.}~\bibnamefont {Kavokin}},\ }\href
  {https://doi.org/10.1103/PhysRevB.97.245312} {\bibfield  {journal} {\bibinfo
  {journal} {Physical Review B}\ }\textbf {\bibinfo {volume} {97}},\ \bibinfo
  {pages} {245312} (\bibinfo {year} {2018})}\BibitemShut {NoStop}%
\bibitem [{\citenamefont {Polimeno}\ \emph
  {et~al.}(2021{\natexlab{a}})\citenamefont {Polimeno}, \citenamefont
  {Fieramosca}, \citenamefont {Lerario}, \citenamefont {De~Marco},
  \citenamefont {De~Giorgi}, \citenamefont {Ballarini}, \citenamefont
  {Dominici}, \citenamefont {Ardizzone}, \citenamefont {Pugliese},
  \citenamefont {Prontera} \emph {et~al.}}]{polimeno2021experimental}%
  \BibitemOpen
  \bibfield  {author} {\bibinfo {author} {\bibfnamefont {L.}~\bibnamefont
  {Polimeno}}, \bibinfo {author} {\bibfnamefont {A.}~\bibnamefont
  {Fieramosca}}, \bibinfo {author} {\bibfnamefont {G.}~\bibnamefont {Lerario}},
  \bibinfo {author} {\bibfnamefont {L.}~\bibnamefont {De~Marco}}, \bibinfo
  {author} {\bibfnamefont {M.}~\bibnamefont {De~Giorgi}}, \bibinfo {author}
  {\bibfnamefont {D.}~\bibnamefont {Ballarini}}, \bibinfo {author}
  {\bibfnamefont {L.}~\bibnamefont {Dominici}}, \bibinfo {author}
  {\bibfnamefont {V.}~\bibnamefont {Ardizzone}}, \bibinfo {author}
  {\bibfnamefont {M.}~\bibnamefont {Pugliese}}, \bibinfo {author}
  {\bibfnamefont {C.}~\bibnamefont {Prontera}}, \emph {et~al.},\ }\href
  {https://doi.org/10.1364/OPTICA.427088} {\bibfield  {journal} {\bibinfo
  {journal} {Optica}\ }\textbf {\bibinfo {volume} {8}},\ \bibinfo {pages}
  {1442} (\bibinfo {year} {2021}{\natexlab{a}})}\BibitemShut {NoStop}%
\bibitem [{\citenamefont {Lovet}\ \emph {et~al.}(2022)\citenamefont {Lovet},
  \citenamefont {Walker}, \citenamefont {Osipov}, \citenamefont {Yulin},
  \citenamefont {Naik}, \citenamefont {Shelykh}, \citenamefont {Skolnick},\
  and\ \citenamefont {Krizhanovskii}}]{lovet2022observation}%
  \BibitemOpen
  \bibfield  {author} {\bibinfo {author} {\bibfnamefont {S.}~\bibnamefont
  {Lovet}}, \bibinfo {author} {\bibfnamefont {P.~M.}\ \bibnamefont {Walker}},
  \bibinfo {author} {\bibfnamefont {A.}~\bibnamefont {Osipov}}, \bibinfo
  {author} {\bibfnamefont {A.}~\bibnamefont {Yulin}}, \bibinfo {author}
  {\bibfnamefont {P.~U.}\ \bibnamefont {Naik}}, \bibinfo {author}
  {\bibfnamefont {I.~A.}\ \bibnamefont {Shelykh}}, \bibinfo {author}
  {\bibfnamefont {M.~S.}\ \bibnamefont {Skolnick}},\ and\ \bibinfo {author}
  {\bibfnamefont {D.~N.}\ \bibnamefont {Krizhanovskii}},\ }\bibfield  {journal}
  {\bibinfo  {journal} {arXiv preprint arXiv:2211.09907}\ }\href
  {https://doi.org/10.48550/arXiv.2211.09907} {10.48550/arXiv.2211.09907}
  (\bibinfo {year} {2022})\BibitemShut {NoStop}%
\bibitem [{\citenamefont {Whittaker}\ \emph {et~al.}(2021)\citenamefont
  {Whittaker}, \citenamefont {Dowling}, \citenamefont {Nalitov}, \citenamefont
  {Yulin}, \citenamefont {Royall}, \citenamefont {Clarke}, \citenamefont
  {Skolnick}, \citenamefont {Shelykh},\ and\ \citenamefont
  {Krizhanovskii}}]{whittaker2021optical}%
  \BibitemOpen
  \bibfield  {author} {\bibinfo {author} {\bibfnamefont {C.}~\bibnamefont
  {Whittaker}}, \bibinfo {author} {\bibfnamefont {T.}~\bibnamefont {Dowling}},
  \bibinfo {author} {\bibfnamefont {A.}~\bibnamefont {Nalitov}}, \bibinfo
  {author} {\bibfnamefont {A.}~\bibnamefont {Yulin}}, \bibinfo {author}
  {\bibfnamefont {B.}~\bibnamefont {Royall}}, \bibinfo {author} {\bibfnamefont
  {E.}~\bibnamefont {Clarke}}, \bibinfo {author} {\bibfnamefont
  {M.}~\bibnamefont {Skolnick}}, \bibinfo {author} {\bibfnamefont
  {I.}~\bibnamefont {Shelykh}},\ and\ \bibinfo {author} {\bibfnamefont
  {D.}~\bibnamefont {Krizhanovskii}},\ }\href
  {https://doi.org/10.1038/s41566-020-00729-z} {\bibfield  {journal} {\bibinfo
  {journal} {Nature Photonics}\ }\textbf {\bibinfo {volume} {15}},\ \bibinfo
  {pages} {193} (\bibinfo {year} {2021})}\BibitemShut {NoStop}%
\bibitem [{\citenamefont {Osterloh}\ \emph {et~al.}(2005)\citenamefont
  {Osterloh}, \citenamefont {Baig}, \citenamefont {Santos}, \citenamefont
  {Zoller},\ and\ \citenamefont {Lewenstein}}]{osterloh2005cold}%
  \BibitemOpen
  \bibfield  {author} {\bibinfo {author} {\bibfnamefont {K.}~\bibnamefont
  {Osterloh}}, \bibinfo {author} {\bibfnamefont {M.}~\bibnamefont {Baig}},
  \bibinfo {author} {\bibfnamefont {L.}~\bibnamefont {Santos}}, \bibinfo
  {author} {\bibfnamefont {P.}~\bibnamefont {Zoller}},\ and\ \bibinfo {author}
  {\bibfnamefont {M.}~\bibnamefont {Lewenstein}},\ }\href
  {https://doi.org/10.1103/PhysRevLett.95.010403} {\bibfield  {journal}
  {\bibinfo  {journal} {Phys. Rev. Lett.}\ }\textbf {\bibinfo {volume} {95}},\
  \bibinfo {pages} {010403} (\bibinfo {year} {2005})}\BibitemShut {NoStop}%
\bibitem [{\citenamefont {Wu}\ \emph {et~al.}(2022)\citenamefont {Wu},
  \citenamefont {Wang}, \citenamefont {Biao}, \citenamefont {Fei},
  \citenamefont {Zhang}, \citenamefont {Yin}, \citenamefont {Hu}, \citenamefont
  {Song}, \citenamefont {Wu}, \citenamefont {Song} \emph {et~al.}}]{wu2022non}%
  \BibitemOpen
  \bibfield  {author} {\bibinfo {author} {\bibfnamefont {J.}~\bibnamefont
  {Wu}}, \bibinfo {author} {\bibfnamefont {Z.}~\bibnamefont {Wang}}, \bibinfo
  {author} {\bibfnamefont {Y.}~\bibnamefont {Biao}}, \bibinfo {author}
  {\bibfnamefont {F.}~\bibnamefont {Fei}}, \bibinfo {author} {\bibfnamefont
  {S.}~\bibnamefont {Zhang}}, \bibinfo {author} {\bibfnamefont
  {Z.}~\bibnamefont {Yin}}, \bibinfo {author} {\bibfnamefont {Y.}~\bibnamefont
  {Hu}}, \bibinfo {author} {\bibfnamefont {Z.}~\bibnamefont {Song}}, \bibinfo
  {author} {\bibfnamefont {T.}~\bibnamefont {Wu}}, \bibinfo {author}
  {\bibfnamefont {F.}~\bibnamefont {Song}}, \emph {et~al.},\ }\href
  {https://doi.org/10.1038/s41928-022-00833-8} {\bibfield  {journal} {\bibinfo
  {journal} {Nature Electronics}\ }\textbf {\bibinfo {volume} {5}},\ \bibinfo
  {pages} {635} (\bibinfo {year} {2022})}\BibitemShut {NoStop}%
\bibitem [{\citenamefont {Noh}\ \emph {et~al.}(2020)\citenamefont {Noh},
  \citenamefont {Schuster}, \citenamefont {Iadecola}, \citenamefont {Huang},
  \citenamefont {Wang}, \citenamefont {Chen}, \citenamefont {Chamon},\ and\
  \citenamefont {Rechtsman}}]{noh2020braiding}%
  \BibitemOpen
  \bibfield  {author} {\bibinfo {author} {\bibfnamefont {J.}~\bibnamefont
  {Noh}}, \bibinfo {author} {\bibfnamefont {T.}~\bibnamefont {Schuster}},
  \bibinfo {author} {\bibfnamefont {T.}~\bibnamefont {Iadecola}}, \bibinfo
  {author} {\bibfnamefont {S.}~\bibnamefont {Huang}}, \bibinfo {author}
  {\bibfnamefont {M.}~\bibnamefont {Wang}}, \bibinfo {author} {\bibfnamefont
  {K.~P.}\ \bibnamefont {Chen}}, \bibinfo {author} {\bibfnamefont
  {C.}~\bibnamefont {Chamon}},\ and\ \bibinfo {author} {\bibfnamefont {M.~C.}\
  \bibnamefont {Rechtsman}},\ }\href
  {https://doi.org/10.1038/s41567-020-1007-5} {\bibfield  {journal} {\bibinfo
  {journal} {Nature Physics}\ }\textbf {\bibinfo {volume} {16}},\ \bibinfo
  {pages} {989} (\bibinfo {year} {2020})}\BibitemShut {NoStop}%
\bibitem [{\citenamefont {Chen}\ \emph
  {et~al.}(2022{\natexlab{a}})\citenamefont {Chen}, \citenamefont {Zhang},
  \citenamefont {Chan},\ and\ \citenamefont {Ma}}]{chen2022classical}%
  \BibitemOpen
  \bibfield  {author} {\bibinfo {author} {\bibfnamefont {Z.-G.}\ \bibnamefont
  {Chen}}, \bibinfo {author} {\bibfnamefont {R.-Y.}\ \bibnamefont {Zhang}},
  \bibinfo {author} {\bibfnamefont {C.~T.}\ \bibnamefont {Chan}},\ and\
  \bibinfo {author} {\bibfnamefont {G.}~\bibnamefont {Ma}},\ }\href
  {https://doi.org/10.1038/s41567-021-01431-9} {\bibfield  {journal} {\bibinfo
  {journal} {Nature Physics}\ }\textbf {\bibinfo {volume} {18}},\ \bibinfo
  {pages} {179} (\bibinfo {year} {2022}{\natexlab{a}})}\BibitemShut {NoStop}%
\bibitem [{\citenamefont {Zhang}\ \emph
  {et~al.}(2022{\natexlab{a}})\citenamefont {Zhang}, \citenamefont {Yu},
  \citenamefont {Chen}, \citenamefont {Tian}, \citenamefont {Chen},
  \citenamefont {Sun},\ and\ \citenamefont {Ma}}]{zhang2022non}%
  \BibitemOpen
  \bibfield  {author} {\bibinfo {author} {\bibfnamefont {X.-L.}\ \bibnamefont
  {Zhang}}, \bibinfo {author} {\bibfnamefont {F.}~\bibnamefont {Yu}}, \bibinfo
  {author} {\bibfnamefont {Z.-G.}\ \bibnamefont {Chen}}, \bibinfo {author}
  {\bibfnamefont {Z.-N.}\ \bibnamefont {Tian}}, \bibinfo {author}
  {\bibfnamefont {Q.-D.}\ \bibnamefont {Chen}}, \bibinfo {author}
  {\bibfnamefont {H.-B.}\ \bibnamefont {Sun}},\ and\ \bibinfo {author}
  {\bibfnamefont {G.}~\bibnamefont {Ma}},\ }\href
  {https://doi.org/10.1038/s41566-022-00976-2} {\bibfield  {journal} {\bibinfo
  {journal} {Nature Photonics}\ }\textbf {\bibinfo {volume} {16}},\ \bibinfo
  {pages} {390} (\bibinfo {year} {2022}{\natexlab{a}})}\BibitemShut {NoStop}%
\bibitem [{\citenamefont {Brosco}\ \emph {et~al.}(2021)\citenamefont {Brosco},
  \citenamefont {Pilozzi}, \citenamefont {Fazio},\ and\ \citenamefont
  {Conti}}]{brosco2021non}%
  \BibitemOpen
  \bibfield  {author} {\bibinfo {author} {\bibfnamefont {V.}~\bibnamefont
  {Brosco}}, \bibinfo {author} {\bibfnamefont {L.}~\bibnamefont {Pilozzi}},
  \bibinfo {author} {\bibfnamefont {R.}~\bibnamefont {Fazio}},\ and\ \bibinfo
  {author} {\bibfnamefont {C.}~\bibnamefont {Conti}},\ }\href
  {https://doi.org/10.1103/PhysRevA.103.063518} {\bibfield  {journal} {\bibinfo
   {journal} {Physical Review A}\ }\textbf {\bibinfo {volume} {103}},\ \bibinfo
  {pages} {063518} (\bibinfo {year} {2021})}\BibitemShut {NoStop}%
\bibitem [{\citenamefont {Sun}\ \emph {et~al.}(2022)\citenamefont {Sun},
  \citenamefont {Zhang}, \citenamefont {Yu}, \citenamefont {Tian},
  \citenamefont {Chen},\ and\ \citenamefont {Sun}}]{sun2022non}%
  \BibitemOpen
  \bibfield  {author} {\bibinfo {author} {\bibfnamefont {Y.-K.}\ \bibnamefont
  {Sun}}, \bibinfo {author} {\bibfnamefont {X.-L.}\ \bibnamefont {Zhang}},
  \bibinfo {author} {\bibfnamefont {F.}~\bibnamefont {Yu}}, \bibinfo {author}
  {\bibfnamefont {Z.-N.}\ \bibnamefont {Tian}}, \bibinfo {author}
  {\bibfnamefont {Q.-D.}\ \bibnamefont {Chen}},\ and\ \bibinfo {author}
  {\bibfnamefont {H.-B.}\ \bibnamefont {Sun}},\ }\href
  {https://doi.org/10.1038/s41567-022-01669-x} {\bibfield  {journal} {\bibinfo
  {journal} {Nature Physics}\ }\textbf {\bibinfo {volume} {18}},\ \bibinfo
  {pages} {1080} (\bibinfo {year} {2022})}\BibitemShut {NoStop}%
\bibitem [{\citenamefont {You}\ \emph {et~al.}(2022)\citenamefont {You},
  \citenamefont {Liang}, \citenamefont {Xie}, \citenamefont {Gao},
  \citenamefont {Ye}, \citenamefont {Zhu},\ and\ \citenamefont
  {Zhang}}]{you2022observation}%
  \BibitemOpen
  \bibfield  {author} {\bibinfo {author} {\bibfnamefont {O.}~\bibnamefont
  {You}}, \bibinfo {author} {\bibfnamefont {S.}~\bibnamefont {Liang}}, \bibinfo
  {author} {\bibfnamefont {B.}~\bibnamefont {Xie}}, \bibinfo {author}
  {\bibfnamefont {W.}~\bibnamefont {Gao}}, \bibinfo {author} {\bibfnamefont
  {W.}~\bibnamefont {Ye}}, \bibinfo {author} {\bibfnamefont {J.}~\bibnamefont
  {Zhu}},\ and\ \bibinfo {author} {\bibfnamefont {S.}~\bibnamefont {Zhang}},\
  }\href {https://doi.org/10.1103/PhysRevLett.128.244302} {\bibfield  {journal}
  {\bibinfo  {journal} {Physical Review Letters}\ }\textbf {\bibinfo {volume}
  {128}},\ \bibinfo {pages} {244302} (\bibinfo {year} {2022})}\BibitemShut
  {NoStop}%
\bibitem [{\citenamefont {Khanikaev}\ \emph {et~al.}(2013)\citenamefont
  {Khanikaev}, \citenamefont {Hossein~Mousavi}, \citenamefont {Tse},
  \citenamefont {Kargarian}, \citenamefont {MacDonald},\ and\ \citenamefont
  {Shvets}}]{khanikaev2013photonic}%
  \BibitemOpen
  \bibfield  {author} {\bibinfo {author} {\bibfnamefont {A.~B.}\ \bibnamefont
  {Khanikaev}}, \bibinfo {author} {\bibfnamefont {S.}~\bibnamefont
  {Hossein~Mousavi}}, \bibinfo {author} {\bibfnamefont {W.-K.}\ \bibnamefont
  {Tse}}, \bibinfo {author} {\bibfnamefont {M.}~\bibnamefont {Kargarian}},
  \bibinfo {author} {\bibfnamefont {A.~H.}\ \bibnamefont {MacDonald}},\ and\
  \bibinfo {author} {\bibfnamefont {G.}~\bibnamefont {Shvets}},\ }\href
  {https://doi.org/10.1038/nmat3520} {\bibfield  {journal} {\bibinfo  {journal}
  {Nature materials}\ }\textbf {\bibinfo {volume} {12}},\ \bibinfo {pages}
  {233} (\bibinfo {year} {2013})}\BibitemShut {NoStop}%
\bibitem [{\citenamefont {He}\ \emph {et~al.}(2016)\citenamefont {He},
  \citenamefont {Sun}, \citenamefont {Liu}, \citenamefont {Lu}, \citenamefont
  {Chen}, \citenamefont {Feng},\ and\ \citenamefont {Chen}}]{he2016photonic}%
  \BibitemOpen
  \bibfield  {author} {\bibinfo {author} {\bibfnamefont {C.}~\bibnamefont
  {He}}, \bibinfo {author} {\bibfnamefont {X.-C.}\ \bibnamefont {Sun}},
  \bibinfo {author} {\bibfnamefont {X.-P.}\ \bibnamefont {Liu}}, \bibinfo
  {author} {\bibfnamefont {M.-H.}\ \bibnamefont {Lu}}, \bibinfo {author}
  {\bibfnamefont {Y.}~\bibnamefont {Chen}}, \bibinfo {author} {\bibfnamefont
  {L.}~\bibnamefont {Feng}},\ and\ \bibinfo {author} {\bibfnamefont {Y.-F.}\
  \bibnamefont {Chen}},\ }\href {https://doi.org/10.1073/pnas.1525502113}
  {\bibfield  {journal} {\bibinfo  {journal} {Proceedings of the National
  Academy of Sciences}\ }\textbf {\bibinfo {volume} {113}},\ \bibinfo {pages}
  {4924} (\bibinfo {year} {2016})}\BibitemShut {NoStop}%
\bibitem [{\citenamefont {Silveirinha}(2017)}]{silveirinha2017p}%
  \BibitemOpen
  \bibfield  {author} {\bibinfo {author} {\bibfnamefont {M.~G.}\ \bibnamefont
  {Silveirinha}},\ }\href {https://doi.org/10.1103/PhysRevB.95.035153}
  {\bibfield  {journal} {\bibinfo  {journal} {Physical Review B}\ }\textbf
  {\bibinfo {volume} {95}},\ \bibinfo {pages} {035153} (\bibinfo {year}
  {2017})}\BibitemShut {NoStop}%
\bibitem [{\citenamefont {Liu}\ and\ \citenamefont {Li}(2015)}]{liu2015gauge}%
  \BibitemOpen
  \bibfield  {author} {\bibinfo {author} {\bibfnamefont {F.}~\bibnamefont
  {Liu}}\ and\ \bibinfo {author} {\bibfnamefont {J.}~\bibnamefont {Li}},\
  }\href {https://doi.org/10.1103/PhysRevLett.114.103902} {\bibfield  {journal}
  {\bibinfo  {journal} {Physical Review Letters}\ }\textbf {\bibinfo {volume}
  {114}},\ \bibinfo {pages} {103902} (\bibinfo {year} {2015})}\BibitemShut
  {NoStop}%
\bibitem [{\citenamefont {Hafezi}\ \emph {et~al.}(2011)\citenamefont {Hafezi},
  \citenamefont {Demler}, \citenamefont {Lukin},\ and\ \citenamefont
  {Taylor}}]{hafezi2011robust}%
  \BibitemOpen
  \bibfield  {author} {\bibinfo {author} {\bibfnamefont {M.}~\bibnamefont
  {Hafezi}}, \bibinfo {author} {\bibfnamefont {E.~A.}\ \bibnamefont {Demler}},
  \bibinfo {author} {\bibfnamefont {M.~D.}\ \bibnamefont {Lukin}},\ and\
  \bibinfo {author} {\bibfnamefont {J.~M.}\ \bibnamefont {Taylor}},\ }\href
  {https://doi.org/10.1038/nphys2063} {\bibfield  {journal} {\bibinfo
  {journal} {Nat. Phys.}\ }\textbf {\bibinfo {volume} {7}},\ \bibinfo {pages}
  {907} (\bibinfo {year} {2011})}\BibitemShut {NoStop}%
\bibitem [{\citenamefont {Hafezi}\ \emph {et~al.}(2013)\citenamefont {Hafezi},
  \citenamefont {Mittal}, \citenamefont {Fan}, \citenamefont {Migdall},\ and\
  \citenamefont {Taylor}}]{hafezi2013imaging}%
  \BibitemOpen
  \bibfield  {author} {\bibinfo {author} {\bibfnamefont {M.}~\bibnamefont
  {Hafezi}}, \bibinfo {author} {\bibfnamefont {S.}~\bibnamefont {Mittal}},
  \bibinfo {author} {\bibfnamefont {J.}~\bibnamefont {Fan}}, \bibinfo {author}
  {\bibfnamefont {A.}~\bibnamefont {Migdall}},\ and\ \bibinfo {author}
  {\bibfnamefont {J.}~\bibnamefont {Taylor}},\ }\href
  {https://doi.org/10.1038/nphoton.2013.274} {\bibfield  {journal} {\bibinfo
  {journal} {Nat. Photon.}\ }\textbf {\bibinfo {volume} {7}},\ \bibinfo {pages}
  {1001} (\bibinfo {year} {2013})}\BibitemShut {NoStop}%
\bibitem [{\citenamefont {Mittal}\ \emph {et~al.}(2018)\citenamefont {Mittal},
  \citenamefont {Goldschmidt},\ and\ \citenamefont
  {Hafezi}}]{mittal2018topological}%
  \BibitemOpen
  \bibfield  {author} {\bibinfo {author} {\bibfnamefont {S.}~\bibnamefont
  {Mittal}}, \bibinfo {author} {\bibfnamefont {E.~A.}\ \bibnamefont
  {Goldschmidt}},\ and\ \bibinfo {author} {\bibfnamefont {M.}~\bibnamefont
  {Hafezi}},\ }\href {https://doi.org/10.1038/s41586-018-0478-3} {\bibfield
  {journal} {\bibinfo  {journal} {Nature}\ }\textbf {\bibinfo {volume} {561}},\
  \bibinfo {pages} {502} (\bibinfo {year} {2018})}\BibitemShut {NoStop}%
\bibitem [{\citenamefont {Mittal}\ \emph {et~al.}(2019)\citenamefont {Mittal},
  \citenamefont {Orre}, \citenamefont {Leykam}, \citenamefont {Chong},\ and\
  \citenamefont {Hafezi}}]{mittal2019photonic}%
  \BibitemOpen
  \bibfield  {author} {\bibinfo {author} {\bibfnamefont {S.}~\bibnamefont
  {Mittal}}, \bibinfo {author} {\bibfnamefont {V.~V.}\ \bibnamefont {Orre}},
  \bibinfo {author} {\bibfnamefont {D.}~\bibnamefont {Leykam}}, \bibinfo
  {author} {\bibfnamefont {Y.~D.}\ \bibnamefont {Chong}},\ and\ \bibinfo
  {author} {\bibfnamefont {M.}~\bibnamefont {Hafezi}},\ }\href
  {https://doi.org/10.1103/PhysRevLett.123.043201} {\bibfield  {journal}
  {\bibinfo  {journal} {Physical review letters}\ }\textbf {\bibinfo {volume}
  {123}},\ \bibinfo {pages} {043201} (\bibinfo {year} {2019})}\BibitemShut
  {NoStop}%
\bibitem [{\citenamefont {Dai}\ \emph {et~al.}(2022)\citenamefont {Dai},
  \citenamefont {Ao}, \citenamefont {Bao}, \citenamefont {Mao}, \citenamefont
  {Chi}, \citenamefont {Fu}, \citenamefont {You}, \citenamefont {Chen},
  \citenamefont {Zhai}, \citenamefont {Tang} \emph
  {et~al.}}]{dai2022topologically}%
  \BibitemOpen
  \bibfield  {author} {\bibinfo {author} {\bibfnamefont {T.}~\bibnamefont
  {Dai}}, \bibinfo {author} {\bibfnamefont {Y.}~\bibnamefont {Ao}}, \bibinfo
  {author} {\bibfnamefont {J.}~\bibnamefont {Bao}}, \bibinfo {author}
  {\bibfnamefont {J.}~\bibnamefont {Mao}}, \bibinfo {author} {\bibfnamefont
  {Y.}~\bibnamefont {Chi}}, \bibinfo {author} {\bibfnamefont {Z.}~\bibnamefont
  {Fu}}, \bibinfo {author} {\bibfnamefont {Y.}~\bibnamefont {You}}, \bibinfo
  {author} {\bibfnamefont {X.}~\bibnamefont {Chen}}, \bibinfo {author}
  {\bibfnamefont {C.}~\bibnamefont {Zhai}}, \bibinfo {author} {\bibfnamefont
  {B.}~\bibnamefont {Tang}}, \emph {et~al.},\ }\href
  {https://doi.org/10.1038/s41566-021-00944-2} {\bibfield  {journal} {\bibinfo
  {journal} {Nature Photonics}\ }\textbf {\bibinfo {volume} {16}},\ \bibinfo
  {pages} {248} (\bibinfo {year} {2022})}\BibitemShut {NoStop}%
\bibitem [{\citenamefont {Lu}\ \emph {et~al.}(2022)\citenamefont {Lu},
  \citenamefont {McClung},\ and\ \citenamefont {Srinivasan}}]{lu2022high}%
  \BibitemOpen
  \bibfield  {author} {\bibinfo {author} {\bibfnamefont {X.}~\bibnamefont
  {Lu}}, \bibinfo {author} {\bibfnamefont {A.}~\bibnamefont {McClung}},\ and\
  \bibinfo {author} {\bibfnamefont {K.}~\bibnamefont {Srinivasan}},\ }\href
  {https://doi.org/10.1038/s41566-021-00912-w} {\bibfield  {journal} {\bibinfo
  {journal} {Nature Photonics}\ }\textbf {\bibinfo {volume} {16}},\ \bibinfo
  {pages} {66} (\bibinfo {year} {2022})}\BibitemShut {NoStop}%
\bibitem [{\citenamefont {Wang}\ \emph {et~al.}(2023)\citenamefont {Wang},
  \citenamefont {Valligatla}, \citenamefont {Yin}, \citenamefont {Schwarz},
  \citenamefont {Medina-S{\'a}nchez}, \citenamefont {Baunack}, \citenamefont
  {Lee}, \citenamefont {Thomale}, \citenamefont {Li}, \citenamefont {Fomin}
  \emph {et~al.}}]{wang2023experimental}%
  \BibitemOpen
  \bibfield  {author} {\bibinfo {author} {\bibfnamefont {J.}~\bibnamefont
  {Wang}}, \bibinfo {author} {\bibfnamefont {S.}~\bibnamefont {Valligatla}},
  \bibinfo {author} {\bibfnamefont {Y.}~\bibnamefont {Yin}}, \bibinfo {author}
  {\bibfnamefont {L.}~\bibnamefont {Schwarz}}, \bibinfo {author} {\bibfnamefont
  {M.}~\bibnamefont {Medina-S{\'a}nchez}}, \bibinfo {author} {\bibfnamefont
  {S.}~\bibnamefont {Baunack}}, \bibinfo {author} {\bibfnamefont {C.~H.}\
  \bibnamefont {Lee}}, \bibinfo {author} {\bibfnamefont {R.}~\bibnamefont
  {Thomale}}, \bibinfo {author} {\bibfnamefont {S.}~\bibnamefont {Li}},
  \bibinfo {author} {\bibfnamefont {V.~M.}\ \bibnamefont {Fomin}}, \emph
  {et~al.},\ }\href {https://doi.org/10.1038/s41566-022-01107-7} {\bibfield
  {journal} {\bibinfo  {journal} {Nature Photonics}\ }\textbf {\bibinfo
  {volume} {17}},\ \bibinfo {pages} {120} (\bibinfo {year} {2023})}\BibitemShut
  {NoStop}%
\bibitem [{\citenamefont {Neef}\ \emph {et~al.}(2023)\citenamefont {Neef},
  \citenamefont {Pinske}, \citenamefont {Klauck}, \citenamefont {Teuber},
  \citenamefont {Kremer}, \citenamefont {Ehrhardt}, \citenamefont {Heinrich},
  \citenamefont {Scheel},\ and\ \citenamefont {Szameit}}]{neef2023three}%
  \BibitemOpen
  \bibfield  {author} {\bibinfo {author} {\bibfnamefont {V.}~\bibnamefont
  {Neef}}, \bibinfo {author} {\bibfnamefont {J.}~\bibnamefont {Pinske}},
  \bibinfo {author} {\bibfnamefont {F.}~\bibnamefont {Klauck}}, \bibinfo
  {author} {\bibfnamefont {L.}~\bibnamefont {Teuber}}, \bibinfo {author}
  {\bibfnamefont {M.}~\bibnamefont {Kremer}}, \bibinfo {author} {\bibfnamefont
  {M.}~\bibnamefont {Ehrhardt}}, \bibinfo {author} {\bibfnamefont
  {M.}~\bibnamefont {Heinrich}}, \bibinfo {author} {\bibfnamefont
  {S.}~\bibnamefont {Scheel}},\ and\ \bibinfo {author} {\bibfnamefont
  {A.}~\bibnamefont {Szameit}},\ }\href
  {https://doi.org/10.1038/s41567-022-01807-5} {\bibfield  {journal} {\bibinfo
  {journal} {Nature Physics}\ }\textbf {\bibinfo {volume} {19}},\ \bibinfo
  {pages} {30} (\bibinfo {year} {2023})}\BibitemShut {NoStop}%
\bibitem [{\citenamefont {Kremer}\ \emph {et~al.}(2019)\citenamefont {Kremer},
  \citenamefont {Teuber}, \citenamefont {Szameit},\ and\ \citenamefont
  {Scheel}}]{kremer2019optimal}%
  \BibitemOpen
  \bibfield  {author} {\bibinfo {author} {\bibfnamefont {M.}~\bibnamefont
  {Kremer}}, \bibinfo {author} {\bibfnamefont {L.}~\bibnamefont {Teuber}},
  \bibinfo {author} {\bibfnamefont {A.}~\bibnamefont {Szameit}},\ and\ \bibinfo
  {author} {\bibfnamefont {S.}~\bibnamefont {Scheel}},\ }\href
  {https://doi.org/10.1103/PhysRevResearch.1.033117} {\bibfield  {journal}
  {\bibinfo  {journal} {Physical Review Research}\ }\textbf {\bibinfo {volume}
  {1}},\ \bibinfo {pages} {033117} (\bibinfo {year} {2019})}\BibitemShut
  {NoStop}%
\bibitem [{\citenamefont {Cheng}\ \emph {et~al.}(2023)\citenamefont {Cheng},
  \citenamefont {Wang},\ and\ \citenamefont {Fan}}]{cheng2023artificial}%
  \BibitemOpen
  \bibfield  {author} {\bibinfo {author} {\bibfnamefont {D.}~\bibnamefont
  {Cheng}}, \bibinfo {author} {\bibfnamefont {K.}~\bibnamefont {Wang}},\ and\
  \bibinfo {author} {\bibfnamefont {S.}~\bibnamefont {Fan}},\ }\href
  {https://doi.org/10.1103/PhysRevLett.130.083601} {\bibfield  {journal}
  {\bibinfo  {journal} {Physical Review Letters}\ }\textbf {\bibinfo {volume}
  {130}},\ \bibinfo {pages} {083601} (\bibinfo {year} {2023})}\BibitemShut
  {NoStop}%
\bibitem [{\citenamefont {Zhao}\ \emph {et~al.}(2020)\citenamefont {Zhao},
  \citenamefont {Huang},\ and\ \citenamefont {Yang}}]{zhao2020z}%
  \BibitemOpen
  \bibfield  {author} {\bibinfo {author} {\bibfnamefont {Y.}~\bibnamefont
  {Zhao}}, \bibinfo {author} {\bibfnamefont {Y.-X.}\ \bibnamefont {Huang}},\
  and\ \bibinfo {author} {\bibfnamefont {S.~A.}\ \bibnamefont {Yang}},\ }\href
  {https://doi.org/10.1103/PhysRevB.102.161117} {\bibfield  {journal} {\bibinfo
   {journal} {Physical Review B}\ }\textbf {\bibinfo {volume} {102}},\ \bibinfo
  {pages} {161117} (\bibinfo {year} {2020})}\BibitemShut {NoStop}%
\bibitem [{\citenamefont {Zhao}\ \emph {et~al.}(2021)\citenamefont {Zhao},
  \citenamefont {Chen}, \citenamefont {Sheng},\ and\ \citenamefont
  {Yang}}]{zhao2021switching}%
  \BibitemOpen
  \bibfield  {author} {\bibinfo {author} {\bibfnamefont {Y.}~\bibnamefont
  {Zhao}}, \bibinfo {author} {\bibfnamefont {C.}~\bibnamefont {Chen}}, \bibinfo
  {author} {\bibfnamefont {X.-L.}\ \bibnamefont {Sheng}},\ and\ \bibinfo
  {author} {\bibfnamefont {S.~A.}\ \bibnamefont {Yang}},\ }\href
  {https://doi.org/10.1103/PhysRevLett.126.196402} {\bibfield  {journal}
  {\bibinfo  {journal} {Physical Review Letters}\ }\textbf {\bibinfo {volume}
  {126}},\ \bibinfo {pages} {196402} (\bibinfo {year} {2021})}\BibitemShut
  {NoStop}%
\bibitem [{\citenamefont {Yang}\ \emph
  {et~al.}(2022{\natexlab{a}})\citenamefont {Yang}, \citenamefont {Po},
  \citenamefont {Liu}, \citenamefont {Joannopoulos}, \citenamefont {Fu},\ and\
  \citenamefont {Solja{\v{c}}i{\'c}}}]{yang2022non}%
  \BibitemOpen
  \bibfield  {author} {\bibinfo {author} {\bibfnamefont {Y.}~\bibnamefont
  {Yang}}, \bibinfo {author} {\bibfnamefont {H.~C.}\ \bibnamefont {Po}},
  \bibinfo {author} {\bibfnamefont {V.}~\bibnamefont {Liu}}, \bibinfo {author}
  {\bibfnamefont {J.~D.}\ \bibnamefont {Joannopoulos}}, \bibinfo {author}
  {\bibfnamefont {L.}~\bibnamefont {Fu}},\ and\ \bibinfo {author}
  {\bibfnamefont {M.}~\bibnamefont {Solja{\v{c}}i{\'c}}},\ }\href
  {https://doi.org/10.1103/PhysRevB.106.L161108} {\bibfield  {journal}
  {\bibinfo  {journal} {Physical Review B}\ }\textbf {\bibinfo {volume}
  {106}},\ \bibinfo {pages} {L161108} (\bibinfo {year}
  {2022}{\natexlab{a}})}\BibitemShut {NoStop}%
\bibitem [{\citenamefont {Xue}\ \emph {et~al.}(2022)\citenamefont {Xue},
  \citenamefont {Wang}, \citenamefont {Huang}, \citenamefont {Cheng},
  \citenamefont {Yu}, \citenamefont {Foo}, \citenamefont {Zhao}, \citenamefont
  {Yang},\ and\ \citenamefont {Zhang}}]{xue2022projectively}%
  \BibitemOpen
  \bibfield  {author} {\bibinfo {author} {\bibfnamefont {H.}~\bibnamefont
  {Xue}}, \bibinfo {author} {\bibfnamefont {Z.}~\bibnamefont {Wang}}, \bibinfo
  {author} {\bibfnamefont {Y.-X.}\ \bibnamefont {Huang}}, \bibinfo {author}
  {\bibfnamefont {Z.}~\bibnamefont {Cheng}}, \bibinfo {author} {\bibfnamefont
  {L.}~\bibnamefont {Yu}}, \bibinfo {author} {\bibfnamefont {Y.}~\bibnamefont
  {Foo}}, \bibinfo {author} {\bibfnamefont {Y.}~\bibnamefont {Zhao}}, \bibinfo
  {author} {\bibfnamefont {S.~A.}\ \bibnamefont {Yang}},\ and\ \bibinfo
  {author} {\bibfnamefont {B.}~\bibnamefont {Zhang}},\ }\href
  {https://doi.org/10.1103/PhysRevLett.128.116802} {\bibfield  {journal}
  {\bibinfo  {journal} {Physical Review Letters}\ }\textbf {\bibinfo {volume}
  {128}},\ \bibinfo {pages} {116802} (\bibinfo {year} {2022})}\BibitemShut
  {NoStop}%
\bibitem [{\citenamefont {Li}\ \emph {et~al.}(2022{\natexlab{a}})\citenamefont
  {Li}, \citenamefont {Du}, \citenamefont {Zhang}, \citenamefont {Li},
  \citenamefont {Fan}, \citenamefont {Zhang},\ and\ \citenamefont
  {Qiu}}]{li2022acoustic}%
  \BibitemOpen
  \bibfield  {author} {\bibinfo {author} {\bibfnamefont {T.}~\bibnamefont
  {Li}}, \bibinfo {author} {\bibfnamefont {J.}~\bibnamefont {Du}}, \bibinfo
  {author} {\bibfnamefont {Q.}~\bibnamefont {Zhang}}, \bibinfo {author}
  {\bibfnamefont {Y.}~\bibnamefont {Li}}, \bibinfo {author} {\bibfnamefont
  {X.}~\bibnamefont {Fan}}, \bibinfo {author} {\bibfnamefont {F.}~\bibnamefont
  {Zhang}},\ and\ \bibinfo {author} {\bibfnamefont {C.}~\bibnamefont {Qiu}},\
  }\href {https://doi.org/10.1103/PhysRevLett.128.116803} {\bibfield  {journal}
  {\bibinfo  {journal} {Physical Review Letters}\ }\textbf {\bibinfo {volume}
  {128}},\ \bibinfo {pages} {116803} (\bibinfo {year}
  {2022}{\natexlab{a}})}\BibitemShut {NoStop}%
\bibitem [{\citenamefont {Meng}\ \emph {et~al.}(2023)\citenamefont {Meng},
  \citenamefont {Lin}, \citenamefont {Shi}, \citenamefont {Wei}, \citenamefont
  {Yang}, \citenamefont {Yan}, \citenamefont {Zhu}, \citenamefont {Xi},
  \citenamefont {Wang}, \citenamefont {Ge} \emph {et~al.}}]{meng2023spinful}%
  \BibitemOpen
  \bibfield  {author} {\bibinfo {author} {\bibfnamefont {Y.}~\bibnamefont
  {Meng}}, \bibinfo {author} {\bibfnamefont {S.}~\bibnamefont {Lin}}, \bibinfo
  {author} {\bibfnamefont {B.-j.}\ \bibnamefont {Shi}}, \bibinfo {author}
  {\bibfnamefont {B.}~\bibnamefont {Wei}}, \bibinfo {author} {\bibfnamefont
  {L.}~\bibnamefont {Yang}}, \bibinfo {author} {\bibfnamefont {B.}~\bibnamefont
  {Yan}}, \bibinfo {author} {\bibfnamefont {Z.}~\bibnamefont {Zhu}}, \bibinfo
  {author} {\bibfnamefont {X.}~\bibnamefont {Xi}}, \bibinfo {author}
  {\bibfnamefont {Y.}~\bibnamefont {Wang}}, \bibinfo {author} {\bibfnamefont
  {Y.}~\bibnamefont {Ge}}, \emph {et~al.},\ }\href
  {https://doi.org/10.1103/PhysRevLett.128.116802} {\bibfield  {journal}
  {\bibinfo  {journal} {Physical Review Letters}\ }\textbf {\bibinfo {volume}
  {130}},\ \bibinfo {pages} {026101} (\bibinfo {year} {2023})}\BibitemShut
  {NoStop}%
\bibitem [{\citenamefont {Bouhon}\ \emph {et~al.}(2020)\citenamefont {Bouhon},
  \citenamefont {Wu}, \citenamefont {Slager}, \citenamefont {Weng},
  \citenamefont {Yazyev},\ and\ \citenamefont {Bzdušek}}]{RN8}%
  \BibitemOpen
  \bibfield  {author} {\bibinfo {author} {\bibfnamefont {A.}~\bibnamefont
  {Bouhon}}, \bibinfo {author} {\bibfnamefont {Q.}~\bibnamefont {Wu}}, \bibinfo
  {author} {\bibfnamefont {R.-J.}\ \bibnamefont {Slager}}, \bibinfo {author}
  {\bibfnamefont {H.}~\bibnamefont {Weng}}, \bibinfo {author} {\bibfnamefont
  {O.~V.}\ \bibnamefont {Yazyev}},\ and\ \bibinfo {author} {\bibfnamefont
  {T.}~\bibnamefont {Bzdušek}},\ }\href
  {https://doi.org/10.1038/s41567-020-0967-9} {\bibfield  {journal} {\bibinfo
  {journal} {Nature Physics}\ }\textbf {\bibinfo {volume} {16}},\ \bibinfo
  {pages} {1137} (\bibinfo {year} {2020})}\BibitemShut {NoStop}%
\bibitem [{\citenamefont {Qiu}\ \emph {et~al.}(2023)\citenamefont {Qiu},
  \citenamefont {Zhang}, \citenamefont {Liu}, \citenamefont {Fan},
  \citenamefont {Zhang},\ and\ \citenamefont {Qiu}}]{RN10}%
  \BibitemOpen
  \bibfield  {author} {\bibinfo {author} {\bibfnamefont {H.}~\bibnamefont
  {Qiu}}, \bibinfo {author} {\bibfnamefont {Q.}~\bibnamefont {Zhang}}, \bibinfo
  {author} {\bibfnamefont {T.}~\bibnamefont {Liu}}, \bibinfo {author}
  {\bibfnamefont {X.}~\bibnamefont {Fan}}, \bibinfo {author} {\bibfnamefont
  {F.}~\bibnamefont {Zhang}},\ and\ \bibinfo {author} {\bibfnamefont
  {C.}~\bibnamefont {Qiu}},\ }\href
  {https://doi.org/10.1038/s41467-023-36952-9} {\bibfield  {journal} {\bibinfo
  {journal} {Nature Communications}\ }\textbf {\bibinfo {volume} {14}},\
  \bibinfo {pages} {1261} (\bibinfo {year} {2023})}\BibitemShut {NoStop}%
\bibitem [{\citenamefont {Klitzing}\ \emph {et~al.}(1980)\citenamefont
  {Klitzing}, \citenamefont {Dorda},\ and\ \citenamefont {Pepper}}]{RN1}%
  \BibitemOpen
  \bibfield  {author} {\bibinfo {author} {\bibfnamefont {K.~v.}\ \bibnamefont
  {Klitzing}}, \bibinfo {author} {\bibfnamefont {G.}~\bibnamefont {Dorda}},\
  and\ \bibinfo {author} {\bibfnamefont {M.}~\bibnamefont {Pepper}},\ }\href
  {https://doi.org/10.1103/PhysRevLett.45.494} {\bibfield  {journal} {\bibinfo
  {journal} {Physical Review Letters}\ }\textbf {\bibinfo {volume} {45}},\
  \bibinfo {pages} {494} (\bibinfo {year} {1980})}\BibitemShut {NoStop}%
\bibitem [{\citenamefont {Thouless}\ \emph {et~al.}(1982)\citenamefont
  {Thouless}, \citenamefont {Kohmoto}, \citenamefont {Nightingale},\ and\
  \citenamefont {den Nijs}}]{RN2}%
  \BibitemOpen
  \bibfield  {author} {\bibinfo {author} {\bibfnamefont {D.~J.}\ \bibnamefont
  {Thouless}}, \bibinfo {author} {\bibfnamefont {M.}~\bibnamefont {Kohmoto}},
  \bibinfo {author} {\bibfnamefont {M.~P.}\ \bibnamefont {Nightingale}},\ and\
  \bibinfo {author} {\bibfnamefont {M.}~\bibnamefont {den Nijs}},\ }\href
  {https://doi.org/10.1103/PhysRevLett.49.405} {\bibfield  {journal} {\bibinfo
  {journal} {Physical Review Letters}\ }\textbf {\bibinfo {volume} {49}},\
  \bibinfo {pages} {405} (\bibinfo {year} {1982})}\BibitemShut {NoStop}%
\bibitem [{\citenamefont {Ozawa}\ \emph {et~al.}(2019)\citenamefont {Ozawa},
  \citenamefont {Price}, \citenamefont {Amo}, \citenamefont {Goldman},
  \citenamefont {Hafezi}, \citenamefont {Lu}, \citenamefont {Rechtsman},
  \citenamefont {Schuster}, \citenamefont {Simon}, \citenamefont {Zilberberg},\
  and\ \citenamefont {Carusotto}}]{RN3}%
  \BibitemOpen
  \bibfield  {author} {\bibinfo {author} {\bibfnamefont {T.}~\bibnamefont
  {Ozawa}}, \bibinfo {author} {\bibfnamefont {H.~M.}\ \bibnamefont {Price}},
  \bibinfo {author} {\bibfnamefont {A.}~\bibnamefont {Amo}}, \bibinfo {author}
  {\bibfnamefont {N.}~\bibnamefont {Goldman}}, \bibinfo {author} {\bibfnamefont
  {M.}~\bibnamefont {Hafezi}}, \bibinfo {author} {\bibfnamefont
  {L.}~\bibnamefont {Lu}}, \bibinfo {author} {\bibfnamefont {M.~C.}\
  \bibnamefont {Rechtsman}}, \bibinfo {author} {\bibfnamefont {D.}~\bibnamefont
  {Schuster}}, \bibinfo {author} {\bibfnamefont {J.}~\bibnamefont {Simon}},
  \bibinfo {author} {\bibfnamefont {O.}~\bibnamefont {Zilberberg}},\ and\
  \bibinfo {author} {\bibfnamefont {I.}~\bibnamefont {Carusotto}},\ }\href
  {https://doi.org/10.1103/RevModPhys.91.015006} {\bibfield  {journal}
  {\bibinfo  {journal} {Reviews of Modern Physics}\ }\textbf {\bibinfo {volume}
  {91}},\ \bibinfo {pages} {015006} (\bibinfo {year} {2019})}\BibitemShut
  {NoStop}%
\bibitem [{\citenamefont {Ma}\ \emph {et~al.}(2019)\citenamefont {Ma},
  \citenamefont {Xiao},\ and\ \citenamefont {Chan}}]{RN4}%
  \BibitemOpen
  \bibfield  {author} {\bibinfo {author} {\bibfnamefont {G.}~\bibnamefont
  {Ma}}, \bibinfo {author} {\bibfnamefont {M.}~\bibnamefont {Xiao}},\ and\
  \bibinfo {author} {\bibfnamefont {C.~T.}\ \bibnamefont {Chan}},\ }\href
  {https://doi.org/10.1038/s42254-019-0030-x} {\bibfield  {journal} {\bibinfo
  {journal} {Nature Reviews Physics}\ }\textbf {\bibinfo {volume} {1}},\
  \bibinfo {pages} {281} (\bibinfo {year} {2019})}\BibitemShut {NoStop}%
\bibitem [{\citenamefont {Schnyder}\ \emph {et~al.}(2008)\citenamefont
  {Schnyder}, \citenamefont {Ryu}, \citenamefont {Furusaki},\ and\
  \citenamefont {Ludwig}}]{RN5}%
  \BibitemOpen
  \bibfield  {author} {\bibinfo {author} {\bibfnamefont {A.~P.}\ \bibnamefont
  {Schnyder}}, \bibinfo {author} {\bibfnamefont {S.}~\bibnamefont {Ryu}},
  \bibinfo {author} {\bibfnamefont {A.}~\bibnamefont {Furusaki}},\ and\
  \bibinfo {author} {\bibfnamefont {A.~W.~W.}\ \bibnamefont {Ludwig}},\ }\href
  {https://doi.org/10.1103/PhysRevB.78.195125} {\bibfield  {journal} {\bibinfo
  {journal} {Physical Review B}\ }\textbf {\bibinfo {volume} {78}},\ \bibinfo
  {pages} {195125} (\bibinfo {year} {2008})}\BibitemShut {NoStop}%
\bibitem [{\citenamefont {Ahn}\ \emph {et~al.}(2019)\citenamefont {Ahn},
  \citenamefont {Park},\ and\ \citenamefont {Yang}}]{RN7}%
  \BibitemOpen
  \bibfield  {author} {\bibinfo {author} {\bibfnamefont {J.}~\bibnamefont
  {Ahn}}, \bibinfo {author} {\bibfnamefont {S.}~\bibnamefont {Park}},\ and\
  \bibinfo {author} {\bibfnamefont {B.-J.}\ \bibnamefont {Yang}},\ }\href
  {https://doi.org/10.1103/PhysRevX.9.021013} {\bibfield  {journal} {\bibinfo
  {journal} {Physical Review X}\ }\textbf {\bibinfo {volume} {9}},\ \bibinfo
  {pages} {021013} (\bibinfo {year} {2019})}\BibitemShut {NoStop}%
\bibitem [{\citenamefont {Peng}\ \emph
  {et~al.}(2022{\natexlab{a}})\citenamefont {Peng}, \citenamefont {Bouhon},
  \citenamefont {Slager},\ and\ \citenamefont
  {Monserrat}}]{PhysRevB.105.085115}%
  \BibitemOpen
  \bibfield  {author} {\bibinfo {author} {\bibfnamefont {B.}~\bibnamefont
  {Peng}}, \bibinfo {author} {\bibfnamefont {A.}~\bibnamefont {Bouhon}},
  \bibinfo {author} {\bibfnamefont {R.-J.}\ \bibnamefont {Slager}},\ and\
  \bibinfo {author} {\bibfnamefont {B.}~\bibnamefont {Monserrat}},\ }\href
  {https://doi.org/10.1103/PhysRevB.105.085115} {\bibfield  {journal} {\bibinfo
   {journal} {Phys. Rev. B}\ }\textbf {\bibinfo {volume} {105}},\ \bibinfo
  {pages} {085115} (\bibinfo {year} {2022}{\natexlab{a}})}\BibitemShut
  {NoStop}%
\bibitem [{\citenamefont {Peng}\ \emph
  {et~al.}(2022{\natexlab{b}})\citenamefont {Peng}, \citenamefont {Bouhon},
  \citenamefont {Monserrat},\ and\ \citenamefont {Slager}}]{PengNC}%
  \BibitemOpen
  \bibfield  {author} {\bibinfo {author} {\bibfnamefont {B.}~\bibnamefont
  {Peng}}, \bibinfo {author} {\bibfnamefont {A.}~\bibnamefont {Bouhon}},
  \bibinfo {author} {\bibfnamefont {B.}~\bibnamefont {Monserrat}},\ and\
  \bibinfo {author} {\bibfnamefont {R.-J.}\ \bibnamefont {Slager}},\ }\href
  {https://doi.org/10.1038/s41467-022-28046-9} {\bibfield  {journal} {\bibinfo
  {journal} {Nature Communications}\ }\textbf {\bibinfo {volume} {13}},\
  \bibinfo {pages} {423} (\bibinfo {year} {2022}{\natexlab{b}})}\BibitemShut
  {NoStop}%
\bibitem [{\citenamefont {Tiwari}\ and\ \citenamefont {Bzdušek}(2020)}]{RN12}%
  \BibitemOpen
  \bibfield  {author} {\bibinfo {author} {\bibfnamefont {A.}~\bibnamefont
  {Tiwari}}\ and\ \bibinfo {author} {\bibfnamefont {T.}~\bibnamefont
  {Bzdušek}},\ }\href {https://doi.org/10.1103/PhysRevB.101.195130} {\bibfield
   {journal} {\bibinfo  {journal} {Physical Review B}\ }\textbf {\bibinfo
  {volume} {101}},\ \bibinfo {pages} {195130} (\bibinfo {year}
  {2020})}\BibitemShut {NoStop}%
\bibitem [{\citenamefont {Park}\ \emph
  {et~al.}(2021{\natexlab{a}})\citenamefont {Park}, \citenamefont {Wong},
  \citenamefont {Zhang},\ and\ \citenamefont {Oh}}]{RN13}%
  \BibitemOpen
  \bibfield  {author} {\bibinfo {author} {\bibfnamefont {H.}~\bibnamefont
  {Park}}, \bibinfo {author} {\bibfnamefont {S.}~\bibnamefont {Wong}}, \bibinfo
  {author} {\bibfnamefont {X.}~\bibnamefont {Zhang}},\ and\ \bibinfo {author}
  {\bibfnamefont {S.~S.}\ \bibnamefont {Oh}},\ }\href
  {https://doi.org/10.1021/acsphotonics.1c00876} {\bibfield  {journal}
  {\bibinfo  {journal} {ACS Photonics}\ }\textbf {\bibinfo {volume} {8}},\
  \bibinfo {pages} {2746} (\bibinfo {year} {2021}{\natexlab{a}})}\BibitemShut
  {NoStop}%
\bibitem [{\citenamefont {Wang}\ \emph
  {et~al.}(2022{\natexlab{a}})\citenamefont {Wang}, \citenamefont {Yang},
  \citenamefont {Wang}, \citenamefont {Zhang}, \citenamefont {Li},
  \citenamefont {Zhang}, \citenamefont {Zhang},\ and\ \citenamefont
  {Chan}}]{RN14}%
  \BibitemOpen
  \bibfield  {author} {\bibinfo {author} {\bibfnamefont {D.}~\bibnamefont
  {Wang}}, \bibinfo {author} {\bibfnamefont {B.}~\bibnamefont {Yang}}, \bibinfo
  {author} {\bibfnamefont {M.}~\bibnamefont {Wang}}, \bibinfo {author}
  {\bibfnamefont {R.-Y.}\ \bibnamefont {Zhang}}, \bibinfo {author}
  {\bibfnamefont {X.}~\bibnamefont {Li}}, \bibinfo {author} {\bibfnamefont
  {Z.~Q.}\ \bibnamefont {Zhang}}, \bibinfo {author} {\bibfnamefont
  {S.}~\bibnamefont {Zhang}},\ and\ \bibinfo {author} {\bibfnamefont {C.~T.}\
  \bibnamefont {Chan}},\ }\href
  {https://doi.org/10.1103/PhysRevLett.129.263604} {\bibfield  {journal}
  {\bibinfo  {journal} {Physical Review Letters}\ }\textbf {\bibinfo {volume}
  {129}},\ \bibinfo {pages} {263604} (\bibinfo {year}
  {2022}{\natexlab{a}})}\BibitemShut {NoStop}%
\bibitem [{\citenamefont {Wang}\ \emph
  {et~al.}(2022{\natexlab{b}})\citenamefont {Wang}, \citenamefont {Liu},
  \citenamefont {Ma}, \citenamefont {Zhang}, \citenamefont {Wang},
  \citenamefont {Guo}, \citenamefont {Yang}, \citenamefont {Ke}, \citenamefont
  {Liu},\ and\ \citenamefont {Chan}}]{RN15}%
  \BibitemOpen
  \bibfield  {author} {\bibinfo {author} {\bibfnamefont {M.}~\bibnamefont
  {Wang}}, \bibinfo {author} {\bibfnamefont {S.}~\bibnamefont {Liu}}, \bibinfo
  {author} {\bibfnamefont {Q.}~\bibnamefont {Ma}}, \bibinfo {author}
  {\bibfnamefont {R.-Y.}\ \bibnamefont {Zhang}}, \bibinfo {author}
  {\bibfnamefont {D.}~\bibnamefont {Wang}}, \bibinfo {author} {\bibfnamefont
  {Q.}~\bibnamefont {Guo}}, \bibinfo {author} {\bibfnamefont {B.}~\bibnamefont
  {Yang}}, \bibinfo {author} {\bibfnamefont {M.}~\bibnamefont {Ke}}, \bibinfo
  {author} {\bibfnamefont {Z.}~\bibnamefont {Liu}},\ and\ \bibinfo {author}
  {\bibfnamefont {C.~T.}\ \bibnamefont {Chan}},\ }\href
  {https://doi.org/10.1103/PhysRevLett.128.246601} {\bibfield  {journal}
  {\bibinfo  {journal} {Physical Review Letters}\ }\textbf {\bibinfo {volume}
  {128}},\ \bibinfo {pages} {246601} (\bibinfo {year}
  {2022}{\natexlab{b}})}\BibitemShut {NoStop}%
\bibitem [{\citenamefont {Park}\ \emph {et~al.}(2022)\citenamefont {Park},
  \citenamefont {Wong}, \citenamefont {Bouhon}, \citenamefont {Slager},\ and\
  \citenamefont {Oh}}]{PhysRevB.105.214108}%
  \BibitemOpen
  \bibfield  {author} {\bibinfo {author} {\bibfnamefont {H.}~\bibnamefont
  {Park}}, \bibinfo {author} {\bibfnamefont {S.}~\bibnamefont {Wong}}, \bibinfo
  {author} {\bibfnamefont {A.}~\bibnamefont {Bouhon}}, \bibinfo {author}
  {\bibfnamefont {R.-J.}\ \bibnamefont {Slager}},\ and\ \bibinfo {author}
  {\bibfnamefont {S.~S.}\ \bibnamefont {Oh}},\ }\href
  {https://doi.org/10.1103/PhysRevB.105.214108} {\bibfield  {journal} {\bibinfo
   {journal} {Phys. Rev. B}\ }\textbf {\bibinfo {volume} {105}},\ \bibinfo
  {pages} {214108} (\bibinfo {year} {2022})}\BibitemShut {NoStop}%
\bibitem [{\citenamefont {Sun}\ \emph {et~al.}(2018)\citenamefont {Sun},
  \citenamefont {Zhang},\ and\ \citenamefont {Bzdušek}}]{RN16}%
  \BibitemOpen
  \bibfield  {author} {\bibinfo {author} {\bibfnamefont {X.-Q.}\ \bibnamefont
  {Sun}}, \bibinfo {author} {\bibfnamefont {S.-C.}\ \bibnamefont {Zhang}},\
  and\ \bibinfo {author} {\bibfnamefont {T.}~\bibnamefont {Bzdušek}},\ }\href
  {https://doi.org/10.1103/PhysRevLett.121.106402} {\bibfield  {journal}
  {\bibinfo  {journal} {Physical Review Letters}\ }\textbf {\bibinfo {volume}
  {121}},\ \bibinfo {pages} {106402} (\bibinfo {year} {2018})}\BibitemShut
  {NoStop}%
\bibitem [{\citenamefont {Chen}\ \emph
  {et~al.}(2022{\natexlab{b}})\citenamefont {Chen}, \citenamefont {Bouhon},
  \citenamefont {Slager},\ and\ \citenamefont
  {Monserrat}}]{PhysRevB.105.L081117}%
  \BibitemOpen
  \bibfield  {author} {\bibinfo {author} {\bibfnamefont {S.}~\bibnamefont
  {Chen}}, \bibinfo {author} {\bibfnamefont {A.}~\bibnamefont {Bouhon}},
  \bibinfo {author} {\bibfnamefont {R.-J.}\ \bibnamefont {Slager}},\ and\
  \bibinfo {author} {\bibfnamefont {B.}~\bibnamefont {Monserrat}},\ }\href
  {https://doi.org/10.1103/PhysRevB.105.L081117} {\bibfield  {journal}
  {\bibinfo  {journal} {Phys. Rev. B}\ }\textbf {\bibinfo {volume} {105}},\
  \bibinfo {pages} {L081117} (\bibinfo {year}
  {2022}{\natexlab{b}})}\BibitemShut {NoStop}%
\bibitem [{\citenamefont {Bzdušek}\ and\ \citenamefont
  {Sigrist}(2017)}]{RN17}%
  \BibitemOpen
  \bibfield  {author} {\bibinfo {author} {\bibfnamefont {T.}~\bibnamefont
  {Bzdušek}}\ and\ \bibinfo {author} {\bibfnamefont {M.}~\bibnamefont
  {Sigrist}},\ }\href {https://doi.org/10.1103/PhysRevB.96.155105} {\bibfield
  {journal} {\bibinfo  {journal} {Physical Review B}\ }\textbf {\bibinfo
  {volume} {96}},\ \bibinfo {pages} {155105} (\bibinfo {year}
  {2017})}\BibitemShut {NoStop}%
\bibitem [{\citenamefont {Ahn}\ \emph {et~al.}(2018)\citenamefont {Ahn},
  \citenamefont {Kim}, \citenamefont {Kim},\ and\ \citenamefont {Yang}}]{RN18}%
  \BibitemOpen
  \bibfield  {author} {\bibinfo {author} {\bibfnamefont {J.}~\bibnamefont
  {Ahn}}, \bibinfo {author} {\bibfnamefont {D.}~\bibnamefont {Kim}}, \bibinfo
  {author} {\bibfnamefont {Y.}~\bibnamefont {Kim}},\ and\ \bibinfo {author}
  {\bibfnamefont {B.-J.}\ \bibnamefont {Yang}},\ }\href
  {https://doi.org/10.1103/PhysRevLett.121.106403} {\bibfield  {journal}
  {\bibinfo  {journal} {Physical Review Letters}\ }\textbf {\bibinfo {volume}
  {121}},\ \bibinfo {pages} {106403} (\bibinfo {year} {2018})}\BibitemShut
  {NoStop}%
\bibitem [{\citenamefont {Lenggenhager}\ \emph {et~al.}(2021)\citenamefont
  {Lenggenhager}, \citenamefont {Liu}, \citenamefont {Tsirkin}, \citenamefont
  {Neupert},\ and\ \citenamefont {Bzdušek}}]{RN19}%
  \BibitemOpen
  \bibfield  {author} {\bibinfo {author} {\bibfnamefont {P.~M.}\ \bibnamefont
  {Lenggenhager}}, \bibinfo {author} {\bibfnamefont {X.}~\bibnamefont {Liu}},
  \bibinfo {author} {\bibfnamefont {S.~S.}\ \bibnamefont {Tsirkin}}, \bibinfo
  {author} {\bibfnamefont {T.}~\bibnamefont {Neupert}},\ and\ \bibinfo {author}
  {\bibfnamefont {T.}~\bibnamefont {Bzdušek}},\ }\href
  {https://doi.org/10.1103/PhysRevB.103.L121101} {\bibfield  {journal}
  {\bibinfo  {journal} {Physical Review B}\ }\textbf {\bibinfo {volume}
  {103}},\ \bibinfo {pages} {L121101} (\bibinfo {year} {2021})}\BibitemShut
  {NoStop}%
\bibitem [{\citenamefont {Park}\ \emph
  {et~al.}(2021{\natexlab{b}})\citenamefont {Park}, \citenamefont {Hwang},
  \citenamefont {Choi},\ and\ \citenamefont {Yang}}]{RN20}%
  \BibitemOpen
  \bibfield  {author} {\bibinfo {author} {\bibfnamefont {S.}~\bibnamefont
  {Park}}, \bibinfo {author} {\bibfnamefont {Y.}~\bibnamefont {Hwang}},
  \bibinfo {author} {\bibfnamefont {H.~C.}\ \bibnamefont {Choi}},\ and\
  \bibinfo {author} {\bibfnamefont {B.-J.}\ \bibnamefont {Yang}},\ }\href
  {https://doi.org/10.1038/s41467-021-27158-y} {\bibfield  {journal} {\bibinfo
  {journal} {Nature Communications}\ }\textbf {\bibinfo {volume} {12}},\
  \bibinfo {pages} {6781} (\bibinfo {year} {2021}{\natexlab{b}})}\BibitemShut
  {NoStop}%
\bibitem [{\citenamefont {Bouhon}\ and\ \citenamefont
  {Slager}(2022)}]{bouhon2022multigap}%
  \BibitemOpen
  \bibfield  {author} {\bibinfo {author} {\bibfnamefont {A.}~\bibnamefont
  {Bouhon}}\ and\ \bibinfo {author} {\bibfnamefont {R.-J.}\ \bibnamefont
  {Slager}},\ }\bibfield  {journal} {\bibinfo  {journal} {arXiv preprint
  arXiv:2203.16741}\ }\href {https://doi.org/10.48550/arXiv.2203.16741}
  {10.48550/arXiv.2203.16741} (\bibinfo {year} {2022})\BibitemShut {NoStop}%
\bibitem [{\citenamefont {Jiang}\ \emph
  {et~al.}(2021{\natexlab{b}})\citenamefont {Jiang}, \citenamefont {Guo},
  \citenamefont {Zhang}, \citenamefont {Zhang}, \citenamefont {Yang},\ and\
  \citenamefont {Chan}}]{RN22}%
  \BibitemOpen
  \bibfield  {author} {\bibinfo {author} {\bibfnamefont {T.}~\bibnamefont
  {Jiang}}, \bibinfo {author} {\bibfnamefont {Q.}~\bibnamefont {Guo}}, \bibinfo
  {author} {\bibfnamefont {R.-Y.}\ \bibnamefont {Zhang}}, \bibinfo {author}
  {\bibfnamefont {Z.-Q.}\ \bibnamefont {Zhang}}, \bibinfo {author}
  {\bibfnamefont {B.}~\bibnamefont {Yang}},\ and\ \bibinfo {author}
  {\bibfnamefont {C.~T.}\ \bibnamefont {Chan}},\ }\href
  {https://doi.org/10.1038/s41467-021-26763-1} {\bibfield  {journal} {\bibinfo
  {journal} {Nature Communications}\ }\textbf {\bibinfo {volume} {12}},\
  \bibinfo {pages} {6471} (\bibinfo {year} {2021}{\natexlab{b}})}\BibitemShut
  {NoStop}%
\bibitem [{\citenamefont {Ding}\ \emph {et~al.}(2022)\citenamefont {Ding},
  \citenamefont {Fang},\ and\ \citenamefont {Ma}}]{RN23}%
  \BibitemOpen
  \bibfield  {author} {\bibinfo {author} {\bibfnamefont {K.}~\bibnamefont
  {Ding}}, \bibinfo {author} {\bibfnamefont {C.}~\bibnamefont {Fang}},\ and\
  \bibinfo {author} {\bibfnamefont {G.}~\bibnamefont {Ma}},\ }\href
  {https://doi.org/10.1038/s42254-022-00516-5} {\bibfield  {journal} {\bibinfo
  {journal} {Nature Reviews Physics}\ }\textbf {\bibinfo {volume} {4}},\
  \bibinfo {pages} {745} (\bibinfo {year} {2022})}\BibitemShut {NoStop}%
\bibitem [{\citenamefont {Fang}\ \emph {et~al.}(2015)\citenamefont {Fang},
  \citenamefont {Chen}, \citenamefont {Kee},\ and\ \citenamefont {Fu}}]{RN24}%
  \BibitemOpen
  \bibfield  {author} {\bibinfo {author} {\bibfnamefont {C.}~\bibnamefont
  {Fang}}, \bibinfo {author} {\bibfnamefont {Y.}~\bibnamefont {Chen}}, \bibinfo
  {author} {\bibfnamefont {H.-Y.}\ \bibnamefont {Kee}},\ and\ \bibinfo {author}
  {\bibfnamefont {L.}~\bibnamefont {Fu}},\ }\href
  {https://doi.org/10.1103/PhysRevB.92.081201} {\bibfield  {journal} {\bibinfo
  {journal} {Physical Review B}\ }\textbf {\bibinfo {volume} {92}},\ \bibinfo
  {pages} {081201} (\bibinfo {year} {2015})}\BibitemShut {NoStop}%
\bibitem [{\citenamefont {Fang}\ \emph {et~al.}(2016)\citenamefont {Fang},
  \citenamefont {Weng}, \citenamefont {Dai},\ and\ \citenamefont
  {Fang}}]{RN25}%
  \BibitemOpen
  \bibfield  {author} {\bibinfo {author} {\bibfnamefont {C.}~\bibnamefont
  {Fang}}, \bibinfo {author} {\bibfnamefont {H.}~\bibnamefont {Weng}}, \bibinfo
  {author} {\bibfnamefont {X.}~\bibnamefont {Dai}},\ and\ \bibinfo {author}
  {\bibfnamefont {Z.}~\bibnamefont {Fang}},\ }\href
  {https://doi.org/10.1088/1674-1056/25/11/117106} {\bibfield  {journal}
  {\bibinfo  {journal} {Chinese Physics B}\ }\textbf {\bibinfo {volume} {25}},\
  \bibinfo {pages} {117106} (\bibinfo {year} {2016})}\BibitemShut {NoStop}%
\bibitem [{\citenamefont {Mermin}(1979)}]{RN26}%
  \BibitemOpen
  \bibfield  {author} {\bibinfo {author} {\bibfnamefont {N.~D.}\ \bibnamefont
  {Mermin}},\ }\href {https://doi.org/10.1103/RevModPhys.51.591} {\bibfield
  {journal} {\bibinfo  {journal} {Reviews of Modern Physics}\ }\textbf
  {\bibinfo {volume} {51}},\ \bibinfo {pages} {591} (\bibinfo {year}
  {1979})}\BibitemShut {NoStop}%
\bibitem [{\citenamefont {Basener}(2006)}]{RN27}%
  \BibitemOpen
  \bibfield  {author} {\bibinfo {author} {\bibfnamefont {W.~F.}\ \bibnamefont
  {Basener}},\ }\href@noop {} {\emph {\bibinfo {title} {Topology and Its
  Applications}}}\ (\bibinfo  {publisher} {Wiley \& Sons Ltd},\ \bibinfo {year}
  {2006})\BibitemShut {NoStop}%
\bibitem [{\citenamefont {Kléman}\ \emph {et~al.}(1977)\citenamefont
  {Kléman}, \citenamefont {Michel},\ and\ \citenamefont {Toulouse}}]{RN28}%
  \BibitemOpen
  \bibfield  {author} {\bibinfo {author} {\bibfnamefont {M.}~\bibnamefont
  {Kléman}}, \bibinfo {author} {\bibfnamefont {L.}~\bibnamefont {Michel}},\
  and\ \bibinfo {author} {\bibfnamefont {G.}~\bibnamefont {Toulouse}},\ }\href
  {https://doi.org/10.1051/jphyslet:019770038010019500} {\bibfield  {journal}
  {\bibinfo  {journal} {J. Physique Lett.}\ }\textbf {\bibinfo {volume} {38}},\
  \bibinfo {pages} {195} (\bibinfo {year} {1977})}\BibitemShut {NoStop}%
\bibitem [{\citenamefont {Fumeron}\ and\ \citenamefont {Berche}(2023)}]{RN29}%
  \BibitemOpen
  \bibfield  {author} {\bibinfo {author} {\bibfnamefont {S.}~\bibnamefont
  {Fumeron}}\ and\ \bibinfo {author} {\bibfnamefont {B.}~\bibnamefont
  {Berche}},\ }\href {https://doi.org/10.1140/epjs/s11734-023-00803-x}
  {\bibfield  {journal} {\bibinfo  {journal} {The European Physical Journal
  Special Topics}\ ,\ \bibinfo {pages} {1}} (\bibinfo {year}
  {2023})}\BibitemShut {NoStop}%
\bibitem [{\citenamefont {Staley}(2010)}]{RN30}%
  \BibitemOpen
  \bibfield  {author} {\bibinfo {author} {\bibfnamefont {M.}~\bibnamefont
  {Staley}},\ }\href {https://doi.org/10.1088/0143-0807/31/3/004} {\bibfield
  {journal} {\bibinfo  {journal} {European Journal of Physics}\ }\textbf
  {\bibinfo {volume} {31}},\ \bibinfo {pages} {467} (\bibinfo {year}
  {2010})}\BibitemShut {NoStop}%
\bibitem [{\citenamefont {Jackiw}\ and\ \citenamefont {Rebbi}(1976)}]{RN31}%
  \BibitemOpen
  \bibfield  {author} {\bibinfo {author} {\bibfnamefont {R.}~\bibnamefont
  {Jackiw}}\ and\ \bibinfo {author} {\bibfnamefont {C.}~\bibnamefont {Rebbi}},\
  }\href {https://doi.org/10.1103/PhysRevD.13.3398} {\bibfield  {journal}
  {\bibinfo  {journal} {Physical Review D}\ }\textbf {\bibinfo {volume} {13}},\
  \bibinfo {pages} {3398} (\bibinfo {year} {1976})}\BibitemShut {NoStop}%
\bibitem [{\citenamefont {Hatsugai}(1993)}]{RN32}%
  \BibitemOpen
  \bibfield  {author} {\bibinfo {author} {\bibfnamefont {Y.}~\bibnamefont
  {Hatsugai}},\ }\href {https://doi.org/10.1103/PhysRevLett.71.3697} {\bibfield
   {journal} {\bibinfo  {journal} {Physical Review Letters}\ }\textbf {\bibinfo
  {volume} {71}},\ \bibinfo {pages} {3697} (\bibinfo {year}
  {1993})}\BibitemShut {NoStop}%
\bibitem [{\citenamefont {Hasan}\ and\ \citenamefont {Kane}(2010)}]{RN33}%
  \BibitemOpen
  \bibfield  {author} {\bibinfo {author} {\bibfnamefont {M.~Z.}\ \bibnamefont
  {Hasan}}\ and\ \bibinfo {author} {\bibfnamefont {C.~L.}\ \bibnamefont
  {Kane}},\ }\href {https://doi.org/10.1103/RevModPhys.82.3045} {\bibfield
  {journal} {\bibinfo  {journal} {Reviews of Modern Physics}\ }\textbf
  {\bibinfo {volume} {82}},\ \bibinfo {pages} {3045} (\bibinfo {year}
  {2010})}\BibitemShut {NoStop}%
\bibitem [{\citenamefont {Graf}\ and\ \citenamefont {Porta}(2013)}]{RN34}%
  \BibitemOpen
  \bibfield  {author} {\bibinfo {author} {\bibfnamefont {G.~M.}\ \bibnamefont
  {Graf}}\ and\ \bibinfo {author} {\bibfnamefont {M.}~\bibnamefont {Porta}},\
  }\href {https://doi.org/10.1007/s00220-013-1819-6} {\bibfield  {journal}
  {\bibinfo  {journal} {Communications in Mathematical Physics}\ }\textbf
  {\bibinfo {volume} {324}},\ \bibinfo {pages} {851} (\bibinfo {year}
  {2013})}\BibitemShut {NoStop}%
\bibitem [{\citenamefont {Huang}\ \emph {et~al.}(2011)\citenamefont {Huang},
  \citenamefont {Lai}, \citenamefont {Hang}, \citenamefont {Zheng},\ and\
  \citenamefont {Chan}}]{RN35}%
  \BibitemOpen
  \bibfield  {author} {\bibinfo {author} {\bibfnamefont {X.}~\bibnamefont
  {Huang}}, \bibinfo {author} {\bibfnamefont {Y.}~\bibnamefont {Lai}}, \bibinfo
  {author} {\bibfnamefont {Z.~H.}\ \bibnamefont {Hang}}, \bibinfo {author}
  {\bibfnamefont {H.}~\bibnamefont {Zheng}},\ and\ \bibinfo {author}
  {\bibfnamefont {C.~T.}\ \bibnamefont {Chan}},\ }\href
  {https://doi.org/10.1038/nmat3030} {\bibfield  {journal} {\bibinfo  {journal}
  {Nature Materials}\ }\textbf {\bibinfo {volume} {10}},\ \bibinfo {pages}
  {582} (\bibinfo {year} {2011})}\BibitemShut {NoStop}%
\bibitem [{\citenamefont {Lange}\ \emph {et~al.}(2022)\citenamefont {Lange},
  \citenamefont {Bouhon}, \citenamefont {Monserrat},\ and\ \citenamefont
  {Slager}}]{PhysRevB.105.064301}%
  \BibitemOpen
  \bibfield  {author} {\bibinfo {author} {\bibfnamefont {G.~F.}\ \bibnamefont
  {Lange}}, \bibinfo {author} {\bibfnamefont {A.}~\bibnamefont {Bouhon}},
  \bibinfo {author} {\bibfnamefont {B.}~\bibnamefont {Monserrat}},\ and\
  \bibinfo {author} {\bibfnamefont {R.-J.}\ \bibnamefont {Slager}},\ }\href
  {https://doi.org/10.1103/PhysRevB.105.064301} {\bibfield  {journal} {\bibinfo
   {journal} {Phys. Rev. B}\ }\textbf {\bibinfo {volume} {105}},\ \bibinfo
  {pages} {064301} (\bibinfo {year} {2022})}\BibitemShut {NoStop}%
\bibitem [{\citenamefont {Lenggenhager}\ \emph {et~al.}(2022)\citenamefont
  {Lenggenhager}, \citenamefont {Liu}, \citenamefont {Neupert},\ and\
  \citenamefont {Bzdušek}}]{RN36}%
  \BibitemOpen
  \bibfield  {author} {\bibinfo {author} {\bibfnamefont {P.~M.}\ \bibnamefont
  {Lenggenhager}}, \bibinfo {author} {\bibfnamefont {X.}~\bibnamefont {Liu}},
  \bibinfo {author} {\bibfnamefont {T.}~\bibnamefont {Neupert}},\ and\ \bibinfo
  {author} {\bibfnamefont {T.}~\bibnamefont {Bzdušek}},\ }\href
  {https://doi.org/10.1103/PhysRevB.106.085129} {\bibfield  {journal} {\bibinfo
   {journal} {Physical Review B}\ }\textbf {\bibinfo {volume} {106}},\ \bibinfo
  {pages} {085129} (\bibinfo {year} {2022})}\BibitemShut {NoStop}%
\bibitem [{\citenamefont {Ünal}\ \emph {et~al.}(2020)\citenamefont {Ünal},
  \citenamefont {Bouhon},\ and\ \citenamefont {Slager}}]{RN37}%
  \BibitemOpen
  \bibfield  {author} {\bibinfo {author} {\bibfnamefont {F.~N.}\ \bibnamefont
  {Ünal}}, \bibinfo {author} {\bibfnamefont {A.}~\bibnamefont {Bouhon}},\ and\
  \bibinfo {author} {\bibfnamefont {R.-J.}\ \bibnamefont {Slager}},\ }\href
  {https://doi.org/10.1103/PhysRevLett.125.053601} {\bibfield  {journal}
  {\bibinfo  {journal} {Physical Review Letters}\ }\textbf {\bibinfo {volume}
  {125}},\ \bibinfo {pages} {053601} (\bibinfo {year} {2020})}\BibitemShut
  {NoStop}%
\bibitem [{\citenamefont {Ezawa}(2021)}]{RN38}%
  \BibitemOpen
  \bibfield  {author} {\bibinfo {author} {\bibfnamefont {M.}~\bibnamefont
  {Ezawa}},\ }\href {https://doi.org/10.1103/PhysRevB.103.205303} {\bibfield
  {journal} {\bibinfo  {journal} {Physical Review B}\ }\textbf {\bibinfo
  {volume} {103}},\ \bibinfo {pages} {205303} (\bibinfo {year}
  {2021})}\BibitemShut {NoStop}%
\bibitem [{\citenamefont {Bouhon}\ \emph {et~al.}(2023)\citenamefont {Bouhon},
  \citenamefont {Zhu}, \citenamefont {Slager},\ and\ \citenamefont
  {Palumbo}}]{bouhon2023second}%
  \BibitemOpen
  \bibfield  {author} {\bibinfo {author} {\bibfnamefont {A.}~\bibnamefont
  {Bouhon}}, \bibinfo {author} {\bibfnamefont {Y.-Q.}\ \bibnamefont {Zhu}},
  \bibinfo {author} {\bibfnamefont {R.-J.}\ \bibnamefont {Slager}},\ and\
  \bibinfo {author} {\bibfnamefont {G.}~\bibnamefont {Palumbo}},\ }\bibfield
  {journal} {\bibinfo  {journal} {arXiv preprint arXiv:2301.08827}\ }\href
  {https://doi.org/10.48550/arXiv.2301.08827} {10.48550/arXiv.2301.08827}
  (\bibinfo {year} {2023})\BibitemShut {NoStop}%
\bibitem [{\citenamefont {Nalitov}\ \emph {et~al.}(2015)\citenamefont
  {Nalitov}, \citenamefont {Malpuech}, \citenamefont {Ter{\c{c}}as},\ and\
  \citenamefont {Solnyshkov}}]{nalitov2015spin}%
  \BibitemOpen
  \bibfield  {author} {\bibinfo {author} {\bibfnamefont {A.}~\bibnamefont
  {Nalitov}}, \bibinfo {author} {\bibfnamefont {G.}~\bibnamefont {Malpuech}},
  \bibinfo {author} {\bibfnamefont {H.}~\bibnamefont {Ter{\c{c}}as}},\ and\
  \bibinfo {author} {\bibfnamefont {D.}~\bibnamefont {Solnyshkov}},\ }\href
  {https://doi.org/10.1103/PhysRevLett.114.026803} {\bibfield  {journal}
  {\bibinfo  {journal} {Physical review letters}\ }\textbf {\bibinfo {volume}
  {114}},\ \bibinfo {pages} {026803} (\bibinfo {year} {2015})}\BibitemShut
  {NoStop}%
\bibitem [{\citenamefont {Yang}\ \emph
  {et~al.}(2020{\natexlab{b}})\citenamefont {Yang}, \citenamefont {Zhen},
  \citenamefont {Joannopoulos},\ and\ \citenamefont
  {Solja{\v{c}}i{\'{c}}}}]{yang20202d}%
  \BibitemOpen
  \bibfield  {author} {\bibinfo {author} {\bibfnamefont {Y.}~\bibnamefont
  {Yang}}, \bibinfo {author} {\bibfnamefont {B.}~\bibnamefont {Zhen}}, \bibinfo
  {author} {\bibfnamefont {J.~D.}\ \bibnamefont {Joannopoulos}},\ and\ \bibinfo
  {author} {\bibfnamefont {M.}~\bibnamefont {Solja{\v{c}}i{\'{c}}}},\ }\href
  {https://doi.org/10.1038/s41377-020-00384-7} {\bibfield  {journal} {\bibinfo
  {journal} {Light: Science {\&} Applications}\ }\textbf {\bibinfo {volume}
  {9}},\ \bibinfo {pages} {177} (\bibinfo {year}
  {2020}{\natexlab{b}})}\BibitemShut {NoStop}%
\bibitem [{\citenamefont {Dalibard}\ \emph {et~al.}(2011)\citenamefont
  {Dalibard}, \citenamefont {Gerbier}, \citenamefont {Juzeli{\=u}nas},\ and\
  \citenamefont {{\"O}hberg}}]{dalibard2011colloquium}%
  \BibitemOpen
  \bibfield  {author} {\bibinfo {author} {\bibfnamefont {J.}~\bibnamefont
  {Dalibard}}, \bibinfo {author} {\bibfnamefont {F.}~\bibnamefont {Gerbier}},
  \bibinfo {author} {\bibfnamefont {G.}~\bibnamefont {Juzeli{\=u}nas}},\ and\
  \bibinfo {author} {\bibfnamefont {P.}~\bibnamefont {{\"O}hberg}},\ }\href
  {https://doi.org/10.1103/RevModPhys.83.1523} {\bibfield  {journal} {\bibinfo
  {journal} {Rev. Mod. Phys.}\ }\textbf {\bibinfo {volume} {83}},\ \bibinfo
  {pages} {1523} (\bibinfo {year} {2011})}\BibitemShut {NoStop}%
\bibitem [{\citenamefont {Goldman}\ \emph {et~al.}(2014)\citenamefont
  {Goldman}, \citenamefont {Juzeli{\=u}nas}, \citenamefont {{\"O}hberg},\ and\
  \citenamefont {Spielman}}]{goldman2014light}%
  \BibitemOpen
  \bibfield  {author} {\bibinfo {author} {\bibfnamefont {N.}~\bibnamefont
  {Goldman}}, \bibinfo {author} {\bibfnamefont {G.}~\bibnamefont
  {Juzeli{\=u}nas}}, \bibinfo {author} {\bibfnamefont {P.}~\bibnamefont
  {{\"O}hberg}},\ and\ \bibinfo {author} {\bibfnamefont {I.~B.}\ \bibnamefont
  {Spielman}},\ }\href {https://doi.org/10.1088/0034-4885/77/12/126401}
  {\bibfield  {journal} {\bibinfo  {journal} {Reports on Progress in Physics}\
  }\textbf {\bibinfo {volume} {77}},\ \bibinfo {pages} {126401} (\bibinfo
  {year} {2014})}\BibitemShut {NoStop}%
\bibitem [{\citenamefont {Wu}\ and\ \citenamefont
  {Yang}(1975)}]{wu1975concept}%
  \BibitemOpen
  \bibfield  {author} {\bibinfo {author} {\bibfnamefont {T.~T.}\ \bibnamefont
  {Wu}}\ and\ \bibinfo {author} {\bibfnamefont {C.~N.}\ \bibnamefont {Yang}},\
  }\href {https://doi.org/10.1103/PhysRevD.12.3845} {\bibfield  {journal}
  {\bibinfo  {journal} {Phys. Rev. D}\ }\textbf {\bibinfo {volume} {12}},\
  \bibinfo {pages} {3845} (\bibinfo {year} {1975})}\BibitemShut {NoStop}%
\bibitem [{\citenamefont {Horv{\'a}thy}(1986)}]{horvathy1986non}%
  \BibitemOpen
  \bibfield  {author} {\bibinfo {author} {\bibfnamefont {P.}~\bibnamefont
  {Horv{\'a}thy}},\ }\href {https://doi.org/10.1103/PhysRevD.33.407} {\bibfield
   {journal} {\bibinfo  {journal} {Phys. Rev. D}\ }\textbf {\bibinfo {volume}
  {33}},\ \bibinfo {pages} {407} (\bibinfo {year} {1986})}\BibitemShut
  {NoStop}%
\bibitem [{\citenamefont {Alford}\ \emph {et~al.}(1990)\citenamefont {Alford},
  \citenamefont {March-Russell},\ and\ \citenamefont
  {Wilczek}}]{alford1990discrete}%
  \BibitemOpen
  \bibfield  {author} {\bibinfo {author} {\bibfnamefont {M.~G.}\ \bibnamefont
  {Alford}}, \bibinfo {author} {\bibfnamefont {J.}~\bibnamefont
  {March-Russell}},\ and\ \bibinfo {author} {\bibfnamefont {F.}~\bibnamefont
  {Wilczek}},\ }\href {https://doi.org/10.1016/0550-3213(90)90512-C} {\bibfield
   {journal} {\bibinfo  {journal} {Nuclear Physics B}\ }\textbf {\bibinfo
  {volume} {337}},\ \bibinfo {pages} {695} (\bibinfo {year}
  {1990})}\BibitemShut {NoStop}%
\bibitem [{\citenamefont {Chaichian}\ \emph {et~al.}(2002)\citenamefont
  {Chaichian}, \citenamefont {Pre{\v{s}}najder}, \citenamefont
  {Sheikh-Jabbari},\ and\ \citenamefont {Tureanu}}]{chaichian2002aharonov}%
  \BibitemOpen
  \bibfield  {author} {\bibinfo {author} {\bibfnamefont {M.}~\bibnamefont
  {Chaichian}}, \bibinfo {author} {\bibfnamefont {P.}~\bibnamefont
  {Pre{\v{s}}najder}}, \bibinfo {author} {\bibfnamefont {M.}~\bibnamefont
  {Sheikh-Jabbari}},\ and\ \bibinfo {author} {\bibfnamefont {A.}~\bibnamefont
  {Tureanu}},\ }\href {https://doi.org/10.1016/S0370-2693(02)01176-0}
  {\bibfield  {journal} {\bibinfo  {journal} {Physics Letters B}\ }\textbf
  {\bibinfo {volume} {527}},\ \bibinfo {pages} {149} (\bibinfo {year}
  {2002})}\BibitemShut {NoStop}%
\bibitem [{\citenamefont {Jacob}\ \emph {et~al.}(2007)\citenamefont {Jacob},
  \citenamefont {{\"O}hberg}, \citenamefont {Juzeli{\=u}nas},\ and\
  \citenamefont {Santos}}]{jacob2007cold}%
  \BibitemOpen
  \bibfield  {author} {\bibinfo {author} {\bibfnamefont {A.}~\bibnamefont
  {Jacob}}, \bibinfo {author} {\bibfnamefont {P.}~\bibnamefont {{\"O}hberg}},
  \bibinfo {author} {\bibfnamefont {G.}~\bibnamefont {Juzeli{\=u}nas}},\ and\
  \bibinfo {author} {\bibfnamefont {L.}~\bibnamefont {Santos}},\ }\href
  {https://doi.org/10.1007/s00340-007-2865-6} {\bibfield  {journal} {\bibinfo
  {journal} {Appl. Phys. B}\ }\textbf {\bibinfo {volume} {89}},\ \bibinfo
  {pages} {439} (\bibinfo {year} {2007})}\BibitemShut {NoStop}%
\bibitem [{\citenamefont {Zhang}\ \emph {et~al.}(2008)\citenamefont {Zhang},
  \citenamefont {Wang}, \citenamefont {Hu}, \citenamefont {Zhang},\ and\
  \citenamefont {Zhu}}]{zhang2008detecting}%
  \BibitemOpen
  \bibfield  {author} {\bibinfo {author} {\bibfnamefont {X.-D.}\ \bibnamefont
  {Zhang}}, \bibinfo {author} {\bibfnamefont {Z.}~\bibnamefont {Wang}},
  \bibinfo {author} {\bibfnamefont {L.-B.}\ \bibnamefont {Hu}}, \bibinfo
  {author} {\bibfnamefont {Z.-M.}\ \bibnamefont {Zhang}},\ and\ \bibinfo
  {author} {\bibfnamefont {S.-L.}\ \bibnamefont {Zhu}},\ }\href
  {https://doi.org/10.1088/1367-2630/10/4/043031} {\bibfield  {journal}
  {\bibinfo  {journal} {New Journal of Physics}\ }\textbf {\bibinfo {volume}
  {10}},\ \bibinfo {pages} {043031} (\bibinfo {year} {2008})}\BibitemShut
  {NoStop}%
\bibitem [{\citenamefont {Fruchart}\ \emph {et~al.}(2020)\citenamefont
  {Fruchart}, \citenamefont {Zhou},\ and\ \citenamefont
  {Vitelli}}]{fruchart2019dualities}%
  \BibitemOpen
  \bibfield  {author} {\bibinfo {author} {\bibfnamefont {M.}~\bibnamefont
  {Fruchart}}, \bibinfo {author} {\bibfnamefont {Y.}~\bibnamefont {Zhou}},\
  and\ \bibinfo {author} {\bibfnamefont {V.}~\bibnamefont {Vitelli}},\ }\href
  {https://doi.org/10.1038/s41586-020-1932-6} {\bibfield  {journal} {\bibinfo
  {journal} {Nature}\ }\textbf {\bibinfo {volume} {577}},\ \bibinfo {pages}
  {636} (\bibinfo {year} {2020})}\BibitemShut {NoStop}%
\bibitem [{\citenamefont {Zawadzki}\ and\ \citenamefont
  {Rusin}(2011)}]{zawadzki2011zitterbewegung}%
  \BibitemOpen
  \bibfield  {author} {\bibinfo {author} {\bibfnamefont {W.}~\bibnamefont
  {Zawadzki}}\ and\ \bibinfo {author} {\bibfnamefont {T.~M.}\ \bibnamefont
  {Rusin}},\ }\href {https://doi.org/10.1088/0953-8984/23/14/143201} {\bibfield
   {journal} {\bibinfo  {journal} {Journal of Physics: Condensed Matter}\
  }\textbf {\bibinfo {volume} {23}},\ \bibinfo {pages} {143201} (\bibinfo
  {year} {2011})}\BibitemShut {NoStop}%
\bibitem [{\citenamefont {Vaishnav}\ and\ \citenamefont
  {Clark}(2008)}]{vaishnav2008observing}%
  \BibitemOpen
  \bibfield  {author} {\bibinfo {author} {\bibfnamefont {J.}~\bibnamefont
  {Vaishnav}}\ and\ \bibinfo {author} {\bibfnamefont {C.~W.}\ \bibnamefont
  {Clark}},\ }\href {https://doi.org/10.1103/PhysRevLett.100.153002} {\bibfield
   {journal} {\bibinfo  {journal} {Physical review letters}\ }\textbf {\bibinfo
  {volume} {100}},\ \bibinfo {pages} {153002} (\bibinfo {year}
  {2008})}\BibitemShut {NoStop}%
\bibitem [{\citenamefont {Hasan}\ \emph {et~al.}(2022)\citenamefont {Hasan},
  \citenamefont {Madasu}, \citenamefont {Rathod}, \citenamefont {Kwong},
  \citenamefont {Miniatura}, \citenamefont {Chevy},\ and\ \citenamefont
  {Wilkowski}}]{hasan2022wave}%
  \BibitemOpen
  \bibfield  {author} {\bibinfo {author} {\bibfnamefont {M.}~\bibnamefont
  {Hasan}}, \bibinfo {author} {\bibfnamefont {C.~S.}\ \bibnamefont {Madasu}},
  \bibinfo {author} {\bibfnamefont {K.~D.}\ \bibnamefont {Rathod}}, \bibinfo
  {author} {\bibfnamefont {C.~C.}\ \bibnamefont {Kwong}}, \bibinfo {author}
  {\bibfnamefont {C.}~\bibnamefont {Miniatura}}, \bibinfo {author}
  {\bibfnamefont {F.}~\bibnamefont {Chevy}},\ and\ \bibinfo {author}
  {\bibfnamefont {D.}~\bibnamefont {Wilkowski}},\ }\href
  {https://doi.org/10.1103/PhysRevLett.129.130402} {\bibfield  {journal}
  {\bibinfo  {journal} {Physical Review Letters}\ }\textbf {\bibinfo {volume}
  {129}},\ \bibinfo {pages} {130402} (\bibinfo {year} {2022})}\BibitemShut
  {NoStop}%
\bibitem [{\citenamefont {Zhang}\ and\ \citenamefont
  {Liu}(2008)}]{zhang2008extremal}%
  \BibitemOpen
  \bibfield  {author} {\bibinfo {author} {\bibfnamefont {X.}~\bibnamefont
  {Zhang}}\ and\ \bibinfo {author} {\bibfnamefont {Z.}~\bibnamefont {Liu}},\
  }\href {https://doi.org/10.1103/PhysRevLett.101.264303} {\bibfield  {journal}
  {\bibinfo  {journal} {Physical review letters}\ }\textbf {\bibinfo {volume}
  {101}},\ \bibinfo {pages} {264303} (\bibinfo {year} {2008})}\BibitemShut
  {NoStop}%
\bibitem [{\citenamefont {Wang}\ \emph {et~al.}(2009)\citenamefont {Wang},
  \citenamefont {Wang},\ and\ \citenamefont {Zhu}}]{wang2009zitterbewegung}%
  \BibitemOpen
  \bibfield  {author} {\bibinfo {author} {\bibfnamefont {L.-G.}\ \bibnamefont
  {Wang}}, \bibinfo {author} {\bibfnamefont {Z.-G.}\ \bibnamefont {Wang}},\
  and\ \bibinfo {author} {\bibfnamefont {S.-Y.}\ \bibnamefont {Zhu}},\ }\href
  {https://doi.org/10.1209/0295-5075/86/47008} {\bibfield  {journal} {\bibinfo
  {journal} {Europhysics Letters}\ }\textbf {\bibinfo {volume} {86}},\ \bibinfo
  {pages} {47008} (\bibinfo {year} {2009})}\BibitemShut {NoStop}%
\bibitem [{\citenamefont {Longhi}(2010)}]{longhi2010photonic}%
  \BibitemOpen
  \bibfield  {author} {\bibinfo {author} {\bibfnamefont {S.}~\bibnamefont
  {Longhi}},\ }\href {https://doi.org/10.1364/OL.35.000235} {\bibfield
  {journal} {\bibinfo  {journal} {Optics letters}\ }\textbf {\bibinfo {volume}
  {35}},\ \bibinfo {pages} {235} (\bibinfo {year} {2010})}\BibitemShut
  {NoStop}%
\bibitem [{\citenamefont {Zhang}(2008)}]{zhang2008observing}%
  \BibitemOpen
  \bibfield  {author} {\bibinfo {author} {\bibfnamefont {X.}~\bibnamefont
  {Zhang}},\ }\href {https://doi.org/10.1103/PhysRevLett.100.113903} {\bibfield
   {journal} {\bibinfo  {journal} {Physical review letters}\ }\textbf {\bibinfo
  {volume} {100}},\ \bibinfo {pages} {113903} (\bibinfo {year}
  {2008})}\BibitemShut {NoStop}%
\bibitem [{\citenamefont {Dreisow}\ \emph {et~al.}(2010)\citenamefont
  {Dreisow}, \citenamefont {Heinrich}, \citenamefont {Keil}, \citenamefont
  {T{\"u}nnermann}, \citenamefont {Nolte}, \citenamefont {Longhi},\ and\
  \citenamefont {Szameit}}]{dreisow2010classical}%
  \BibitemOpen
  \bibfield  {author} {\bibinfo {author} {\bibfnamefont {F.}~\bibnamefont
  {Dreisow}}, \bibinfo {author} {\bibfnamefont {M.}~\bibnamefont {Heinrich}},
  \bibinfo {author} {\bibfnamefont {R.}~\bibnamefont {Keil}}, \bibinfo {author}
  {\bibfnamefont {A.}~\bibnamefont {T{\"u}nnermann}}, \bibinfo {author}
  {\bibfnamefont {S.}~\bibnamefont {Nolte}}, \bibinfo {author} {\bibfnamefont
  {S.}~\bibnamefont {Longhi}},\ and\ \bibinfo {author} {\bibfnamefont
  {A.}~\bibnamefont {Szameit}},\ }\href
  {https://doi.org/10.1103/PhysRevLett.105.143902} {\bibfield  {journal}
  {\bibinfo  {journal} {Physical Review Letters}\ }\textbf {\bibinfo {volume}
  {105}},\ \bibinfo {pages} {143902} (\bibinfo {year} {2010})}\BibitemShut
  {NoStop}%
\bibitem [{\citenamefont {Shelykh}\ \emph {et~al.}(2009)\citenamefont
  {Shelykh}, \citenamefont {Kavokin}, \citenamefont {Rubo}, \citenamefont
  {Liew},\ and\ \citenamefont {Malpuech}}]{shelykh2009polariton}%
  \BibitemOpen
  \bibfield  {author} {\bibinfo {author} {\bibfnamefont {I.~A.}\ \bibnamefont
  {Shelykh}}, \bibinfo {author} {\bibfnamefont {A.~V.}\ \bibnamefont
  {Kavokin}}, \bibinfo {author} {\bibfnamefont {Y.~G.}\ \bibnamefont {Rubo}},
  \bibinfo {author} {\bibfnamefont {T.}~\bibnamefont {Liew}},\ and\ \bibinfo
  {author} {\bibfnamefont {G.}~\bibnamefont {Malpuech}},\ }\href
  {https://doi.org/10.1088/0268-1242/25/1/013001} {\bibfield  {journal}
  {\bibinfo  {journal} {Semiconductor Science and Technology}\ }\textbf
  {\bibinfo {volume} {25}},\ \bibinfo {pages} {013001} (\bibinfo {year}
  {2009})}\BibitemShut {NoStop}%
\bibitem [{\citenamefont {Solnyshkov}(2018)}]{bleu2018measuring}%
  \BibitemOpen
  \bibfield  {author} {\bibinfo {author} {\bibfnamefont {D.}~\bibnamefont
  {Solnyshkov}},\ }\href {https://doi.org/10.1103/PhysRevB.97.195422}
  {\bibfield  {journal} {\bibinfo  {journal} {Physical Review B}\ }\textbf
  {\bibinfo {volume} {97}},\ \bibinfo {pages} {195422} (\bibinfo {year}
  {2018})}\BibitemShut {NoStop}%
\bibitem [{\citenamefont {Gianfrate}\ \emph {et~al.}(2020)\citenamefont
  {Gianfrate}, \citenamefont {Bleu}, \citenamefont {Dominici}, \citenamefont
  {Ardizzone}, \citenamefont {De~Giorgi}, \citenamefont {Ballarini},
  \citenamefont {Lerario}, \citenamefont {West}, \citenamefont {Pfeiffer},
  \citenamefont {Solnyshkov}, \citenamefont {Sanvitto},\ and\ \citenamefont
  {Malpuech}}]{gianfrate2020direct}%
  \BibitemOpen
  \bibfield  {author} {\bibinfo {author} {\bibfnamefont {A.}~\bibnamefont
  {Gianfrate}}, \bibinfo {author} {\bibfnamefont {O.}~\bibnamefont {Bleu}},
  \bibinfo {author} {\bibfnamefont {L.}~\bibnamefont {Dominici}}, \bibinfo
  {author} {\bibfnamefont {V.}~\bibnamefont {Ardizzone}}, \bibinfo {author}
  {\bibfnamefont {M.}~\bibnamefont {De~Giorgi}}, \bibinfo {author}
  {\bibfnamefont {D.}~\bibnamefont {Ballarini}}, \bibinfo {author}
  {\bibfnamefont {G.}~\bibnamefont {Lerario}}, \bibinfo {author} {\bibfnamefont
  {K.~W.}\ \bibnamefont {West}}, \bibinfo {author} {\bibfnamefont {L.~N.}\
  \bibnamefont {Pfeiffer}}, \bibinfo {author} {\bibfnamefont {D.~D.}\
  \bibnamefont {Solnyshkov}}, \bibinfo {author} {\bibfnamefont
  {D.}~\bibnamefont {Sanvitto}},\ and\ \bibinfo {author} {\bibfnamefont
  {G.}~\bibnamefont {Malpuech}},\ }\href
  {https://doi.org/10.1038/s41586-020-1989-2} {\bibfield  {journal} {\bibinfo
  {journal} {Nature}\ }\textbf {\bibinfo {volume} {578}},\ \bibinfo {pages}
  {381} (\bibinfo {year} {2020})}\BibitemShut {NoStop}%
\bibitem [{\citenamefont {Jacqmin}\ \emph {et~al.}(2014)\citenamefont
  {Jacqmin}, \citenamefont {Carusotto}, \citenamefont {Sagnes}, \citenamefont
  {Abbarchi}, \citenamefont {Solnyshkov}, \citenamefont {Malpuech},
  \citenamefont {Galopin}, \citenamefont {Lema{\^\i}tre}, \citenamefont
  {Bloch},\ and\ \citenamefont {Amo}}]{jacqmin2014direct}%
  \BibitemOpen
  \bibfield  {author} {\bibinfo {author} {\bibfnamefont {T.}~\bibnamefont
  {Jacqmin}}, \bibinfo {author} {\bibfnamefont {I.}~\bibnamefont {Carusotto}},
  \bibinfo {author} {\bibfnamefont {I.}~\bibnamefont {Sagnes}}, \bibinfo
  {author} {\bibfnamefont {M.}~\bibnamefont {Abbarchi}}, \bibinfo {author}
  {\bibfnamefont {D.}~\bibnamefont {Solnyshkov}}, \bibinfo {author}
  {\bibfnamefont {G.}~\bibnamefont {Malpuech}}, \bibinfo {author}
  {\bibfnamefont {E.}~\bibnamefont {Galopin}}, \bibinfo {author} {\bibfnamefont
  {A.}~\bibnamefont {Lema{\^\i}tre}}, \bibinfo {author} {\bibfnamefont
  {J.}~\bibnamefont {Bloch}},\ and\ \bibinfo {author} {\bibfnamefont
  {A.}~\bibnamefont {Amo}},\ }\href
  {https://doi.org/10.1103/PhysRevLett.112.116402} {\bibfield  {journal}
  {\bibinfo  {journal} {Physical review letters}\ }\textbf {\bibinfo {volume}
  {112}},\ \bibinfo {pages} {116402} (\bibinfo {year} {2014})}\BibitemShut
  {NoStop}%
\bibitem [{\citenamefont {Sala}\ \emph {et~al.}(2015)\citenamefont {Sala},
  \citenamefont {Solnyshkov}, \citenamefont {Carusotto}, \citenamefont
  {Jacqmin}, \citenamefont {Lema{\^\i}tre}, \citenamefont {Ter{\c{c}}as},
  \citenamefont {Nalitov}, \citenamefont {Abbarchi}, \citenamefont {Galopin},
  \citenamefont {Sagnes} \emph {et~al.}}]{sala2015spin}%
  \BibitemOpen
  \bibfield  {author} {\bibinfo {author} {\bibfnamefont {V.}~\bibnamefont
  {Sala}}, \bibinfo {author} {\bibfnamefont {D.}~\bibnamefont {Solnyshkov}},
  \bibinfo {author} {\bibfnamefont {I.}~\bibnamefont {Carusotto}}, \bibinfo
  {author} {\bibfnamefont {T.}~\bibnamefont {Jacqmin}}, \bibinfo {author}
  {\bibfnamefont {A.}~\bibnamefont {Lema{\^\i}tre}}, \bibinfo {author}
  {\bibfnamefont {H.}~\bibnamefont {Ter{\c{c}}as}}, \bibinfo {author}
  {\bibfnamefont {A.}~\bibnamefont {Nalitov}}, \bibinfo {author} {\bibfnamefont
  {M.}~\bibnamefont {Abbarchi}}, \bibinfo {author} {\bibfnamefont
  {E.}~\bibnamefont {Galopin}}, \bibinfo {author} {\bibfnamefont
  {I.}~\bibnamefont {Sagnes}}, \emph {et~al.},\ }\href
  {https://doi.org/10.1103/PhysRevX.5.011034} {\bibfield  {journal} {\bibinfo
  {journal} {Physical Review X}\ }\textbf {\bibinfo {volume} {5}},\ \bibinfo
  {pages} {011034} (\bibinfo {year} {2015})}\BibitemShut {NoStop}%
\bibitem [{\citenamefont {Leyder}\ \emph {et~al.}(2007)\citenamefont {Leyder},
  \citenamefont {Romanelli}, \citenamefont {Karr}, \citenamefont {Giacobino},
  \citenamefont {Liew}, \citenamefont {Glazov}, \citenamefont {Kavokin},
  \citenamefont {Malpuech},\ and\ \citenamefont
  {Bramati}}]{leyder2007observation}%
  \BibitemOpen
  \bibfield  {author} {\bibinfo {author} {\bibfnamefont {C.}~\bibnamefont
  {Leyder}}, \bibinfo {author} {\bibfnamefont {M.}~\bibnamefont {Romanelli}},
  \bibinfo {author} {\bibfnamefont {J.~P.}\ \bibnamefont {Karr}}, \bibinfo
  {author} {\bibfnamefont {E.}~\bibnamefont {Giacobino}}, \bibinfo {author}
  {\bibfnamefont {T.~C.}\ \bibnamefont {Liew}}, \bibinfo {author}
  {\bibfnamefont {M.~M.}\ \bibnamefont {Glazov}}, \bibinfo {author}
  {\bibfnamefont {A.~V.}\ \bibnamefont {Kavokin}}, \bibinfo {author}
  {\bibfnamefont {G.}~\bibnamefont {Malpuech}},\ and\ \bibinfo {author}
  {\bibfnamefont {A.}~\bibnamefont {Bramati}},\ }\href
  {https://doi.org/10.1038/nphys676} {\bibfield  {journal} {\bibinfo  {journal}
  {Nature Physics}\ }\textbf {\bibinfo {volume} {3}},\ \bibinfo {pages} {628}
  (\bibinfo {year} {2007})}\BibitemShut {NoStop}%
\bibitem [{\citenamefont {Spencer}\ \emph {et~al.}(2021)\citenamefont
  {Spencer}, \citenamefont {Fu}, \citenamefont {Schlaus}, \citenamefont
  {Hwang}, \citenamefont {Dai}, \citenamefont {Smith}, \citenamefont
  {Gamelin},\ and\ \citenamefont {Zhu}}]{spencer2021spin}%
  \BibitemOpen
  \bibfield  {author} {\bibinfo {author} {\bibfnamefont {M.~S.}\ \bibnamefont
  {Spencer}}, \bibinfo {author} {\bibfnamefont {Y.}~\bibnamefont {Fu}},
  \bibinfo {author} {\bibfnamefont {A.~P.}\ \bibnamefont {Schlaus}}, \bibinfo
  {author} {\bibfnamefont {D.}~\bibnamefont {Hwang}}, \bibinfo {author}
  {\bibfnamefont {Y.}~\bibnamefont {Dai}}, \bibinfo {author} {\bibfnamefont
  {M.~D.}\ \bibnamefont {Smith}}, \bibinfo {author} {\bibfnamefont {D.~R.}\
  \bibnamefont {Gamelin}},\ and\ \bibinfo {author} {\bibfnamefont {X.-Y.}\
  \bibnamefont {Zhu}},\ }\href {https://doi.org/10.1126/sciadv.abj7667}
  {\bibfield  {journal} {\bibinfo  {journal} {Science Advances}\ }\textbf
  {\bibinfo {volume} {7}},\ \bibinfo {pages} {eabj7667} (\bibinfo {year}
  {2021})}\BibitemShut {NoStop}%
\bibitem [{\citenamefont {{\L}empicka-Mirek}\ \emph {et~al.}(2022)\citenamefont
  {{\L}empicka-Mirek}, \citenamefont {Kr{\'o}l}, \citenamefont {Sigurdsson},
  \citenamefont {Wincukiewicz}, \citenamefont {Morawiak}, \citenamefont
  {Mazur}, \citenamefont {Muszy{\'n}ski}, \citenamefont {Piecek}, \citenamefont
  {Kula}, \citenamefont {Stefaniuk} \emph {et~al.}}]{lempicka2022electrically}%
  \BibitemOpen
  \bibfield  {author} {\bibinfo {author} {\bibfnamefont {K.}~\bibnamefont
  {{\L}empicka-Mirek}}, \bibinfo {author} {\bibfnamefont {M.}~\bibnamefont
  {Kr{\'o}l}}, \bibinfo {author} {\bibfnamefont {H.}~\bibnamefont
  {Sigurdsson}}, \bibinfo {author} {\bibfnamefont {A.}~\bibnamefont
  {Wincukiewicz}}, \bibinfo {author} {\bibfnamefont {P.}~\bibnamefont
  {Morawiak}}, \bibinfo {author} {\bibfnamefont {R.}~\bibnamefont {Mazur}},
  \bibinfo {author} {\bibfnamefont {M.}~\bibnamefont {Muszy{\'n}ski}}, \bibinfo
  {author} {\bibfnamefont {W.}~\bibnamefont {Piecek}}, \bibinfo {author}
  {\bibfnamefont {P.}~\bibnamefont {Kula}}, \bibinfo {author} {\bibfnamefont
  {T.}~\bibnamefont {Stefaniuk}}, \emph {et~al.},\ }\href
  {https://doi.org/10.1126/sciadv.abq7533} {\bibfield  {journal} {\bibinfo
  {journal} {Science Advances}\ }\textbf {\bibinfo {volume} {8}},\ \bibinfo
  {pages} {eabq7533} (\bibinfo {year} {2022})}\BibitemShut {NoStop}%
\bibitem [{\citenamefont {Li}\ \emph {et~al.}(2022{\natexlab{b}})\citenamefont
  {Li}, \citenamefont {Ma}, \citenamefont {Zhai}, \citenamefont {Gao},
  \citenamefont {Dai}, \citenamefont {Schumacher},\ and\ \citenamefont
  {Gao}}]{li2022manipulating}%
  \BibitemOpen
  \bibfield  {author} {\bibinfo {author} {\bibfnamefont {Y.}~\bibnamefont
  {Li}}, \bibinfo {author} {\bibfnamefont {X.}~\bibnamefont {Ma}}, \bibinfo
  {author} {\bibfnamefont {X.}~\bibnamefont {Zhai}}, \bibinfo {author}
  {\bibfnamefont {M.}~\bibnamefont {Gao}}, \bibinfo {author} {\bibfnamefont
  {H.}~\bibnamefont {Dai}}, \bibinfo {author} {\bibfnamefont {S.}~\bibnamefont
  {Schumacher}},\ and\ \bibinfo {author} {\bibfnamefont {T.}~\bibnamefont
  {Gao}},\ }\href {https://doi.org/10.1038/s41467-022-32032-6} {\bibfield
  {journal} {\bibinfo  {journal} {nature communications}\ }\textbf {\bibinfo
  {volume} {13}},\ \bibinfo {pages} {3785} (\bibinfo {year}
  {2022}{\natexlab{b}})}\BibitemShut {NoStop}%
\bibitem [{\citenamefont {Polimeno}\ \emph
  {et~al.}(2021{\natexlab{b}})\citenamefont {Polimeno}, \citenamefont
  {Lerario}, \citenamefont {De~Giorgi}, \citenamefont {De~Marco}, \citenamefont
  {Dominici}, \citenamefont {Todisco}, \citenamefont {Coriolano}, \citenamefont
  {Ardizzone}, \citenamefont {Pugliese}, \citenamefont {Prontera} \emph
  {et~al.}}]{polimeno2021tuning}%
  \BibitemOpen
  \bibfield  {author} {\bibinfo {author} {\bibfnamefont {L.}~\bibnamefont
  {Polimeno}}, \bibinfo {author} {\bibfnamefont {G.}~\bibnamefont {Lerario}},
  \bibinfo {author} {\bibfnamefont {M.}~\bibnamefont {De~Giorgi}}, \bibinfo
  {author} {\bibfnamefont {L.}~\bibnamefont {De~Marco}}, \bibinfo {author}
  {\bibfnamefont {L.}~\bibnamefont {Dominici}}, \bibinfo {author}
  {\bibfnamefont {F.}~\bibnamefont {Todisco}}, \bibinfo {author} {\bibfnamefont
  {A.}~\bibnamefont {Coriolano}}, \bibinfo {author} {\bibfnamefont
  {V.}~\bibnamefont {Ardizzone}}, \bibinfo {author} {\bibfnamefont
  {M.}~\bibnamefont {Pugliese}}, \bibinfo {author} {\bibfnamefont {C.~T.}\
  \bibnamefont {Prontera}}, \emph {et~al.},\ }\href
  {https://doi.org/10.1038/s41565-021-01046-4} {\bibfield  {journal} {\bibinfo
  {journal} {Nature nanotechnology}\ }\textbf {\bibinfo {volume} {16}},\
  \bibinfo {pages} {1349} (\bibinfo {year} {2021}{\natexlab{b}})}\BibitemShut
  {NoStop}%
\bibitem [{\citenamefont {Ren}\ \emph {et~al.}(2021)\citenamefont {Ren},
  \citenamefont {Liao}, \citenamefont {Li}, \citenamefont {Li}, \citenamefont
  {Bleu}, \citenamefont {Malpuech}, \citenamefont {Yao}, \citenamefont {Fu},\
  and\ \citenamefont {Solnyshkov}}]{ren2021nontrivial}%
  \BibitemOpen
  \bibfield  {author} {\bibinfo {author} {\bibfnamefont {J.}~\bibnamefont
  {Ren}}, \bibinfo {author} {\bibfnamefont {Q.}~\bibnamefont {Liao}}, \bibinfo
  {author} {\bibfnamefont {F.}~\bibnamefont {Li}}, \bibinfo {author}
  {\bibfnamefont {Y.}~\bibnamefont {Li}}, \bibinfo {author} {\bibfnamefont
  {O.}~\bibnamefont {Bleu}}, \bibinfo {author} {\bibfnamefont {G.}~\bibnamefont
  {Malpuech}}, \bibinfo {author} {\bibfnamefont {J.}~\bibnamefont {Yao}},
  \bibinfo {author} {\bibfnamefont {H.}~\bibnamefont {Fu}},\ and\ \bibinfo
  {author} {\bibfnamefont {D.}~\bibnamefont {Solnyshkov}},\ }\href
  {https://doi.org/10.1038/s41467-020-20845-2} {\bibfield  {journal} {\bibinfo
  {journal} {Nature communications}\ }\textbf {\bibinfo {volume} {12}},\
  \bibinfo {pages} {689} (\bibinfo {year} {2021})}\BibitemShut {NoStop}%
\bibitem [{\citenamefont {Ren}\ \emph {et~al.}(2022)\citenamefont {Ren},
  \citenamefont {Liao}, \citenamefont {Ma}, \citenamefont {Schumacher},
  \citenamefont {Yao},\ and\ \citenamefont {Fu}}]{ren2022realization}%
  \BibitemOpen
  \bibfield  {author} {\bibinfo {author} {\bibfnamefont {J.}~\bibnamefont
  {Ren}}, \bibinfo {author} {\bibfnamefont {Q.}~\bibnamefont {Liao}}, \bibinfo
  {author} {\bibfnamefont {X.}~\bibnamefont {Ma}}, \bibinfo {author}
  {\bibfnamefont {S.}~\bibnamefont {Schumacher}}, \bibinfo {author}
  {\bibfnamefont {J.}~\bibnamefont {Yao}},\ and\ \bibinfo {author}
  {\bibfnamefont {H.}~\bibnamefont {Fu}},\ }\href
  {https://doi.org/10.1002/lpor.202100252} {\bibfield  {journal} {\bibinfo
  {journal} {Laser \& Photonics Reviews}\ }\textbf {\bibinfo {volume} {16}},\
  \bibinfo {pages} {2100252} (\bibinfo {year} {2022})}\BibitemShut {NoStop}%
\bibitem [{\citenamefont {Rechci{\'n}ska}\ \emph {et~al.}(2019)\citenamefont
  {Rechci{\'n}ska}, \citenamefont {Kr{\'o}l}, \citenamefont {Mazur},
  \citenamefont {Morawiak}, \citenamefont {Mirek}, \citenamefont {{\L}empicka},
  \citenamefont {Bardyszewski}, \citenamefont {Matuszewski}, \citenamefont
  {Kula}, \citenamefont {Piecek}, \citenamefont {Lagoudakis}, \citenamefont
  {Pietka},\ and\ \citenamefont {Szczytko}}]{rechcinska2019engineering}%
  \BibitemOpen
  \bibfield  {author} {\bibinfo {author} {\bibfnamefont {K.}~\bibnamefont
  {Rechci{\'n}ska}}, \bibinfo {author} {\bibfnamefont {M.}~\bibnamefont
  {Kr{\'o}l}}, \bibinfo {author} {\bibfnamefont {R.}~\bibnamefont {Mazur}},
  \bibinfo {author} {\bibfnamefont {P.}~\bibnamefont {Morawiak}}, \bibinfo
  {author} {\bibfnamefont {R.}~\bibnamefont {Mirek}}, \bibinfo {author}
  {\bibfnamefont {K.}~\bibnamefont {{\L}empicka}}, \bibinfo {author}
  {\bibfnamefont {W.}~\bibnamefont {Bardyszewski}}, \bibinfo {author}
  {\bibfnamefont {M.}~\bibnamefont {Matuszewski}}, \bibinfo {author}
  {\bibfnamefont {P.}~\bibnamefont {Kula}}, \bibinfo {author} {\bibfnamefont
  {W.}~\bibnamefont {Piecek}}, \bibinfo {author} {\bibfnamefont {P.~G.}\
  \bibnamefont {Lagoudakis}}, \bibinfo {author} {\bibfnamefont
  {B.}~\bibnamefont {Pietka}},\ and\ \bibinfo {author} {\bibfnamefont
  {J.}~\bibnamefont {Szczytko}},\ }\href
  {https://doi.org/10.1126/science.aay4182} {\bibfield  {journal} {\bibinfo
  {journal} {Science}\ }\textbf {\bibinfo {volume} {366}},\ \bibinfo {pages}
  {727} (\bibinfo {year} {2019})}\BibitemShut {NoStop}%
\bibitem [{\citenamefont {Qi}\ \emph {et~al.}(2006)\citenamefont {Qi},
  \citenamefont {Wu},\ and\ \citenamefont {Zhang}}]{qi2006topological}%
  \BibitemOpen
  \bibfield  {author} {\bibinfo {author} {\bibfnamefont {X.-L.}\ \bibnamefont
  {Qi}}, \bibinfo {author} {\bibfnamefont {Y.-S.}\ \bibnamefont {Wu}},\ and\
  \bibinfo {author} {\bibfnamefont {S.-C.}\ \bibnamefont {Zhang}},\ }\href
  {https://doi.org/10.1103/PhysRevB.74.085308} {\bibfield  {journal} {\bibinfo
  {journal} {Physical Review B}\ }\textbf {\bibinfo {volume} {74}},\ \bibinfo
  {pages} {085308} (\bibinfo {year} {2006})}\BibitemShut {NoStop}%
\bibitem [{\citenamefont {Ozawa}\ \emph {et~al.}(2016)\citenamefont {Ozawa},
  \citenamefont {Price}, \citenamefont {Goldman}, \citenamefont {Zilberberg},\
  and\ \citenamefont {Carusotto}}]{ozawa2016synthetic}%
  \BibitemOpen
  \bibfield  {author} {\bibinfo {author} {\bibfnamefont {T.}~\bibnamefont
  {Ozawa}}, \bibinfo {author} {\bibfnamefont {H.~M.}\ \bibnamefont {Price}},
  \bibinfo {author} {\bibfnamefont {N.}~\bibnamefont {Goldman}}, \bibinfo
  {author} {\bibfnamefont {O.}~\bibnamefont {Zilberberg}},\ and\ \bibinfo
  {author} {\bibfnamefont {I.}~\bibnamefont {Carusotto}},\ }\href
  {https://doi.org/10.1103/PhysRevA.93.043827} {\bibfield  {journal} {\bibinfo
  {journal} {Phys. Rev. A}\ }\textbf {\bibinfo {volume} {93}},\ \bibinfo
  {pages} {043827} (\bibinfo {year} {2016})}\BibitemShut {NoStop}%
\bibitem [{\citenamefont {Lohse}\ \emph {et~al.}(2018)\citenamefont {Lohse},
  \citenamefont {Schweizer}, \citenamefont {Price}, \citenamefont
  {Zilberberg},\ and\ \citenamefont {Bloch}}]{lohse2018exploring}%
  \BibitemOpen
  \bibfield  {author} {\bibinfo {author} {\bibfnamefont {M.}~\bibnamefont
  {Lohse}}, \bibinfo {author} {\bibfnamefont {C.}~\bibnamefont {Schweizer}},
  \bibinfo {author} {\bibfnamefont {H.~M.}\ \bibnamefont {Price}}, \bibinfo
  {author} {\bibfnamefont {O.}~\bibnamefont {Zilberberg}},\ and\ \bibinfo
  {author} {\bibfnamefont {I.}~\bibnamefont {Bloch}},\ }\href
  {https://doi.org/10.1038/nature25000} {\bibfield  {journal} {\bibinfo
  {journal} {Nature}\ }\textbf {\bibinfo {volume} {553}},\ \bibinfo {pages}
  {55} (\bibinfo {year} {2018})}\BibitemShut {NoStop}%
\bibitem [{\citenamefont {Zilberberg}\ \emph {et~al.}(2018)\citenamefont
  {Zilberberg}, \citenamefont {Huang}, \citenamefont {Guglielmon},
  \citenamefont {Wang}, \citenamefont {Chen}, \citenamefont {Kraus},\ and\
  \citenamefont {Rechtsman}}]{zilberberg2018photonic}%
  \BibitemOpen
  \bibfield  {author} {\bibinfo {author} {\bibfnamefont {O.}~\bibnamefont
  {Zilberberg}}, \bibinfo {author} {\bibfnamefont {S.}~\bibnamefont {Huang}},
  \bibinfo {author} {\bibfnamefont {J.}~\bibnamefont {Guglielmon}}, \bibinfo
  {author} {\bibfnamefont {M.}~\bibnamefont {Wang}}, \bibinfo {author}
  {\bibfnamefont {K.~P.}\ \bibnamefont {Chen}}, \bibinfo {author}
  {\bibfnamefont {Y.~E.}\ \bibnamefont {Kraus}},\ and\ \bibinfo {author}
  {\bibfnamefont {M.~C.}\ \bibnamefont {Rechtsman}},\ }\href
  {https://doi.org/10.1038/nature25011} {\bibfield  {journal} {\bibinfo
  {journal} {Nature}\ }\textbf {\bibinfo {volume} {553}},\ \bibinfo {pages}
  {59} (\bibinfo {year} {2018})}\BibitemShut {NoStop}%
\bibitem [{\citenamefont {Chen}\ \emph {et~al.}(2021)\citenamefont {Chen},
  \citenamefont {Zhu}, \citenamefont {Tan}, \citenamefont {Wang},\ and\
  \citenamefont {Ma}}]{chen2021acoustic}%
  \BibitemOpen
  \bibfield  {author} {\bibinfo {author} {\bibfnamefont {Z.-G.}\ \bibnamefont
  {Chen}}, \bibinfo {author} {\bibfnamefont {W.}~\bibnamefont {Zhu}}, \bibinfo
  {author} {\bibfnamefont {Y.}~\bibnamefont {Tan}}, \bibinfo {author}
  {\bibfnamefont {L.}~\bibnamefont {Wang}},\ and\ \bibinfo {author}
  {\bibfnamefont {G.}~\bibnamefont {Ma}},\ }\href
  {https://doi.org/10.1103/PhysRevX.11.011016} {\bibfield  {journal} {\bibinfo
  {journal} {Physical Review X}\ }\textbf {\bibinfo {volume} {11}},\ \bibinfo
  {pages} {011016} (\bibinfo {year} {2021})}\BibitemShut {NoStop}%
\bibitem [{\citenamefont {Zhang}\ \emph
  {et~al.}(2023{\natexlab{a}})\citenamefont {Zhang}, \citenamefont {Di},
  \citenamefont {Zheng}, \citenamefont {Sun},\ and\ \citenamefont
  {Zhang}}]{zhang2023hyperbolic}%
  \BibitemOpen
  \bibfield  {author} {\bibinfo {author} {\bibfnamefont {W.}~\bibnamefont
  {Zhang}}, \bibinfo {author} {\bibfnamefont {F.}~\bibnamefont {Di}}, \bibinfo
  {author} {\bibfnamefont {X.}~\bibnamefont {Zheng}}, \bibinfo {author}
  {\bibfnamefont {H.}~\bibnamefont {Sun}},\ and\ \bibinfo {author}
  {\bibfnamefont {X.}~\bibnamefont {Zhang}},\ }\href
  {https://doi.org/10.1038/s41467-023-36767-8} {\bibfield  {journal} {\bibinfo
  {journal} {Nature Communications}\ }\textbf {\bibinfo {volume} {14}},\
  \bibinfo {pages} {1083} (\bibinfo {year} {2023}{\natexlab{a}})}\BibitemShut
  {NoStop}%
\bibitem [{\citenamefont {Boross}\ \emph {et~al.}(2019)\citenamefont {Boross},
  \citenamefont {Asb{\'o}th}, \citenamefont {Sz{\'e}chenyi}, \citenamefont
  {Oroszl{\'a}ny},\ and\ \citenamefont {P{\'a}lyi}}]{boross2019poor}%
  \BibitemOpen
  \bibfield  {author} {\bibinfo {author} {\bibfnamefont {P.}~\bibnamefont
  {Boross}}, \bibinfo {author} {\bibfnamefont {J.~K.}\ \bibnamefont
  {Asb{\'o}th}}, \bibinfo {author} {\bibfnamefont {G.}~\bibnamefont
  {Sz{\'e}chenyi}}, \bibinfo {author} {\bibfnamefont {L.}~\bibnamefont
  {Oroszl{\'a}ny}},\ and\ \bibinfo {author} {\bibfnamefont {A.}~\bibnamefont
  {P{\'a}lyi}},\ }\href {https://doi.org/10.1103/PhysRevB.100.045414}
  {\bibfield  {journal} {\bibinfo  {journal} {Physical Review B}\ }\textbf
  {\bibinfo {volume} {100}},\ \bibinfo {pages} {045414} (\bibinfo {year}
  {2019})}\BibitemShut {NoStop}%
\bibitem [{\citenamefont {Wu}\ \emph {et~al.}(2020)\citenamefont {Wu},
  \citenamefont {Liu}, \citenamefont {Liu}, \citenamefont {Jiang},\ and\
  \citenamefont {Xie}}]{wu2020double}%
  \BibitemOpen
  \bibfield  {author} {\bibinfo {author} {\bibfnamefont {Y.}~\bibnamefont
  {Wu}}, \bibinfo {author} {\bibfnamefont {H.}~\bibnamefont {Liu}}, \bibinfo
  {author} {\bibfnamefont {J.}~\bibnamefont {Liu}}, \bibinfo {author}
  {\bibfnamefont {H.}~\bibnamefont {Jiang}},\ and\ \bibinfo {author}
  {\bibfnamefont {X.}~\bibnamefont {Xie}},\ }\href
  {https://doi.org/10.1093/nsr/nwz189} {\bibfield  {journal} {\bibinfo
  {journal} {National Science Review}\ }\textbf {\bibinfo {volume} {7}},\
  \bibinfo {pages} {572} (\bibinfo {year} {2020})}\BibitemShut {NoStop}%
\bibitem [{\citenamefont {Brown}\ \emph {et~al.}(2022)\citenamefont {Brown},
  \citenamefont {Chang}, \citenamefont {Schwarz}, \citenamefont {Leung},
  \citenamefont {Kozii}, \citenamefont {Avdoshkin}, \citenamefont {Moore},\
  and\ \citenamefont {Stamper-Kurn}}]{brown2022direct}%
  \BibitemOpen
  \bibfield  {author} {\bibinfo {author} {\bibfnamefont {C.~D.}\ \bibnamefont
  {Brown}}, \bibinfo {author} {\bibfnamefont {S.-W.}\ \bibnamefont {Chang}},
  \bibinfo {author} {\bibfnamefont {M.~N.}\ \bibnamefont {Schwarz}}, \bibinfo
  {author} {\bibfnamefont {T.-H.}\ \bibnamefont {Leung}}, \bibinfo {author}
  {\bibfnamefont {V.}~\bibnamefont {Kozii}}, \bibinfo {author} {\bibfnamefont
  {A.}~\bibnamefont {Avdoshkin}}, \bibinfo {author} {\bibfnamefont {J.~E.}\
  \bibnamefont {Moore}},\ and\ \bibinfo {author} {\bibfnamefont
  {D.}~\bibnamefont {Stamper-Kurn}},\ }\href
  {https://doi.org/10.1126/science.abm6442} {\bibfield  {journal} {\bibinfo
  {journal} {Science}\ }\textbf {\bibinfo {volume} {377}},\ \bibinfo {pages}
  {1319} (\bibinfo {year} {2022})}\BibitemShut {NoStop}%
\bibitem [{\citenamefont {Nayak}\ \emph {et~al.}(2008)\citenamefont {Nayak},
  \citenamefont {Simon}, \citenamefont {Stern}, \citenamefont {Freedman},\ and\
  \citenamefont {Sarma}}]{nayak2008non}%
  \BibitemOpen
  \bibfield  {author} {\bibinfo {author} {\bibfnamefont {C.}~\bibnamefont
  {Nayak}}, \bibinfo {author} {\bibfnamefont {S.~H.}\ \bibnamefont {Simon}},
  \bibinfo {author} {\bibfnamefont {A.}~\bibnamefont {Stern}}, \bibinfo
  {author} {\bibfnamefont {M.}~\bibnamefont {Freedman}},\ and\ \bibinfo
  {author} {\bibfnamefont {S.~D.}\ \bibnamefont {Sarma}},\ }\href
  {https://doi.org/10.1103/RevModPhys.80.1083} {\bibfield  {journal} {\bibinfo
  {journal} {Rev. Mod. Phys.}\ }\textbf {\bibinfo {volume} {80}},\ \bibinfo
  {pages} {1083} (\bibinfo {year} {2008})}\BibitemShut {NoStop}%
\bibitem [{\citenamefont {Hou}\ \emph {et~al.}(2007)\citenamefont {Hou},
  \citenamefont {Chamon},\ and\ \citenamefont {Mudry}}]{hou2007electron}%
  \BibitemOpen
  \bibfield  {author} {\bibinfo {author} {\bibfnamefont {C.-Y.}\ \bibnamefont
  {Hou}}, \bibinfo {author} {\bibfnamefont {C.}~\bibnamefont {Chamon}},\ and\
  \bibinfo {author} {\bibfnamefont {C.}~\bibnamefont {Mudry}},\ }\href
  {https://doi.org/10.1103/PhysRevLett.98.186809} {\bibfield  {journal}
  {\bibinfo  {journal} {Physical review letters}\ }\textbf {\bibinfo {volume}
  {98}},\ \bibinfo {pages} {186809} (\bibinfo {year} {2007})}\BibitemShut
  {NoStop}%
\bibitem [{\citenamefont {Gao}\ \emph {et~al.}(2020)\citenamefont {Gao},
  \citenamefont {Yang}, \citenamefont {Lin}, \citenamefont {Zhang},
  \citenamefont {Li}, \citenamefont {Bo}, \citenamefont {Wang},\ and\
  \citenamefont {Lu}}]{gao2020dirac}%
  \BibitemOpen
  \bibfield  {author} {\bibinfo {author} {\bibfnamefont {X.}~\bibnamefont
  {Gao}}, \bibinfo {author} {\bibfnamefont {L.}~\bibnamefont {Yang}}, \bibinfo
  {author} {\bibfnamefont {H.}~\bibnamefont {Lin}}, \bibinfo {author}
  {\bibfnamefont {L.}~\bibnamefont {Zhang}}, \bibinfo {author} {\bibfnamefont
  {J.}~\bibnamefont {Li}}, \bibinfo {author} {\bibfnamefont {F.}~\bibnamefont
  {Bo}}, \bibinfo {author} {\bibfnamefont {Z.}~\bibnamefont {Wang}},\ and\
  \bibinfo {author} {\bibfnamefont {L.}~\bibnamefont {Lu}},\ }\href
  {https://doi.org/10.1038/s41565-020-0773-7} {\bibfield  {journal} {\bibinfo
  {journal} {Nature Nanotechnology}\ }\textbf {\bibinfo {volume} {15}},\
  \bibinfo {pages} {1012} (\bibinfo {year} {2020})}\BibitemShut {NoStop}%
\bibitem [{\citenamefont {Chen}\ \emph
  {et~al.}(2019{\natexlab{b}})\citenamefont {Chen}, \citenamefont {Lera},
  \citenamefont {Chaunsali}, \citenamefont {Torrent}, \citenamefont {Alvarez},
  \citenamefont {Yang}, \citenamefont {San-Jose},\ and\ \citenamefont
  {Christensen}}]{chen2019mechanical}%
  \BibitemOpen
  \bibfield  {author} {\bibinfo {author} {\bibfnamefont {C.-W.}\ \bibnamefont
  {Chen}}, \bibinfo {author} {\bibfnamefont {N.}~\bibnamefont {Lera}}, \bibinfo
  {author} {\bibfnamefont {R.}~\bibnamefont {Chaunsali}}, \bibinfo {author}
  {\bibfnamefont {D.}~\bibnamefont {Torrent}}, \bibinfo {author} {\bibfnamefont
  {J.~V.}\ \bibnamefont {Alvarez}}, \bibinfo {author} {\bibfnamefont
  {J.}~\bibnamefont {Yang}}, \bibinfo {author} {\bibfnamefont {P.}~\bibnamefont
  {San-Jose}},\ and\ \bibinfo {author} {\bibfnamefont {J.}~\bibnamefont
  {Christensen}},\ }\href {https://doi.org/10.1002/adma.201904386} {\bibfield
  {journal} {\bibinfo  {journal} {Advanced Materials}\ }\textbf {\bibinfo
  {volume} {31}},\ \bibinfo {pages} {1904386} (\bibinfo {year}
  {2019}{\natexlab{b}})}\BibitemShut {NoStop}%
\bibitem [{\citenamefont {Iadecola}\ \emph {et~al.}(2016)\citenamefont
  {Iadecola}, \citenamefont {Schuster},\ and\ \citenamefont
  {Chamon}}]{iadecola2016non}%
  \BibitemOpen
  \bibfield  {author} {\bibinfo {author} {\bibfnamefont {T.}~\bibnamefont
  {Iadecola}}, \bibinfo {author} {\bibfnamefont {T.}~\bibnamefont {Schuster}},\
  and\ \bibinfo {author} {\bibfnamefont {C.}~\bibnamefont {Chamon}},\ }\href
  {https://doi.org/10.1103/PhysRevLett.117.073901} {\bibfield  {journal}
  {\bibinfo  {journal} {Phys. Rev. Lett.}\ }\textbf {\bibinfo {volume} {117}},\
  \bibinfo {pages} {073901} (\bibinfo {year} {2016})}\BibitemShut {NoStop}%
\bibitem [{\citenamefont {Prodan}\ \emph {et~al.}(2017)\citenamefont {Prodan},
  \citenamefont {Dobiszewski}, \citenamefont {Kanwal}, \citenamefont
  {Palmieri},\ and\ \citenamefont {Prodan}}]{prodan2017dynamical}%
  \BibitemOpen
  \bibfield  {author} {\bibinfo {author} {\bibfnamefont {E.}~\bibnamefont
  {Prodan}}, \bibinfo {author} {\bibfnamefont {K.}~\bibnamefont {Dobiszewski}},
  \bibinfo {author} {\bibfnamefont {A.}~\bibnamefont {Kanwal}}, \bibinfo
  {author} {\bibfnamefont {J.}~\bibnamefont {Palmieri}},\ and\ \bibinfo
  {author} {\bibfnamefont {C.}~\bibnamefont {Prodan}},\ }\href
  {https://doi.org/10.1038/ncomms14587} {\bibfield  {journal} {\bibinfo
  {journal} {Nature communications}\ }\textbf {\bibinfo {volume} {8}},\
  \bibinfo {pages} {14587} (\bibinfo {year} {2017})}\BibitemShut {NoStop}%
\bibitem [{\citenamefont {Barlas}\ and\ \citenamefont
  {Prodan}(2020)}]{barlas2020topological}%
  \BibitemOpen
  \bibfield  {author} {\bibinfo {author} {\bibfnamefont {Y.}~\bibnamefont
  {Barlas}}\ and\ \bibinfo {author} {\bibfnamefont {E.}~\bibnamefont
  {Prodan}},\ }\href {https://doi.org/10.1103/PhysRevLett.124.146801}
  {\bibfield  {journal} {\bibinfo  {journal} {Physical Review Letters}\
  }\textbf {\bibinfo {volume} {124}},\ \bibinfo {pages} {146801} (\bibinfo
  {year} {2020})}\BibitemShut {NoStop}%
\bibitem [{\citenamefont {Tang}\ \emph {et~al.}(2020)\citenamefont {Tang},
  \citenamefont {Jiang}, \citenamefont {Ding}, \citenamefont {Xiao},
  \citenamefont {Zhang}, \citenamefont {Chan},\ and\ \citenamefont
  {Ma}}]{tang2020exceptional}%
  \BibitemOpen
  \bibfield  {author} {\bibinfo {author} {\bibfnamefont {W.}~\bibnamefont
  {Tang}}, \bibinfo {author} {\bibfnamefont {X.}~\bibnamefont {Jiang}},
  \bibinfo {author} {\bibfnamefont {K.}~\bibnamefont {Ding}}, \bibinfo {author}
  {\bibfnamefont {Y.-X.}\ \bibnamefont {Xiao}}, \bibinfo {author}
  {\bibfnamefont {Z.-Q.}\ \bibnamefont {Zhang}}, \bibinfo {author}
  {\bibfnamefont {C.~T.}\ \bibnamefont {Chan}},\ and\ \bibinfo {author}
  {\bibfnamefont {G.}~\bibnamefont {Ma}},\ }\href
  {https://doi.org/10.1126/science.abd8872} {\bibfield  {journal} {\bibinfo
  {journal} {Science}\ }\textbf {\bibinfo {volume} {370}},\ \bibinfo {pages}
  {1077} (\bibinfo {year} {2020})}\BibitemShut {NoStop}%
\bibitem [{\citenamefont {Tang}\ \emph {et~al.}(2022)\citenamefont {Tang},
  \citenamefont {Ding},\ and\ \citenamefont {Ma}}]{tang2022experimental}%
  \BibitemOpen
  \bibfield  {author} {\bibinfo {author} {\bibfnamefont {W.}~\bibnamefont
  {Tang}}, \bibinfo {author} {\bibfnamefont {K.}~\bibnamefont {Ding}},\ and\
  \bibinfo {author} {\bibfnamefont {G.}~\bibnamefont {Ma}},\ }\href
  {https://doi.org/10.1093/nsr/nwac010} {\bibfield  {journal} {\bibinfo
  {journal} {National Science Review}\ }\textbf {\bibinfo {volume} {9}},\
  \bibinfo {pages} {nwac010} (\bibinfo {year} {2022})}\BibitemShut {NoStop}%
\bibitem [{\citenamefont {Hu}\ and\ \citenamefont {Zhao}(2021)}]{hu2021knots}%
  \BibitemOpen
  \bibfield  {author} {\bibinfo {author} {\bibfnamefont {H.}~\bibnamefont
  {Hu}}\ and\ \bibinfo {author} {\bibfnamefont {E.}~\bibnamefont {Zhao}},\
  }\href {https://doi.org/10.1103/PhysRevLett.126.010401} {\bibfield  {journal}
  {\bibinfo  {journal} {Physical Review Letters}\ }\textbf {\bibinfo {volume}
  {126}},\ \bibinfo {pages} {010401} (\bibinfo {year} {2021})}\BibitemShut
  {NoStop}%
\bibitem [{\citenamefont {Wang}\ \emph {et~al.}(2021)\citenamefont {Wang},
  \citenamefont {Dutt}, \citenamefont {Wojcik},\ and\ \citenamefont
  {Fan}}]{RN45}%
  \BibitemOpen
  \bibfield  {author} {\bibinfo {author} {\bibfnamefont {K.}~\bibnamefont
  {Wang}}, \bibinfo {author} {\bibfnamefont {A.}~\bibnamefont {Dutt}}, \bibinfo
  {author} {\bibfnamefont {C.~C.}\ \bibnamefont {Wojcik}},\ and\ \bibinfo
  {author} {\bibfnamefont {S.}~\bibnamefont {Fan}},\ }\href
  {https://doi.org/10.1038/s41586-021-03848-x} {\bibfield  {journal} {\bibinfo
  {journal} {Nature}\ }\textbf {\bibinfo {volume} {598}},\ \bibinfo {pages}
  {59} (\bibinfo {year} {2021})}\BibitemShut {NoStop}%
\bibitem [{\citenamefont {Patil}\ \emph {et~al.}(2022)\citenamefont {Patil},
  \citenamefont {Höller}, \citenamefont {Henry}, \citenamefont {Guria},
  \citenamefont {Zhang}, \citenamefont {Jiang}, \citenamefont {Kralj},
  \citenamefont {Read},\ and\ \citenamefont {Harris}}]{RN42}%
  \BibitemOpen
  \bibfield  {author} {\bibinfo {author} {\bibfnamefont {Y.~S.~S.}\
  \bibnamefont {Patil}}, \bibinfo {author} {\bibfnamefont {J.}~\bibnamefont
  {Höller}}, \bibinfo {author} {\bibfnamefont {P.~A.}\ \bibnamefont {Henry}},
  \bibinfo {author} {\bibfnamefont {C.}~\bibnamefont {Guria}}, \bibinfo
  {author} {\bibfnamefont {Y.}~\bibnamefont {Zhang}}, \bibinfo {author}
  {\bibfnamefont {L.}~\bibnamefont {Jiang}}, \bibinfo {author} {\bibfnamefont
  {N.}~\bibnamefont {Kralj}}, \bibinfo {author} {\bibfnamefont
  {N.}~\bibnamefont {Read}},\ and\ \bibinfo {author} {\bibfnamefont {J.~G.~E.}\
  \bibnamefont {Harris}},\ }\href {https://doi.org/10.1038/s41586-022-04796-w}
  {\bibfield  {journal} {\bibinfo  {journal} {Nature}\ }\textbf {\bibinfo
  {volume} {607}},\ \bibinfo {pages} {271} (\bibinfo {year}
  {2022})}\BibitemShut {NoStop}%
\bibitem [{\citenamefont {Parto}\ \emph {et~al.}(2023)\citenamefont {Parto},
  \citenamefont {Leefmans}, \citenamefont {Williams}, \citenamefont {Nori},\
  and\ \citenamefont {Marandi}}]{parto2023non}%
  \BibitemOpen
  \bibfield  {author} {\bibinfo {author} {\bibfnamefont {M.}~\bibnamefont
  {Parto}}, \bibinfo {author} {\bibfnamefont {C.}~\bibnamefont {Leefmans}},
  \bibinfo {author} {\bibfnamefont {J.}~\bibnamefont {Williams}}, \bibinfo
  {author} {\bibfnamefont {F.}~\bibnamefont {Nori}},\ and\ \bibinfo {author}
  {\bibfnamefont {A.}~\bibnamefont {Marandi}},\ }\href
  {https://doi.org/10.1038/s41467-023-37065-z} {\bibfield  {journal} {\bibinfo
  {journal} {Nature Communications}\ }\textbf {\bibinfo {volume} {14}},\
  \bibinfo {pages} {1440} (\bibinfo {year} {2023})}\BibitemShut {NoStop}%
\bibitem [{\citenamefont {Kawabata}\ \emph {et~al.}(2019)\citenamefont
  {Kawabata}, \citenamefont {Shiozaki}, \citenamefont {Ueda},\ and\
  \citenamefont {Sato}}]{kawabata2019symmetry}%
  \BibitemOpen
  \bibfield  {author} {\bibinfo {author} {\bibfnamefont {K.}~\bibnamefont
  {Kawabata}}, \bibinfo {author} {\bibfnamefont {K.}~\bibnamefont {Shiozaki}},
  \bibinfo {author} {\bibfnamefont {M.}~\bibnamefont {Ueda}},\ and\ \bibinfo
  {author} {\bibfnamefont {M.}~\bibnamefont {Sato}},\ }\href
  {https://doi.org/10.1103/PhysRevX.9.041015} {\bibfield  {journal} {\bibinfo
  {journal} {Physical Review X}\ }\textbf {\bibinfo {volume} {9}},\ \bibinfo
  {pages} {041015} (\bibinfo {year} {2019})}\BibitemShut {NoStop}%
\bibitem [{\citenamefont {Ashida}\ \emph {et~al.}(2020)\citenamefont {Ashida},
  \citenamefont {Gong},\ and\ \citenamefont {Ueda}}]{ashida2020non}%
  \BibitemOpen
  \bibfield  {author} {\bibinfo {author} {\bibfnamefont {Y.}~\bibnamefont
  {Ashida}}, \bibinfo {author} {\bibfnamefont {Z.}~\bibnamefont {Gong}},\ and\
  \bibinfo {author} {\bibfnamefont {M.}~\bibnamefont {Ueda}},\ }\href
  {https://doi.org/10.1080/00018732.2021.1876991} {\bibfield  {journal}
  {\bibinfo  {journal} {Advances in Physics}\ }\textbf {\bibinfo {volume}
  {69}},\ \bibinfo {pages} {249} (\bibinfo {year} {2020})}\BibitemShut
  {NoStop}%
\bibitem [{\citenamefont {Bergholtz}\ \emph {et~al.}(2021)\citenamefont
  {Bergholtz}, \citenamefont {Budich},\ and\ \citenamefont
  {Kunst}}]{bergholtz2021exceptional}%
  \BibitemOpen
  \bibfield  {author} {\bibinfo {author} {\bibfnamefont {E.~J.}\ \bibnamefont
  {Bergholtz}}, \bibinfo {author} {\bibfnamefont {J.~C.}\ \bibnamefont
  {Budich}},\ and\ \bibinfo {author} {\bibfnamefont {F.~K.}\ \bibnamefont
  {Kunst}},\ }\href {https://doi.org/10.1103/RevModPhys.93.015005} {\bibfield
  {journal} {\bibinfo  {journal} {Reviews of Modern Physics}\ }\textbf
  {\bibinfo {volume} {93}},\ \bibinfo {pages} {015005} (\bibinfo {year}
  {2021})}\BibitemShut {NoStop}%
\bibitem [{\citenamefont {Wojcik}\ \emph {et~al.}(2020)\citenamefont {Wojcik},
  \citenamefont {Sun}, \citenamefont {Bzdu{\v{s}}ek},\ and\ \citenamefont
  {Fan}}]{wojcik2020homotopy}%
  \BibitemOpen
  \bibfield  {author} {\bibinfo {author} {\bibfnamefont {C.~C.}\ \bibnamefont
  {Wojcik}}, \bibinfo {author} {\bibfnamefont {X.-Q.}\ \bibnamefont {Sun}},
  \bibinfo {author} {\bibfnamefont {T.}~\bibnamefont {Bzdu{\v{s}}ek}},\ and\
  \bibinfo {author} {\bibfnamefont {S.}~\bibnamefont {Fan}},\ }\href
  {https://doi.org/10.1103/PhysRevB.101.205417} {\bibfield  {journal} {\bibinfo
   {journal} {Physical Review B}\ }\textbf {\bibinfo {volume} {101}},\ \bibinfo
  {pages} {205417} (\bibinfo {year} {2020})}\BibitemShut {NoStop}%
\bibitem [{\citenamefont {Li}\ and\ \citenamefont
  {Mong}(2021)}]{li2021homotopical}%
  \BibitemOpen
  \bibfield  {author} {\bibinfo {author} {\bibfnamefont {Z.}~\bibnamefont
  {Li}}\ and\ \bibinfo {author} {\bibfnamefont {R.~S.}\ \bibnamefont {Mong}},\
  }\href {https://doi.org/10.1103/PhysRevB.103.155129} {\bibfield  {journal}
  {\bibinfo  {journal} {Physical Review B}\ }\textbf {\bibinfo {volume}
  {103}},\ \bibinfo {pages} {155129} (\bibinfo {year} {2021})}\BibitemShut
  {NoStop}%
\bibitem [{\citenamefont {Sun}\ \emph {et~al.}(2020)\citenamefont {Sun},
  \citenamefont {Wojcik}, \citenamefont {Fan},\ and\ \citenamefont
  {Bzdušek}}]{RN46}%
  \BibitemOpen
  \bibfield  {author} {\bibinfo {author} {\bibfnamefont {X.-Q.}\ \bibnamefont
  {Sun}}, \bibinfo {author} {\bibfnamefont {C.~C.}\ \bibnamefont {Wojcik}},
  \bibinfo {author} {\bibfnamefont {S.}~\bibnamefont {Fan}},\ and\ \bibinfo
  {author} {\bibfnamefont {T.}~\bibnamefont {Bzdušek}},\ }\href
  {https://doi.org/10.1103/PhysRevResearch.2.023226} {\bibfield  {journal}
  {\bibinfo  {journal} {Physical Review Research}\ }\textbf {\bibinfo {volume}
  {2}},\ \bibinfo {pages} {023226} (\bibinfo {year} {2020})}\BibitemShut
  {NoStop}%
\bibitem [{\citenamefont {Zhang}\ \emph
  {et~al.}(2023{\natexlab{b}})\citenamefont {Zhang}, \citenamefont {Li},
  \citenamefont {Sun}, \citenamefont {Liu}, \citenamefont {Zhao}, \citenamefont
  {Feng}, \citenamefont {Fan},\ and\ \citenamefont
  {Qiu}}]{zhang2023observation}%
  \BibitemOpen
  \bibfield  {author} {\bibinfo {author} {\bibfnamefont {Q.}~\bibnamefont
  {Zhang}}, \bibinfo {author} {\bibfnamefont {Y.}~\bibnamefont {Li}}, \bibinfo
  {author} {\bibfnamefont {H.}~\bibnamefont {Sun}}, \bibinfo {author}
  {\bibfnamefont {X.}~\bibnamefont {Liu}}, \bibinfo {author} {\bibfnamefont
  {L.}~\bibnamefont {Zhao}}, \bibinfo {author} {\bibfnamefont {X.}~\bibnamefont
  {Feng}}, \bibinfo {author} {\bibfnamefont {X.}~\bibnamefont {Fan}},\ and\
  \bibinfo {author} {\bibfnamefont {C.}~\bibnamefont {Qiu}},\ }\href
  {https://doi.org/10.1103/PhysRevLett.130.017201} {\bibfield  {journal}
  {\bibinfo  {journal} {Physical Review Letters}\ }\textbf {\bibinfo {volume}
  {130}},\ \bibinfo {pages} {017201} (\bibinfo {year}
  {2023}{\natexlab{b}})}\BibitemShut {NoStop}%
\bibitem [{\citenamefont {Li}\ \emph {et~al.}(2023)\citenamefont {Li},
  \citenamefont {Ding},\ and\ \citenamefont {Ma}}]{li2023eigenvalue}%
  \BibitemOpen
  \bibfield  {author} {\bibinfo {author} {\bibfnamefont {Z.}~\bibnamefont
  {Li}}, \bibinfo {author} {\bibfnamefont {K.}~\bibnamefont {Ding}},\ and\
  \bibinfo {author} {\bibfnamefont {G.}~\bibnamefont {Ma}},\ }\href
  {https://doi.org/10.1103/PhysRevResearch.5.023038} {\bibfield  {journal}
  {\bibinfo  {journal} {Physical Review Research}\ }\textbf {\bibinfo {volume}
  {5}},\ \bibinfo {pages} {023038} (\bibinfo {year} {2023})}\BibitemShut
  {NoStop}%
\bibitem [{\citenamefont {Dembowski}\ \emph {et~al.}(2001)\citenamefont
  {Dembowski}, \citenamefont {Gr{\"a}f}, \citenamefont {Harney}, \citenamefont
  {Heine}, \citenamefont {Heiss}, \citenamefont {Rehfeld},\ and\ \citenamefont
  {Richter}}]{dembowski2001experimental}%
  \BibitemOpen
  \bibfield  {author} {\bibinfo {author} {\bibfnamefont {C.}~\bibnamefont
  {Dembowski}}, \bibinfo {author} {\bibfnamefont {H.-D.}\ \bibnamefont
  {Gr{\"a}f}}, \bibinfo {author} {\bibfnamefont {H.}~\bibnamefont {Harney}},
  \bibinfo {author} {\bibfnamefont {A.}~\bibnamefont {Heine}}, \bibinfo
  {author} {\bibfnamefont {W.}~\bibnamefont {Heiss}}, \bibinfo {author}
  {\bibfnamefont {H.}~\bibnamefont {Rehfeld}},\ and\ \bibinfo {author}
  {\bibfnamefont {A.}~\bibnamefont {Richter}},\ }\href
  {https://doi.org/10.1103/PhysRevLett.86.787} {\bibfield  {journal} {\bibinfo
  {journal} {Physical review letters}\ }\textbf {\bibinfo {volume} {86}},\
  \bibinfo {pages} {787} (\bibinfo {year} {2001})}\BibitemShut {NoStop}%
\bibitem [{\citenamefont {Dembowski}\ \emph {et~al.}(2004)\citenamefont
  {Dembowski}, \citenamefont {Dietz}, \citenamefont {Gr{\"a}f}, \citenamefont
  {Harney}, \citenamefont {Heine}, \citenamefont {Heiss},\ and\ \citenamefont
  {Richter}}]{dembowski2004encircling}%
  \BibitemOpen
  \bibfield  {author} {\bibinfo {author} {\bibfnamefont {C.}~\bibnamefont
  {Dembowski}}, \bibinfo {author} {\bibfnamefont {B.}~\bibnamefont {Dietz}},
  \bibinfo {author} {\bibfnamefont {H.-D.}\ \bibnamefont {Gr{\"a}f}}, \bibinfo
  {author} {\bibfnamefont {H.}~\bibnamefont {Harney}}, \bibinfo {author}
  {\bibfnamefont {A.}~\bibnamefont {Heine}}, \bibinfo {author} {\bibfnamefont
  {W.}~\bibnamefont {Heiss}},\ and\ \bibinfo {author} {\bibfnamefont
  {A.}~\bibnamefont {Richter}},\ }\href
  {https://doi.org/10.1103/PhysRevE.69.056216} {\bibfield  {journal} {\bibinfo
  {journal} {Physical Review E}\ }\textbf {\bibinfo {volume} {69}},\ \bibinfo
  {pages} {056216} (\bibinfo {year} {2004})}\BibitemShut {NoStop}%
\bibitem [{\citenamefont {Gao}\ \emph {et~al.}(2015)\citenamefont {Gao},
  \citenamefont {Estrecho}, \citenamefont {Bliokh}, \citenamefont {Liew},
  \citenamefont {Fraser}, \citenamefont {Brodbeck}, \citenamefont {Kamp},
  \citenamefont {Schneider}, \citenamefont {H{\"o}fling}, \citenamefont
  {Yamamoto} \emph {et~al.}}]{gao2015observation}%
  \BibitemOpen
  \bibfield  {author} {\bibinfo {author} {\bibfnamefont {T.}~\bibnamefont
  {Gao}}, \bibinfo {author} {\bibfnamefont {E.}~\bibnamefont {Estrecho}},
  \bibinfo {author} {\bibfnamefont {K.}~\bibnamefont {Bliokh}}, \bibinfo
  {author} {\bibfnamefont {T.}~\bibnamefont {Liew}}, \bibinfo {author}
  {\bibfnamefont {M.}~\bibnamefont {Fraser}}, \bibinfo {author} {\bibfnamefont
  {S.}~\bibnamefont {Brodbeck}}, \bibinfo {author} {\bibfnamefont
  {M.}~\bibnamefont {Kamp}}, \bibinfo {author} {\bibfnamefont {C.}~\bibnamefont
  {Schneider}}, \bibinfo {author} {\bibfnamefont {S.}~\bibnamefont
  {H{\"o}fling}}, \bibinfo {author} {\bibfnamefont {Y.}~\bibnamefont
  {Yamamoto}}, \emph {et~al.},\ }\href {https://doi.org/10.1038/nature15522}
  {\bibfield  {journal} {\bibinfo  {journal} {Nature}\ }\textbf {\bibinfo
  {volume} {526}},\ \bibinfo {pages} {554} (\bibinfo {year}
  {2015})}\BibitemShut {NoStop}%
\bibitem [{\citenamefont {Ding}\ \emph {et~al.}(2016)\citenamefont {Ding},
  \citenamefont {Ma}, \citenamefont {Xiao}, \citenamefont {Zhang},\ and\
  \citenamefont {Chan}}]{ding2016emergence}%
  \BibitemOpen
  \bibfield  {author} {\bibinfo {author} {\bibfnamefont {K.}~\bibnamefont
  {Ding}}, \bibinfo {author} {\bibfnamefont {G.}~\bibnamefont {Ma}}, \bibinfo
  {author} {\bibfnamefont {M.}~\bibnamefont {Xiao}}, \bibinfo {author}
  {\bibfnamefont {Z.}~\bibnamefont {Zhang}},\ and\ \bibinfo {author}
  {\bibfnamefont {C.~T.}\ \bibnamefont {Chan}},\ }\href
  {https://doi.org/10.1103/PhysRevX.6.021007} {\bibfield  {journal} {\bibinfo
  {journal} {Physical Review X}\ }\textbf {\bibinfo {volume} {6}},\ \bibinfo
  {pages} {021007} (\bibinfo {year} {2016})}\BibitemShut {NoStop}%
\bibitem [{\citenamefont {Schindler}\ and\ \citenamefont
  {Bender}(2017)}]{schindler2017winding}%
  \BibitemOpen
  \bibfield  {author} {\bibinfo {author} {\bibfnamefont {S.~T.}\ \bibnamefont
  {Schindler}}\ and\ \bibinfo {author} {\bibfnamefont {C.~M.}\ \bibnamefont
  {Bender}},\ }\href {https://doi.org/10.1088/1751-8121/aa9faf} {\bibfield
  {journal} {\bibinfo  {journal} {Journal of Physics A: Mathematical and
  Theoretical}\ }\textbf {\bibinfo {volume} {51}},\ \bibinfo {pages} {055201}
  (\bibinfo {year} {2017})}\BibitemShut {NoStop}%
\bibitem [{\citenamefont {Tang}\ \emph {et~al.}(2021)\citenamefont {Tang},
  \citenamefont {Ding},\ and\ \citenamefont {Ma}}]{tang2021direct}%
  \BibitemOpen
  \bibfield  {author} {\bibinfo {author} {\bibfnamefont {W.}~\bibnamefont
  {Tang}}, \bibinfo {author} {\bibfnamefont {K.}~\bibnamefont {Ding}},\ and\
  \bibinfo {author} {\bibfnamefont {G.}~\bibnamefont {Ma}},\ }\href
  {https://doi.org/10.1103/PhysRevLett.127.034301} {\bibfield  {journal}
  {\bibinfo  {journal} {Physical Review Letters}\ }\textbf {\bibinfo {volume}
  {127}},\ \bibinfo {pages} {034301} (\bibinfo {year} {2021})}\BibitemShut
  {NoStop}%
\bibitem [{\citenamefont {Zhong}\ \emph {et~al.}(2018)\citenamefont {Zhong},
  \citenamefont {Khajavikhan}, \citenamefont {Christodoulides},\ and\
  \citenamefont {El-Ganainy}}]{zhong2018winding}%
  \BibitemOpen
  \bibfield  {author} {\bibinfo {author} {\bibfnamefont {Q.}~\bibnamefont
  {Zhong}}, \bibinfo {author} {\bibfnamefont {M.}~\bibnamefont {Khajavikhan}},
  \bibinfo {author} {\bibfnamefont {D.~N.}\ \bibnamefont {Christodoulides}},\
  and\ \bibinfo {author} {\bibfnamefont {R.}~\bibnamefont {El-Ganainy}},\
  }\href {https://doi.org/10.1038/s41467-018-07105-0} {\bibfield  {journal}
  {\bibinfo  {journal} {Nature Communications}\ }\textbf {\bibinfo {volume}
  {9}},\ \bibinfo {pages} {4808} (\bibinfo {year} {2018})}\BibitemShut
  {NoStop}%
\bibitem [{\citenamefont {Berry}\ and\ \citenamefont
  {Uzdin}(2011)}]{berry2011slow}%
  \BibitemOpen
  \bibfield  {author} {\bibinfo {author} {\bibfnamefont {M.}~\bibnamefont
  {Berry}}\ and\ \bibinfo {author} {\bibfnamefont {R.}~\bibnamefont {Uzdin}},\
  }\href {https://doi.org/10.1088/1751-8113/44/43/435303} {\bibfield  {journal}
  {\bibinfo  {journal} {Journal of Physics A: Mathematical and Theoretical}\
  }\textbf {\bibinfo {volume} {44}},\ \bibinfo {pages} {435303} (\bibinfo
  {year} {2011})}\BibitemShut {NoStop}%
\bibitem [{\citenamefont {Doppler}\ \emph {et~al.}(2016)\citenamefont
  {Doppler}, \citenamefont {Mailybaev}, \citenamefont {B{\"o}hm}, \citenamefont
  {Kuhl}, \citenamefont {Girschik}, \citenamefont {Libisch}, \citenamefont
  {Milburn}, \citenamefont {Rabl}, \citenamefont {Moiseyev},\ and\
  \citenamefont {Rotter}}]{doppler2016dynamically}%
  \BibitemOpen
  \bibfield  {author} {\bibinfo {author} {\bibfnamefont {J.}~\bibnamefont
  {Doppler}}, \bibinfo {author} {\bibfnamefont {A.~A.}\ \bibnamefont
  {Mailybaev}}, \bibinfo {author} {\bibfnamefont {J.}~\bibnamefont {B{\"o}hm}},
  \bibinfo {author} {\bibfnamefont {U.}~\bibnamefont {Kuhl}}, \bibinfo {author}
  {\bibfnamefont {A.}~\bibnamefont {Girschik}}, \bibinfo {author}
  {\bibfnamefont {F.}~\bibnamefont {Libisch}}, \bibinfo {author} {\bibfnamefont
  {T.~J.}\ \bibnamefont {Milburn}}, \bibinfo {author} {\bibfnamefont
  {P.}~\bibnamefont {Rabl}}, \bibinfo {author} {\bibfnamefont {N.}~\bibnamefont
  {Moiseyev}},\ and\ \bibinfo {author} {\bibfnamefont {S.}~\bibnamefont
  {Rotter}},\ }\href {https://doi.org/10.1038/nature18605} {\bibfield
  {journal} {\bibinfo  {journal} {Nature}\ }\textbf {\bibinfo {volume} {537}},\
  \bibinfo {pages} {76} (\bibinfo {year} {2016})}\BibitemShut {NoStop}%
\bibitem [{\citenamefont {Xu}\ \emph {et~al.}(2016)\citenamefont {Xu},
  \citenamefont {Mason}, \citenamefont {Jiang},\ and\ \citenamefont
  {Harris}}]{xu2016topological}%
  \BibitemOpen
  \bibfield  {author} {\bibinfo {author} {\bibfnamefont {H.}~\bibnamefont
  {Xu}}, \bibinfo {author} {\bibfnamefont {D.}~\bibnamefont {Mason}}, \bibinfo
  {author} {\bibfnamefont {L.}~\bibnamefont {Jiang}},\ and\ \bibinfo {author}
  {\bibfnamefont {J.}~\bibnamefont {Harris}},\ }\href
  {https://doi.org/10.1038/nature18604} {\bibfield  {journal} {\bibinfo
  {journal} {Nature}\ }\textbf {\bibinfo {volume} {537}},\ \bibinfo {pages}
  {80} (\bibinfo {year} {2016})}\BibitemShut {NoStop}%
\bibitem [{\citenamefont {Slager}\ \emph {et~al.}(2022)\citenamefont {Slager},
  \citenamefont {Bouhon},\ and\ \citenamefont {Ünal}}]{slager2022floquet}%
  \BibitemOpen
  \bibfield  {author} {\bibinfo {author} {\bibfnamefont {R.-J.}\ \bibnamefont
  {Slager}}, \bibinfo {author} {\bibfnamefont {A.}~\bibnamefont {Bouhon}},\
  and\ \bibinfo {author} {\bibfnamefont {F.~N.}\ \bibnamefont {Ünal}},\
  }\bibfield  {journal} {\bibinfo  {journal} {arXiv preprint arXiv:2208.12824}\
  }\href {https://doi.org/10.48550/arXiv.2208.12824}
  {10.48550/arXiv.2208.12824} (\bibinfo {year} {2022})\BibitemShut {NoStop}%
\bibitem [{\citenamefont {Jin}\ and\ \citenamefont {Song}(2019)}]{jin2019bulk}%
  \BibitemOpen
  \bibfield  {author} {\bibinfo {author} {\bibfnamefont {L.}~\bibnamefont
  {Jin}}\ and\ \bibinfo {author} {\bibfnamefont {Z.}~\bibnamefont {Song}},\
  }\href {https://doi.org/10.1103/PhysRevB.99.081103} {\bibfield  {journal}
  {\bibinfo  {journal} {Physical Review B}\ }\textbf {\bibinfo {volume} {99}},\
  \bibinfo {pages} {081103} (\bibinfo {year} {2019})}\BibitemShut {NoStop}%
\bibitem [{\citenamefont {Lee}\ and\ \citenamefont
  {Thomale}(2019)}]{lee2019anatomy}%
  \BibitemOpen
  \bibfield  {author} {\bibinfo {author} {\bibfnamefont {C.~H.}\ \bibnamefont
  {Lee}}\ and\ \bibinfo {author} {\bibfnamefont {R.}~\bibnamefont {Thomale}},\
  }\href {https://doi.org/10.1103/PhysRevB.99.201103} {\bibfield  {journal}
  {\bibinfo  {journal} {Physical Review B}\ }\textbf {\bibinfo {volume} {99}},\
  \bibinfo {pages} {201103} (\bibinfo {year} {2019})}\BibitemShut {NoStop}%
\bibitem [{\citenamefont {Pang}\ \emph {et~al.}(2023)\citenamefont {Pang},
  \citenamefont {Hu},\ and\ \citenamefont {Yang}}]{pang2023synthetic}%
  \BibitemOpen
  \bibfield  {author} {\bibinfo {author} {\bibfnamefont {Z.}~\bibnamefont
  {Pang}}, \bibinfo {author} {\bibfnamefont {J.}~\bibnamefont {Hu}},\ and\
  \bibinfo {author} {\bibfnamefont {Y.}~\bibnamefont {Yang}},\ }\bibfield
  {journal} {\bibinfo  {journal} {arXiv preprint arXiv:2304.01876}\ }\href
  {https://doi.org/10.48550/arXiv.2304.01876} {10.48550/arXiv.2304.01876}
  (\bibinfo {year} {2023})\BibitemShut {NoStop}%
\bibitem [{\citenamefont {Burrello}\ and\ \citenamefont
  {Trombettoni}(2010)}]{burrello2010non}%
  \BibitemOpen
  \bibfield  {author} {\bibinfo {author} {\bibfnamefont {M.}~\bibnamefont
  {Burrello}}\ and\ \bibinfo {author} {\bibfnamefont {A.}~\bibnamefont
  {Trombettoni}},\ }\href {https://doi.org/10.1103/PhysRevLett.105.125304}
  {\bibfield  {journal} {\bibinfo  {journal} {Phys. Rev. Lett.}\ }\textbf
  {\bibinfo {volume} {105}},\ \bibinfo {pages} {125304} (\bibinfo {year}
  {2010})}\BibitemShut {NoStop}%
\bibitem [{\citenamefont {Keilmann}\ \emph {et~al.}(2011)\citenamefont
  {Keilmann}, \citenamefont {Lanzmich}, \citenamefont {McCulloch},\ and\
  \citenamefont {Roncaglia}}]{keilmann2011statistically}%
  \BibitemOpen
  \bibfield  {author} {\bibinfo {author} {\bibfnamefont {T.}~\bibnamefont
  {Keilmann}}, \bibinfo {author} {\bibfnamefont {S.}~\bibnamefont {Lanzmich}},
  \bibinfo {author} {\bibfnamefont {I.}~\bibnamefont {McCulloch}},\ and\
  \bibinfo {author} {\bibfnamefont {M.}~\bibnamefont {Roncaglia}},\ }\href
  {https://doi.org/10.1038/ncomms1353} {\bibfield  {journal} {\bibinfo
  {journal} {Nat. Commun.}\ }\textbf {\bibinfo {volume} {2}},\ \bibinfo {pages}
  {361} (\bibinfo {year} {2011})}\BibitemShut {NoStop}%
\bibitem [{\citenamefont {Kapit}\ \emph {et~al.}(2012)\citenamefont {Kapit},
  \citenamefont {Ginsparg},\ and\ \citenamefont {Mueller}}]{kapit2012non}%
  \BibitemOpen
  \bibfield  {author} {\bibinfo {author} {\bibfnamefont {E.}~\bibnamefont
  {Kapit}}, \bibinfo {author} {\bibfnamefont {P.}~\bibnamefont {Ginsparg}},\
  and\ \bibinfo {author} {\bibfnamefont {E.}~\bibnamefont {Mueller}},\ }\href
  {https://doi.org/10.1103/PhysRevLett.108.066802} {\bibfield  {journal}
  {\bibinfo  {journal} {Physical Review Letters}\ }\textbf {\bibinfo {volume}
  {108}},\ \bibinfo {pages} {066802} (\bibinfo {year} {2012})}\BibitemShut
  {NoStop}%
\bibitem [{\citenamefont {Yuan}\ \emph {et~al.}(2017)\citenamefont {Yuan},
  \citenamefont {Xiao}, \citenamefont {Xu},\ and\ \citenamefont
  {Fan}}]{yuan2017creating}%
  \BibitemOpen
  \bibfield  {author} {\bibinfo {author} {\bibfnamefont {L.}~\bibnamefont
  {Yuan}}, \bibinfo {author} {\bibfnamefont {M.}~\bibnamefont {Xiao}}, \bibinfo
  {author} {\bibfnamefont {S.}~\bibnamefont {Xu}},\ and\ \bibinfo {author}
  {\bibfnamefont {S.}~\bibnamefont {Fan}},\ }\href
  {https://doi.org/10.1103/PhysRevA.96.043864} {\bibfield  {journal} {\bibinfo
  {journal} {Phys. Rev. A}\ }\textbf {\bibinfo {volume} {96}},\ \bibinfo
  {pages} {043864} (\bibinfo {year} {2017})}\BibitemShut {NoStop}%
\bibitem [{\citenamefont {Carusotto}\ and\ \citenamefont
  {Ciuti}(2013)}]{carusotto2013quantum}%
  \BibitemOpen
  \bibfield  {author} {\bibinfo {author} {\bibfnamefont {I.}~\bibnamefont
  {Carusotto}}\ and\ \bibinfo {author} {\bibfnamefont {C.}~\bibnamefont
  {Ciuti}},\ }\href {https://doi.org/10.1103/RevModPhys.85.299} {\bibfield
  {journal} {\bibinfo  {journal} {Reviews of Modern Physics}\ }\textbf
  {\bibinfo {volume} {85}},\ \bibinfo {pages} {299} (\bibinfo {year}
  {2013})}\BibitemShut {NoStop}%
\bibitem [{\citenamefont {Luo}\ \emph {et~al.}(2015)\citenamefont {Luo},
  \citenamefont {Zhou}, \citenamefont {Li}, \citenamefont {Xu}, \citenamefont
  {Guo},\ and\ \citenamefont {Zhou}}]{luo2015quantum}%
  \BibitemOpen
  \bibfield  {author} {\bibinfo {author} {\bibfnamefont {X.-W.}\ \bibnamefont
  {Luo}}, \bibinfo {author} {\bibfnamefont {X.}~\bibnamefont {Zhou}}, \bibinfo
  {author} {\bibfnamefont {C.-F.}\ \bibnamefont {Li}}, \bibinfo {author}
  {\bibfnamefont {J.-S.}\ \bibnamefont {Xu}}, \bibinfo {author} {\bibfnamefont
  {G.-C.}\ \bibnamefont {Guo}},\ and\ \bibinfo {author} {\bibfnamefont {Z.-W.}\
  \bibnamefont {Zhou}},\ }\href {https://doi.org/10.1038/ncomms8704} {\bibfield
   {journal} {\bibinfo  {journal} {Nature communications}\ }\textbf {\bibinfo
  {volume} {6}},\ \bibinfo {pages} {7704} (\bibinfo {year} {2015})}\BibitemShut
  {NoStop}%
\bibitem [{\citenamefont {Zhou}\ \emph {et~al.}(2017)\citenamefont {Zhou},
  \citenamefont {Luo}, \citenamefont {Wang}, \citenamefont {Guo}, \citenamefont
  {Zhou}, \citenamefont {Pu},\ and\ \citenamefont
  {Zhou}}]{zhou2017dynamically}%
  \BibitemOpen
  \bibfield  {author} {\bibinfo {author} {\bibfnamefont {X.-F.}\ \bibnamefont
  {Zhou}}, \bibinfo {author} {\bibfnamefont {X.-W.}\ \bibnamefont {Luo}},
  \bibinfo {author} {\bibfnamefont {S.}~\bibnamefont {Wang}}, \bibinfo {author}
  {\bibfnamefont {G.-C.}\ \bibnamefont {Guo}}, \bibinfo {author} {\bibfnamefont
  {X.}~\bibnamefont {Zhou}}, \bibinfo {author} {\bibfnamefont {H.}~\bibnamefont
  {Pu}},\ and\ \bibinfo {author} {\bibfnamefont {Z.-W.}\ \bibnamefont {Zhou}},\
  }\href {https://doi.org/10.1103/PhysRevLett.118.083603} {\bibfield  {journal}
  {\bibinfo  {journal} {Physical review letters}\ }\textbf {\bibinfo {volume}
  {118}},\ \bibinfo {pages} {083603} (\bibinfo {year} {2017})}\BibitemShut
  {NoStop}%
\bibitem [{\citenamefont {Cardano}\ \emph {et~al.}(2015)\citenamefont
  {Cardano}, \citenamefont {Massa}, \citenamefont {Qassim}, \citenamefont
  {Karimi}, \citenamefont {Slussarenko}, \citenamefont {Paparo}, \citenamefont
  {de~Lisio}, \citenamefont {Sciarrino}, \citenamefont {Santamato},
  \citenamefont {Boyd} \emph {et~al.}}]{cardano2015quantum}%
  \BibitemOpen
  \bibfield  {author} {\bibinfo {author} {\bibfnamefont {F.}~\bibnamefont
  {Cardano}}, \bibinfo {author} {\bibfnamefont {F.}~\bibnamefont {Massa}},
  \bibinfo {author} {\bibfnamefont {H.}~\bibnamefont {Qassim}}, \bibinfo
  {author} {\bibfnamefont {E.}~\bibnamefont {Karimi}}, \bibinfo {author}
  {\bibfnamefont {S.}~\bibnamefont {Slussarenko}}, \bibinfo {author}
  {\bibfnamefont {D.}~\bibnamefont {Paparo}}, \bibinfo {author} {\bibfnamefont
  {C.}~\bibnamefont {de~Lisio}}, \bibinfo {author} {\bibfnamefont
  {F.}~\bibnamefont {Sciarrino}}, \bibinfo {author} {\bibfnamefont
  {E.}~\bibnamefont {Santamato}}, \bibinfo {author} {\bibfnamefont {R.~W.}\
  \bibnamefont {Boyd}}, \emph {et~al.},\ }\href
  {https://doi.org/10.1126/sciadv.1500087} {\bibfield  {journal} {\bibinfo
  {journal} {Science advances}\ }\textbf {\bibinfo {volume} {1}},\ \bibinfo
  {pages} {e1500087} (\bibinfo {year} {2015})}\BibitemShut {NoStop}%
\bibitem [{\citenamefont {Cardano}\ \emph {et~al.}(2017)\citenamefont
  {Cardano}, \citenamefont {D’Errico}, \citenamefont {Dauphin}, \citenamefont
  {Maffei}, \citenamefont {Piccirillo}, \citenamefont {de~Lisio}, \citenamefont
  {De~Filippis}, \citenamefont {Cataudella}, \citenamefont {Santamato},
  \citenamefont {Marrucci} \emph {et~al.}}]{cardano2017detection}%
  \BibitemOpen
  \bibfield  {author} {\bibinfo {author} {\bibfnamefont {F.}~\bibnamefont
  {Cardano}}, \bibinfo {author} {\bibfnamefont {A.}~\bibnamefont {D’Errico}},
  \bibinfo {author} {\bibfnamefont {A.}~\bibnamefont {Dauphin}}, \bibinfo
  {author} {\bibfnamefont {M.}~\bibnamefont {Maffei}}, \bibinfo {author}
  {\bibfnamefont {B.}~\bibnamefont {Piccirillo}}, \bibinfo {author}
  {\bibfnamefont {C.}~\bibnamefont {de~Lisio}}, \bibinfo {author}
  {\bibfnamefont {G.}~\bibnamefont {De~Filippis}}, \bibinfo {author}
  {\bibfnamefont {V.}~\bibnamefont {Cataudella}}, \bibinfo {author}
  {\bibfnamefont {E.}~\bibnamefont {Santamato}}, \bibinfo {author}
  {\bibfnamefont {L.}~\bibnamefont {Marrucci}}, \emph {et~al.},\ }\href
  {https://doi.org/10.1038/ncomms15516} {\bibfield  {journal} {\bibinfo
  {journal} {Nature communications}\ }\textbf {\bibinfo {volume} {8}},\
  \bibinfo {pages} {15516} (\bibinfo {year} {2017})}\BibitemShut {NoStop}%
\bibitem [{\citenamefont {Wang}\ \emph {et~al.}(2018)\citenamefont {Wang},
  \citenamefont {Chen},\ and\ \citenamefont {Zhang}}]{wang2018experimental}%
  \BibitemOpen
  \bibfield  {author} {\bibinfo {author} {\bibfnamefont {B.}~\bibnamefont
  {Wang}}, \bibinfo {author} {\bibfnamefont {T.}~\bibnamefont {Chen}},\ and\
  \bibinfo {author} {\bibfnamefont {X.}~\bibnamefont {Zhang}},\ }\href
  {https://doi.org/10.1103/PhysRevLett.121.100501} {\bibfield  {journal}
  {\bibinfo  {journal} {Physical Review Letters}\ }\textbf {\bibinfo {volume}
  {121}},\ \bibinfo {pages} {100501} (\bibinfo {year} {2018})}\BibitemShut
  {NoStop}%
\bibitem [{\citenamefont {Yang}\ \emph
  {et~al.}(2022{\natexlab{b}})\citenamefont {Yang}, \citenamefont {Zhang},
  \citenamefont {Liao}, \citenamefont {Liu}, \citenamefont {Zhou},
  \citenamefont {Zhou}, \citenamefont {Xu}, \citenamefont {Han}, \citenamefont
  {Li},\ and\ \citenamefont {Guo}}]{yang2022topological}%
  \BibitemOpen
  \bibfield  {author} {\bibinfo {author} {\bibfnamefont {M.}~\bibnamefont
  {Yang}}, \bibinfo {author} {\bibfnamefont {H.-Q.}\ \bibnamefont {Zhang}},
  \bibinfo {author} {\bibfnamefont {Y.-W.}\ \bibnamefont {Liao}}, \bibinfo
  {author} {\bibfnamefont {Z.-H.}\ \bibnamefont {Liu}}, \bibinfo {author}
  {\bibfnamefont {Z.-W.}\ \bibnamefont {Zhou}}, \bibinfo {author}
  {\bibfnamefont {X.-X.}\ \bibnamefont {Zhou}}, \bibinfo {author}
  {\bibfnamefont {J.-S.}\ \bibnamefont {Xu}}, \bibinfo {author} {\bibfnamefont
  {Y.-J.}\ \bibnamefont {Han}}, \bibinfo {author} {\bibfnamefont {C.-F.}\
  \bibnamefont {Li}},\ and\ \bibinfo {author} {\bibfnamefont {G.-C.}\
  \bibnamefont {Guo}},\ }\href {https://doi.org/10.1038/s41467-022-29779-3}
  {\bibfield  {journal} {\bibinfo  {journal} {Nature Communications}\ }\textbf
  {\bibinfo {volume} {13}},\ \bibinfo {pages} {2040} (\bibinfo {year}
  {2022}{\natexlab{b}})}\BibitemShut {NoStop}%
\bibitem [{\citenamefont {Yang}\ \emph
  {et~al.}(2023{\natexlab{a}})\citenamefont {Yang}, \citenamefont {Zhang},
  \citenamefont {Liao}, \citenamefont {Liu}, \citenamefont {Zhou},
  \citenamefont {Zhou}, \citenamefont {Xu}, \citenamefont {Han}, \citenamefont
  {Li},\ and\ \citenamefont {Guo}}]{yang2023realization}%
  \BibitemOpen
  \bibfield  {author} {\bibinfo {author} {\bibfnamefont {M.}~\bibnamefont
  {Yang}}, \bibinfo {author} {\bibfnamefont {H.-Q.}\ \bibnamefont {Zhang}},
  \bibinfo {author} {\bibfnamefont {Y.-W.}\ \bibnamefont {Liao}}, \bibinfo
  {author} {\bibfnamefont {Z.-H.}\ \bibnamefont {Liu}}, \bibinfo {author}
  {\bibfnamefont {Z.-W.}\ \bibnamefont {Zhou}}, \bibinfo {author}
  {\bibfnamefont {X.-X.}\ \bibnamefont {Zhou}}, \bibinfo {author}
  {\bibfnamefont {J.-S.}\ \bibnamefont {Xu}}, \bibinfo {author} {\bibfnamefont
  {Y.-J.}\ \bibnamefont {Han}}, \bibinfo {author} {\bibfnamefont {C.-F.}\
  \bibnamefont {Li}},\ and\ \bibinfo {author} {\bibfnamefont {G.-C.}\
  \bibnamefont {Guo}},\ }\href {https://doi.org/10.1126/sciadv.abp8943}
  {\bibfield  {journal} {\bibinfo  {journal} {Science Advances}\ }\textbf
  {\bibinfo {volume} {9}},\ \bibinfo {pages} {eabp8943} (\bibinfo {year}
  {2023}{\natexlab{a}})}\BibitemShut {NoStop}%
\bibitem [{\citenamefont {Zhang}\ \emph
  {et~al.}(2022{\natexlab{b}})\citenamefont {Zhang}, \citenamefont {Zhao},
  \citenamefont {Wu}, \citenamefont {Wu}, \citenamefont {Qiao}, \citenamefont
  {Gao}, \citenamefont {Agarwal}, \citenamefont {Longhi}, \citenamefont
  {Litchinitser}, \citenamefont {Ge} \emph {et~al.}}]{zhang2022spin}%
  \BibitemOpen
  \bibfield  {author} {\bibinfo {author} {\bibfnamefont {Z.}~\bibnamefont
  {Zhang}}, \bibinfo {author} {\bibfnamefont {H.}~\bibnamefont {Zhao}},
  \bibinfo {author} {\bibfnamefont {S.}~\bibnamefont {Wu}}, \bibinfo {author}
  {\bibfnamefont {T.}~\bibnamefont {Wu}}, \bibinfo {author} {\bibfnamefont
  {X.}~\bibnamefont {Qiao}}, \bibinfo {author} {\bibfnamefont {Z.}~\bibnamefont
  {Gao}}, \bibinfo {author} {\bibfnamefont {R.}~\bibnamefont {Agarwal}},
  \bibinfo {author} {\bibfnamefont {S.}~\bibnamefont {Longhi}}, \bibinfo
  {author} {\bibfnamefont {N.~M.}\ \bibnamefont {Litchinitser}}, \bibinfo
  {author} {\bibfnamefont {L.}~\bibnamefont {Ge}}, \emph {et~al.},\ }\href
  {https://doi.org/10.1038/s41586-022-05339-z} {\bibfield  {journal} {\bibinfo
  {journal} {Nature}\ }\textbf {\bibinfo {volume} {612}},\ \bibinfo {pages}
  {246} (\bibinfo {year} {2022}{\natexlab{b}})}\BibitemShut {NoStop}%
\bibitem [{\citenamefont {Sounas}\ and\ \citenamefont
  {Al{\`u}}(2017)}]{sounas2017non}%
  \BibitemOpen
  \bibfield  {author} {\bibinfo {author} {\bibfnamefont {D.~L.}\ \bibnamefont
  {Sounas}}\ and\ \bibinfo {author} {\bibfnamefont {A.}~\bibnamefont
  {Al{\`u}}},\ }\href {https://doi.org/10.1038/s41566-017-0051-x} {\bibfield
  {journal} {\bibinfo  {journal} {Nat. Photon.}\ }\textbf {\bibinfo {volume}
  {11}},\ \bibinfo {pages} {774} (\bibinfo {year} {2017})}\BibitemShut
  {NoStop}%
\bibitem [{\citenamefont {Guo}\ \emph {et~al.}(2019)\citenamefont {Guo},
  \citenamefont {Ding}, \citenamefont {Duan},\ and\ \citenamefont
  {Ni}}]{guo2019nonreciprocal}%
  \BibitemOpen
  \bibfield  {author} {\bibinfo {author} {\bibfnamefont {X.}~\bibnamefont
  {Guo}}, \bibinfo {author} {\bibfnamefont {Y.}~\bibnamefont {Ding}}, \bibinfo
  {author} {\bibfnamefont {Y.}~\bibnamefont {Duan}},\ and\ \bibinfo {author}
  {\bibfnamefont {X.}~\bibnamefont {Ni}},\ }\href
  {https://doi.org/10.1038/s41377-019-0225-z} {\bibfield  {journal} {\bibinfo
  {journal} {Light: Science \& Applications}\ }\textbf {\bibinfo {volume}
  {8}},\ \bibinfo {pages} {123} (\bibinfo {year} {2019})}\BibitemShut {NoStop}%
\bibitem [{\citenamefont {Yang}\ \emph
  {et~al.}(2023{\natexlab{b}})\citenamefont {Yang}, \citenamefont {Qin},
  \citenamefont {Long}, \citenamefont {Yan}, \citenamefont {Yang},
  \citenamefont {Li}, \citenamefont {Li}, \citenamefont {Hu}, \citenamefont
  {Deng}, \citenamefont {Du} \emph {et~al.}}]{yang2023self}%
  \BibitemOpen
  \bibfield  {author} {\bibinfo {author} {\bibfnamefont {W.}~\bibnamefont
  {Yang}}, \bibinfo {author} {\bibfnamefont {J.}~\bibnamefont {Qin}}, \bibinfo
  {author} {\bibfnamefont {J.}~\bibnamefont {Long}}, \bibinfo {author}
  {\bibfnamefont {W.}~\bibnamefont {Yan}}, \bibinfo {author} {\bibfnamefont
  {Y.}~\bibnamefont {Yang}}, \bibinfo {author} {\bibfnamefont {C.}~\bibnamefont
  {Li}}, \bibinfo {author} {\bibfnamefont {E.}~\bibnamefont {Li}}, \bibinfo
  {author} {\bibfnamefont {J.}~\bibnamefont {Hu}}, \bibinfo {author}
  {\bibfnamefont {L.}~\bibnamefont {Deng}}, \bibinfo {author} {\bibfnamefont
  {Q.}~\bibnamefont {Du}}, \emph {et~al.},\ }\href
  {https://doi.org/10.1038/s41928-023-00936-w} {\bibfield  {journal} {\bibinfo
  {journal} {Nature Electronics}\ }\textbf {\bibinfo {volume} {6}},\ \bibinfo
  {pages} {225} (\bibinfo {year} {2023}{\natexlab{b}})}\BibitemShut {NoStop}%
\end{thebibliography}%

\end{document}